\documentclass[12pt]{article}
\usepackage{titling}
\usepackage{amsmath}
\usepackage{slashed}
\usepackage{amssymb}
\usepackage{epsfig}
\usepackage{graphicx}
\usepackage{multirow}
\usepackage{color}
\usepackage{subcaption}
\usepackage{rotating}
\usepackage{hyperref}
\usepackage[margin=1.0in]{geometry}
\usepackage[table]{xcolor}
\usepackage{enumitem}
\usepackage[utf8x]{inputenc}
\usepackage[compress,numbers,sort]{natbib}
\usepackage{authblk}
\usepackage{colortbl}
\usepackage{pdflscape}
\usepackage{color}
\usepackage{float}
\usepackage{siunitx}
\usepackage[compat=1.1.0]{tikz-feynman}
\usepackage{lscape}
\usepackage{mathtools}
\usepackage{amsfonts}
\usepackage{bbold}
\usepackage{soul}

\definecolor{nicered}{rgb}{0.6,0.1,0.1}
\definecolor{nicegreen}{rgb}{0.1,0.5,0.1}
\definecolor{mediumcandyapplered}{rgb}{0.99, 0.12, 0.07}
\definecolor{red}{rgb}{1.0, 0, 0}
\hypersetup{colorlinks,citecolor= nicegreen,linkcolor= nicered}

\renewcommand{\bar}{\overline}

\definecolor{LightCyan}{rgb}{0.88,1,1}
\definecolor{piggypink}{rgb}{0.99, 0.87, 0.9}
\definecolor{applegreen}{rgb}{0.55, 0.71, 0.0}
\definecolor{darkpastelgreen}{rgb}{0.01, 0.75, 0.24}
\definecolor{green-yellow}{rgb}{0.68, 1.0, 0.18}

\newcommand{\beq}{\begin{equation}}
\newcommand{\eeq}{\end{equation}}
\newcommand{\bea}{\begin{eqnarray}}
\newcommand{\eea}{\end{eqnarray}}
\newcommand{\lUQone}{\ensuremath{\lambda_{UQ_1}}}
\newcommand{\lU}{\ensuremath{\lambda_U}}
\newcommand{\luQone}{\ensuremath{\lambda^u_{Q_1}}}
\newcommand{\ldQone}{\ensuremath{\lambda^d_{Q_1}}}

\newcommand{\cC}{\mathcal{C}}
\newcommand{\cO}{\mathcal{O}}

\title{\bf{How large can the Light Quark Yukawa couplings be?}}

\author[1,2]
{Barbara Anna Erdelyi \thanks{barbaraanna.erdelyi@pd.infn.it}}
\author[1,2]
{Ramona Gr\"ober \thanks{ramona.groeber@pd.infn.it}}
\author[2]
{Nud\v zeim Selimovi\'c \thanks{nudzeim.selimovic@pd.infn.it}}

\affil[1]{\emph{\normalsize Dipartimento di Fisica e Astronomia ``G. Galilei'', Universit\`a di Padova, Padova, Italy}}
\affil[2]{\emph{\normalsize Istituto Nazionale di Fisica Nucleare, Sezione di Padova, Padova, Italy}}

\date{}

\begin{document}

\maketitle
\begin{abstract}
\normalsize
We investigate models that can induce significant modifications to the couplings of first- and second-generation quarks with Higgs bosons. Specifically, we identify all simplified models featuring two vector-like quark states which can lead to substantial enhancements in these couplings.
In addition, these models generate operators in Standard Model Effective Field Theory, both at tree-level and one-loop, that are constrained by electroweak precision and Higgs data.
We show how to evade constraints from flavour physics and consider direct searches for vector-like quarks. Ultimately, we demonstrate that viable ultraviolet models can be found with first-generation quark Yukawa couplings enhanced by several hundred times their Standard Model value, while the Higgs couplings to charm (strange) quarks can be increased by factors of a few (few tens). 
Given the importance of electroweak precision data in constraining these models, we also discuss projections for future measurements at the Tera-$Z$ FCC-ee machine. 
\end{abstract}

\clearpage

{
  \hypersetup{linkcolor=black}
  \tableofcontents
}

\clearpage

\section{Introduction}
Since the landmark discovery of the Higgs boson at the Large Hadron Collider in 2012 \cite{ATLAS:2012yve,CMS:2012qbp}, significant progress has been made in understanding the properties and interactions of this fundamental particle \cite{ATLAS:2022vkf,CMS:2022dwd}. Despite these advancements, some of the Higgs boson couplings remain elusive, as is the case for the Yukawa couplings of the Higgs boson to the first two generations. While the third-generation couplings have been observed and studied, significant challenges remain in probing the Higgs couplings to the first and second-generation fermions, including the recently measured Higgs coupling to muons \cite{ATLAS:2020fzp,CMS:2020xwi}. This is on the one hand due to their smallness but for what regards the quarks also due to the difficulty to tag them. 
\par
Nevertheless, for the charm quark Yukawa coupling, it is possible to set bounds from $Vh$ production with subsequent decay $h\to c \bar{c}$ and tagging the final state charm quarks \cite{ATLAS:2022ers, CMS:2019hve}.\footnote{In Refs.~\cite{Delaunay:2013pja, Perez:2015aoa, Perez:2015lra} it was pointed out that combined signal strengths of Higgs decays to bottom and charm could be helpful to constrain the charm Yukawa coupling.} Doing so allows setting a limit on the charm quark mass of $|\kappa_c/\kappa_b|<4.5$ \cite{ATLAS:2022ers} with $\kappa_f=y_f/y_f^{\rm SM}$ where $y_f$ stands for the coupling of the quark $f$ to the Higgs boson and $y_f^{\rm SM}$ is its Standard Model (SM) value. This bound is slightly stronger than the ratio between charm and bottom mass, indicating indeed that the second-generation quark Yukawa couplings are weaker than the bottom Yukawa coupling.
Additional probes of the charm Yukawa coupling are exclusive Higgs decays to vector mesons \cite{Bodwin:2013gca, Kagan:2014ila, Alte:2016yuw} that have been used to constrain the charm Yukawa coupling in Refs.~\cite{ATLAS:2018xfc,CMS:2022fsq}. The change in the $p_T$ spectrum as proposed in Ref.~\cite{Bishara:2016jga} was used in \cite{ATLAS:2022fnp} in $h\to \gamma\gamma$, to set a bound on $\kappa_c\in [-2.5,2.3]$ mostly stemming though from normalisation of the rate.  Other proposals include Higgs+charm quark production \cite{Brivio:2015fxa,ATLAS:2024ext} or $VVcj$ production \cite{Vignaroli:2022fqh}. The strange Yukawa coupling is even more challenging but strange tagging might be possible at future $e^+e^-$ colliders \cite{Duarte-Campderros:2018ouv,Kamenik:2023hvi}, as well as identification of the jet origin allowing to obtain bounds on Higgs decays to first generation quarks \cite{Liang:2023yyi}. Moreover, by angularity measurements in Higgs decays, light quarks can be distinguished from gluons~\cite{Yan:2023xsd}.

In order to measure the first generation quark Yukawa couplings, one can either make use of the fact that the untagged branching ratio of the Higgs boson changes in the presence of strongly enhanced light quark Yukawa couplings \cite{Carpenter:2016mwd}, hence constraining them by a global fit which leads to the HL-LHC projections of $|\kappa_u|<560$ and $|\kappa_d|<260$~\cite{deBlas:2019rxi}\footnote{For the strange and charm Yukawa couplings these projections give $|\kappa_s|<13$ and $|\kappa_c|<1.2$.}, or use that in case of an enhancement the Higgs boson might be produced directly from the incoming partons. The latter possibility has been explored in the context of the Higgs $p_T$-spectrum~\cite{Soreq:2016rae}, the $W^{\pm} h$ charge asymmetry~\cite{Yu:2017vul, Yu:2016rvv}, $h+\gamma$ production~\cite{Aguilar-Saavedra:2020rgo}, tri-vector boson production in the high-energy limit~\cite{Falkowski:2020znk, CMS:2023rcv}, Higgs pair production \cite{Alasfar:2019pmn, Alasfar:2022vqw}, and the off-shell Higgs measurement~\cite{Zhou:2015wra, Balzani:2023jas}. The latter reference shows the most promising projections for the down (up) Yukawa coupling, constraining them up to a factor of 156 (260) times the Standard Model value at the high-luminosity LHC assuming only experimental systematic uncertainties. Recently, a first analysis has been performed for light quark Yukawa couplings in $h\to 4 \ell$, ruling out the hypothesis that the first or second generation have the same Yukawa couplings as those in the third generation, with a CL greater than 95\% \cite{CMS-PAS-HIG-23-011}.
\par
Given that even at the HL-LHC the projections of the light quark Yukawa couplings are, in particular for the first generations, still orders of magnitude away from the SM prediction, we will address in this work the question on \textit{how large the light quark Yukawa couplings can be} considering concrete UV scenarios in terms of simplified models. We identify and study \emph{all possible} models with minimal field content consisting of two extra vector-like quark (VLQ) states that provide unsuppressed contributions to the Yukawa couplings.
We check that under the light of various constraints from electroweak precision tests, Higgs data, direct searches, and flavour physics, it is still possible to obtain naturally large deviations in the Yukawa couplings of the light generations, motivating their further probes.

The new quark states which we introduce are vector-like under the Standard Model (SM) gauge group, allowing their mass generation to be decoupled from the Higgs mechanism. Consequently, their mass can naturally be larger than the electroweak symmetry-breaking scale, enabling us to treat their effects in the framework of Standard Model Effective Field Theory (SMEFT). Such particles are ubiquitous across the broad beyond the Standard Model landscape, appearing in various models addressing the hierarchy problem~\cite{Arkani-Hamed:2002ikv,Han:2003wu, Agashe:2004rs,Contino:2006qr,Anastasiou:2009rv,Matsedonskyi:2012ym,Delaunay:2013iia,Gillioz:2013pba}, the flavour puzzle~\cite{Bordone:2017bld,Alonso:2018bcg,Fuentes-Martin:2022xnb}, the strong CP problem~\cite{PhysRevLett.53.329, Bento:1991ez}, as well as in the models explaining the origin of CP violation~\cite{BENTO1990599}. Moreover, vector-like quarks appear in supersymmetric theories~\cite{Kang:2007ib,Martin:2009bg,Martin:2010dc,Fischler:2013tva} and as Kaluza-Klein modes in models with extra dimensions~\cite{Randall:1999ee,Contino:2006nn, Carena:2007ua}. 
Given their commonness in various new physics (NP) scenarios, it is worth exploring their low-energy effects in the context of their significant contributions to the light quark Yukawa couplings. As these couplings are predicted to be extremely small in the SM, the relative deviation in the new physics models could be extremely large, making light quark Yukawa couplings a promising place to look for the imprints of vector-like quarks.
\par
The paper is structured as follows: in Sec.~\ref{sec:EFT_models} we present the simplified models with vector-like quarks and how they enhance the light quark Yukawa couplings. The phenomenological implications for flavour, LHC, electroweak, and Higgs physics are studied in Sec.~\ref{sec:constraints}. The results answering the question posed in the title are discussed in Sec.~\ref{sec:results}, while their future projections for the FCC-ee machine are shown in Sec.~\ref{sec:future}. We conclude in Sec.~\ref{sec:conclusion}. Technical details about the one-loop matching procedure, electroweak fit, and branching ratios for the various states that determine the direct search bounds are discussed in Apps.~\ref{app:CHD}--\ref{app:GeneralisedBR}.

\section{Effective Field Theory and Models}
\label{sec:EFT_models}
In this section, we identify and discuss the minimal SM extensions that enhance the effective couplings of light quarks to the Higgs boson. To evade the experimental limits on the new states, we assume a scale separation between the electroweak scale and the masses of new particles, encoding new physics effects through Standard Model Effective Field Theory (SMEFT). In SMEFT, all possible higher-dimensional operators respecting the SM symmetries are added to the SM Lagrangian $\mathcal{L}_{SM}$, such that the SMEFT Lagrangian is typically written as 
\begin{equation}
    \mathcal{L}_{\rm SMEFT}=\mathcal{L}_{\rm SM}+\sum_{d=5}^{\infty} \frac{\hat{\cC}_k^{(d)}}{\Lambda^{d-4}} \cO_k^{(d)}\,.
\end{equation}
Here, $d$ is the mass dimension of the operator $\cO_k^{(d)}$ associated with the dimensionless Wilson coefficient $\hat{\cC}_k^{(d)}$, and the subscript $k$ runs through all independent operators at a given dimension. The operators are suppressed by the energy scale $\Lambda$ reflecting the effective mass of the new states and the validity range for the EFT description. However, for our study it proved to be more convenient to adopt an alternative notation in which the energy scale is factored into the Wilson coefficients $\cC^{(d)}_k$, leading to the following expression for the SMEFT Lagrangian
\begin{equation}
    \mathcal{L}_{\rm SMEFT}=\mathcal{L}_{\rm SM}+\sum_{d=5}^{\infty} \cC_k^{(d)} \cO_k^{(d)}\,,\quad \text{where} \quad \left[\cC_k^{(d)}\right]=4-d\,.
\end{equation}
To set our notation, we write the SM Lagrangian as
\begin{align}\label{eq:LagSM}
\mathcal{L}_{\rm SM} &=\, -\frac{1}{4}B_{\mu\nu}B^{\mu\nu} - \frac{1}{4}W_{\mu\nu}^I W^{I\mu\nu} - \frac{1}{4}G_{\mu\nu}^A G^{A\mu\nu}+D_\mu \phi^\dag D^\mu \phi -V(\phi)
\nonumber\\
&\phantom{aai}+\sum_{\psi} \bar\psi i\slashed{D}\psi-\left[y_d \,\bar{q}_L \phi d_R+ y_u\, \bar q_L  \tilde \phi  u_R +  y_e\,\bar  \ell_L \phi e_R+\text{h.c.}\right]\,,\\
V(\phi) &= -\mu^2 \phi^\dag \phi +\lambda (\phi^\dag \phi)^2\,,
\end{align}
where $G_{\mu\nu}$, $W_{\mu\nu}$, and $B_{\mu\nu}$ are the field strength tensors reflecting the gauge structure of the SM gauge group $G_{\rm SM}=SU(3)\times SU(2)\times U(1)$, respectively. The SM field content is denoted by $q_L$ for the left-handed quark fields, $\ell_L$ for the left-handed lepton fields, with $u_R$, $d_R$, and $e_R$ being the right-handed up-type, down-type, and lepton fields. The sum in $\psi$ runs over all these fermion fields. The Yukawa couplings $y_i$ with $i=u,d,e$ are complex $3\times 3$ matrices in flavour space encoding the interactions between fermions and the $SU(2)$ Higgs doublet $\phi$ ($\tilde{\phi}=i\sigma_2 \phi^*$) which develops a vacuum expectation value (vev), $v=|\mu|/\sqrt{\lambda}=0.246$ TeV.

The part of $\mathcal{L}_{\rm SMEFT}$ which describes the leading contribution to the effective Higgs-fermion interactions after the electroweak symmetry-breaking is encoded in $d=6$ terms. In the Warsaw basis \cite{Grzadkowski:2010es}, the modification to the effective quark Yukawa couplings is due to the following operators
\begin{equation}
\mathcal{O}_{u\phi}=\bar{q}_{L} \tilde{\phi}u_{R} \phi^{\dagger}\,\phi \quad\text{and} \quad \mathcal{O}_{d\phi}=\bar{q}_L \phi d_R \,\phi^{\dagger}\phi\,.
\label{eq:Ouphi}
\end{equation}
Together with the operators 
\begin{equation}
\mathcal{O}_{\phi\Box}= (\phi^{\dagger}\phi) \Box (\phi^{\dagger}\phi)\quad\text{and} \quad \mathcal{O}_{\phi D}=(\phi^{\dagger} D_{\mu} \phi)^* (\phi^{\dagger} D^{\mu} \phi)\,,
\label{eq:OphiBox}
\end{equation} 
which modify the kinetic term of the Higgs field requiring a field redefinition for a canonical normalisation
\begin{equation}
\phi=\frac{1}{\sqrt{2}}\left(\begin{matrix} 0 \\ v+h(1+ v^2 \cC_{\phi,{\rm kin}})\end{matrix}\right) \quad\text{with} \quad \cC_{\phi,{\rm kin}}=\left(\cC_{\phi\Box}-\frac{1}{4}C_{\phi D} \right),
\label{eq:C_kin}
\end{equation}
they modify the Higgs boson couplings to quarks as follows
\begin{equation}
-\mathcal{L}\supset g_{hq_{i}\bar{q}_{j}} \bar{q}_j q_i h
\quad\text{with}\quad
g_{hq_{i}\bar{q}_{j}}= \frac{m_q}{v}\delta_{ij} \left[ 1+ v^2 \cC_{\phi,{\rm kin}}\right]-\frac{v^2}{\sqrt{2}} {(\tilde{\cC}_{q\phi})_{ij}}\,. \label{eq:ghqq}
\end{equation}
We have introduced here the Wilson coefficients with tilde which are understood to be in the mass basis, noting that the definition of the up- and down-type masses in terms of the renormalisable SM Yukawa terms gets modified as
\begin{align}
&M_{i j}^u=\frac{v}{\sqrt{2}}\left((y_u)_{i j}-\frac{v^2}{2} (\cC_{u\phi})_{i j}\right), \\
&M_{i j}^d=\frac{v}{\sqrt{2}}\left((y_d)_{i j}-\frac{v^2}{2} (\cC_{d\phi})_{i j}\right) .
\end{align}
The mass matrices are diagonalized by a new set of bi-unitary transformations 
\begin{equation}
m_{q_i}=\left(\left(V_L^{u/d}\right)^\dagger\! M^{u/d}\,V_R^{u/d}\right)_{ii},
\end{equation}  
in which the CKM matrix is defined as $V=\left(V_L^u\right)^\dagger V_L^d$. One can then rewrite $ (\cC_{q\phi})_{i j}$ in terms of $(\tilde{\cC}_{q\phi})_{ij}$ which are now in the mass basis
\begin{equation}
(\tilde{\cC}_{q\phi})_{ij}=\left(V_L^q\right)_{n i}^* (\cC_{q\phi})_{n m} \left(V_R^q\right)_{m j}, \ \ \ \ \ \text{with} \ \ \ q=u, d\,.
\end{equation}
Finally, for the flavour diagonal case one can write Eq.~\eqref{eq:ghqq} in the $\kappa$-formalism as
\begin{equation}
g_{hq\bar{q}}= \kappa_q\, g_{hq\bar{q}}^{\rm SM}\,.
\end{equation}
Since we are considering the light quark Yukawa couplings one needs to pay attention to what exactly is meant by the SM value of the coupling. We will define it here and in the following in the limit of $\Lambda\to \infty$ and with respect to the reference mass values $m_u = 2.2\text{ MeV}$, $m_d = 4.7 
\text{ MeV}$, $m_c = 1.27\text{ GeV}$, and $m_s = 95
\text{ MeV}$ considered as constant, i.e. non-running, values.
We emphasize that using these mass values affects only the interpretation of $\kappa_q$ and does not alter our results, as we have set the kinematic masses, namely the masses in propagators, of the light quarks to zero. For instance, if our results were to be interpreted using running mass values at a scale such as the Higgs mass, then the $\kappa_q$ values would need to be rescaled by a factor of 0.55.
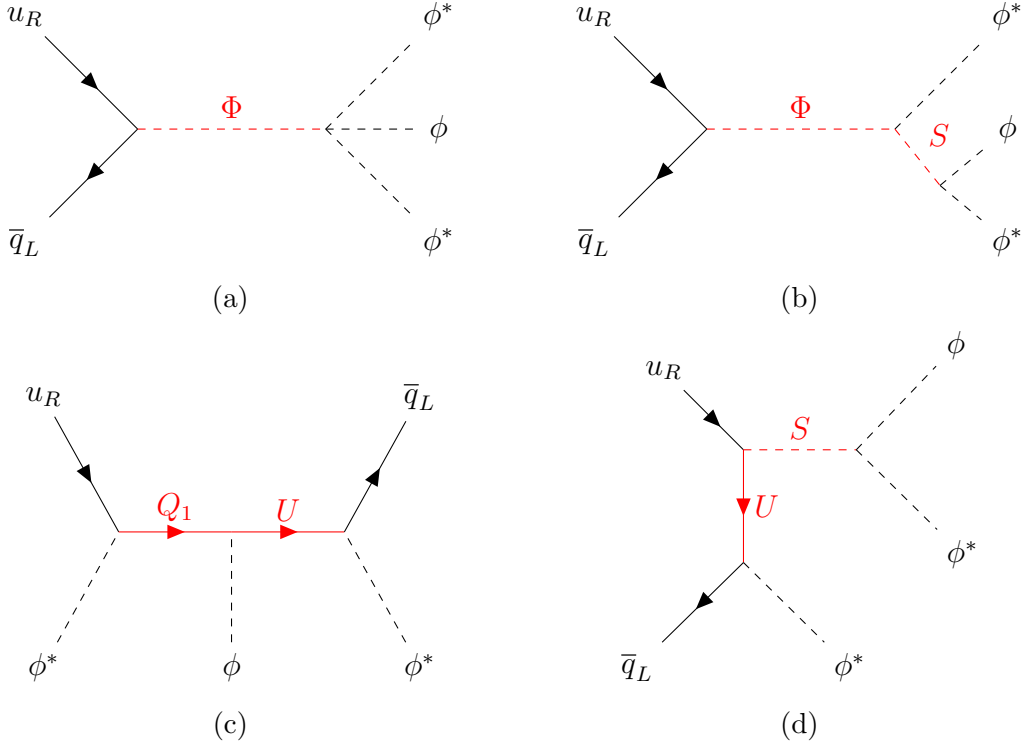
\begin{figure}[t]
\centering
\begin{subfigure}[t]{0.45\linewidth}\centering
\begin{tikzpicture}
    \begin{feynman}
        \vertex (a) at (0,0);
        \vertex(b) at (2.5,0);
        \vertex (qL) at (-1.5,-1.5) {\(\bar{q}_L\)};
        \vertex (uR) at (-1.5,1.5) {\(u_R\)};
        \vertex (phi1) at (4,1.5) {\(\phi^*\)};
        \vertex (phi2) at (4,0) {\(\phi\)};
        \vertex (phi3) at (4,-1.5) {\(\phi^*\)};
        \diagram*{
        (uR) -- [fermion] (a) -- [fermion] (qL),
        (a) -- [scalar, red, edge label = \(\Phi\)] (b);
        (b) -- [scalar] (phi1);
        (b) -- [scalar] (phi2);
        (b) -- [scalar] (phi3);
        };
    \end{feynman}
\end{tikzpicture}
\caption{}\label{subfig:UVmodelssecondHiggs}
\end{subfigure}
\begin{subfigure}[t]{0.45\linewidth}\centering
\begin{tikzpicture}
    \begin{feynman}
        \vertex (a) at (0,0);
        \vertex (b) at (2.5,0);
        \vertex (qL) at (-1.5,-1.5) {\(\bar{q}_L\)};
        \vertex (uR) at (-1.5,1.5) {\(u_R\)};
        \vertex (phi1) at (4,1.5) {\(\phi^*\)};
        \vertex (S) at (3.1,-0.75);
        \vertex (phi2) at (4,-0.) {\(\phi\)};
        \vertex (phi3) at (4,-1.5) {\(\phi^*\)};
        \diagram*{
        (uR) -- [fermion] (a) -- [fermion] (qL),
        (a) -- [scalar, red, edge label = \(\Phi\)] (b);
        (b) -- [scalar] (phi1);
        (b) -- [scalar, red, edge label= \(S\)] (S);
        (S) -- [scalar] (phi2);
        (S) -- [scalar] (phi3);
        };
    \end{feynman}
\end{tikzpicture}
\caption{}\label{subfig:UVmodelssecondHiggsandscalar}
\end{subfigure}\\
\begin{subfigure}[b]{0.45\linewidth}\centering
\begin{tikzpicture}
    \begin{feynman}
	   \vertex (lamuQ1) at (0,0);
	   \vertex (phi1) at (-1,-1.8) {\(\phi^*\)};
	   \vertex (lamUQ1) at (1.5,0);
	   \vertex (phi2) at (1.5,-1.8) {\(\phi\)};
	   \vertex (lamUq) at (3,0);
	   \vertex (phi3) at (4,-1.8) {\(\phi^*\)};
	   \vertex (uR) at (-1,1.8) {\(u_R\)};
	   \vertex (qL) at (4,1.8) {\(\bar{q}_L\)};
       \diagram*{
		  (phi1) -- [scalar] (lamuQ1);
		  (phi2) -- [scalar] (lamUQ1);
		  (phi3) -- [scalar] (lamUq);
		  (uR) -- [fermion] (lamuQ1) -- [fermion, red, edge label = \(Q_1\)] (lamUQ1) -- [fermion, red, edge label = \(U\)] (lamUq) -- [fermion] (qL);
				};
		\end{feynman}			
\end{tikzpicture}
\caption{}\label{subfig:UVmodelsVLQpairs}
\end{subfigure}
\begin{subfigure}[b]{0.45\linewidth}\centering
\begin{tikzpicture}
    \begin{feynman}
        \vertex (uR) at (0,0) {\(u_R\)};
        \vertex[below right= of uR] (v1);
        \vertex[right= of v1] (v2);
        \vertex[above right= of v2] (phi) {\(\phi\)};
        \vertex[below right= of v2] (phistar) {\(\phi^*\)};
        \vertex[below= of v1] (v3);
        \vertex[below left= of v3] (qL) {\(\bar{q}_L\)};
        \vertex[below right = of v3] (phistar2) {\(\phi^*\)};
        \diagram*{
        (uR) -- [fermion] (v1) -- [fermion, red, edge label = \(U\)] (v3) -- [fermion] (qL);
        (v1) -- [scalar, red, edge label = \(S\)] (v2) -- [scalar] (phi);
        (v2) -- [scalar] (phistar);
        (v3) -- [scalar] (phistar2);
        };
    \end{feynman}
\end{tikzpicture} 
\caption{}\label{subfig:UVmodelsVLQandscalar}
\end{subfigure}
\caption{Feynman diagrams showing how new states can generate an operator of type $\bar{q}_{L} \tilde{\phi}u_{R} \,\phi^{\dagger}\phi$ at tree level. \textbf{(a)} A second Higgs doublet $\Phi$. \textbf{(b)} A second Higgs doublet $\Phi$ and a scalar singlet or triplet $S$. \textbf{(c)} A pair of VLQs $Q_1$ and $U$ as in Model 1. \textbf{(d)} A scalar $S$ and a VLQ $U$.}
\label{fig:UVmodel}
\end{figure}

As can be seen from Eq.~\eqref{eq:ghqq}, for the new states to influence the effective light fermion-Higgs couplings, they need to contribute to the operators $\mathcal{O}_{u\phi}$ and $\mathcal{O}_{d\phi}$ in Eq.~\eqref{eq:Ouphi}. Interestingly, there are only a couple of minimal SM extensions with at most two new states that achieve this goal at tree level. Looking at the tree-level matching dictionary describing the general extensions of the Standard Model provided by~\cite{deBlas:2017xtg}, we identify three types of models that generate the required operators at tree-level not suppressed by the light quark Yukawa coupling. \\

Those models are characterised by
\begin{itemize}
    \item Additional scalar fields: this can be either provided by a single new multiplet, namely a second Higgs doublet, which has been studied in~\cite{Egana-Ugrinovic:2019dqu, Giannakopoulou:2024unn} in the context of enhanced light quark Yukawa couplings, or models with a second Higgs doublet plus an additional scalar field (singlet or triplet). The mechanism by which new scalars could contribute to $g_{hq\bar q}$ is shown in Fig.~\ref{subfig:UVmodelssecondHiggs} and~\ref{subfig:UVmodelssecondHiggsandscalar}; 
    \item Models with two additional representations of VLQs (studied in \cite{Bar-Shalom:2018rjs} in a framework to enhance universally all light quark Yukawa couplings to the value of the bottom quark coupling). The mechanism by which new VLQs could contribute to $g_{hq\bar q}$ is shown in Fig.~\ref{subfig:UVmodelsVLQpairs};
    \item Models with an additional VLQ multiplet and an additional scalar (studied in the context of the electron Yukawa coupling in \cite{Davoudiasl:2023huk,Erdelyi:2025axy}). The mechanism by which a combination of new VLQs and scalars could contribute to $g_{hq\bar q}$ is shown in Fig.~\ref{subfig:UVmodelsVLQandscalar}.
\end{itemize}

In this work, we will be interested in the second case of adding two VLQ representations to the SM Lagrangian. The reason is that the case of the two Higgs doublet models leads to a new $s$-channel resonance (from the second Higgs doublet) that would decay into the light quarks and hence undergo tight constraints from di-jet searches. As pointed out in \cite{Egana-Ugrinovic:2021uew}, these types of models would be likely discovered by direct searches for the new scalar. The last case in principle is interesting and has not been studied well in literature yet. However, it requires rather large couplings of mass dimension one to overcome the suppression by an extra heavy mass scale. We leave the study of this case for future work and concentrate for now on the case of the vector-like quarks. 

Regarding the simplified models with only pairs of vector-like quarks, we identify \emph{eight} combinations of states defined in Tab.~\ref{tab:VLQnamesandirreps}
that generate the operators in Eq.~\eqref{eq:Ouphi} at tree level.
\begin{table}[t]\centering
\begin{tabular}{c|c|c|c|c|c|c|c}
Name	&	$U$	&	$D$	&	$Q_1$	&	$Q_5$	&	$Q_7$	& $T_1$ &	$T_2$	\\
\hline
Irrep under $G_{\rm SM}$		& $(\mathbf{3},\mathbf{1})_{\frac{2}{3}}$	& $(\mathbf{3},\mathbf{1})_{-\frac{1}{3}}$	& $(\mathbf{3},\mathbf{2})_{\frac{1}{6}}$	&	$(\mathbf{3},\mathbf{2})_{-\frac{5}{6}}$ & $(\mathbf{3},\mathbf{2})_{\frac{7}{6}}$	&	$(\mathbf{3},\mathbf{3})_{-\frac{1}{3}}$	&	$(\mathbf{3},\mathbf{3})_{\frac{2}{3}}$
\end{tabular}
\caption{Seven vector-like quarks and their transformation properties under the SM gauge group $G_{\rm SM}$ which allow for the couplings to the Higgs doublet and the SM fermions. Ultimately, the combination of two different states results in the generation of dimension-six contributions to the effective Yukawa couplings at tree level, Eqs.~\eqref{eq:Lag_M1}--\eqref{eq:Lag_M8}.}
\label{tab:VLQnamesandirreps}
\end{table}
The Lagrangian of Model M takes the form 
\begin{equation}
    \mathcal{L}_{\rm M} = \mathcal{L}_{\rm SM} + \mathcal{L}_{\rm M}^{\rm quad} + \mathcal{L}^{\rm int}_{\rm M}\,,
\end{equation}
in which $\mathcal{L}_{\rm M}^{\rm quad}$ denotes the kinetic and mass terms for the vector-like quarks, schematically written as
\begin{equation}
\mathcal{L}_{\rm M}^{\rm quad}=i\bar{\Psi}\slashed{D}\Psi-M_{\Psi}\bar{\Psi}\Psi,
\end{equation}
where $\Psi$ stands for the vector-like quark fields $Q_1$, $Q_5$, $Q_7$, $T_1$, $T_2$, $U$, and $D$.
The interaction parts of the simplified models are described by the following Lagrangian densities:
\begin{itemize}
\item Model 1 (adding $U+Q_1$):
\begin{equation}
-\mathcal{L}_{1}^{\rm int} = \lambda_U \bar{U}_R \tilde{\phi}^\dagger q_{L} + \lambda^u_{Q_1}\bar{Q}_{1L}\tilde{\phi}\, u_{R} +\lambda^d_{Q_1}\bar{Q}_{1L} \phi\, d_{R} + \lambda_{UQ_1} \bar{U}\tilde{\phi}^\dagger\,Q_{1}+\text{h.c.}\,,
\label{eq:Lag_M1}
\end{equation}
\item Model 2 (adding $D+Q_1$):
\begin{equation}
    -\mathcal{L}_{2}^{\rm int} = \lambda_D\bar{D}_R \phi^\dagger q_{L} + \lambda^u_{Q_1} \bar{Q}_{1L}\tilde{\phi}\, u_{R} + \lambda^d_{Q_1} \bar{Q}_{1L} \phi\, d_{R} +\lambda_{DQ_1} \bar{D} \phi^\dagger Q_{1}+\text{h.c.}\,,
\end{equation}
\item Model 3 (adding $U+Q_7$):
\begin{equation}
     -\mathcal{L}_{3}^{\rm int} = \lambda_U \bar{U}_R \tilde{\phi}^\dagger\, q_{L} + \lambda_{Q_7}\bar{Q}_{7L}\phi\, u_{R} + \lambda_{UQ_7}\bar{U}\phi^\dagger\, Q_{7} +\text{h.c.}\,,
\end{equation}
\item Model 4 (adding $D+Q_5$):
\begin{equation}
    -\mathcal{L}_{4}^{\rm int} = \lambda_D \bar{D}_R \phi^\dagger q_{L} + \lambda_{Q_5} \bar{Q}_{5L} \tilde{\phi}\, d_{R} + \lambda_{DQ_5}\bar{D}\tilde{\phi}^\dagger\, Q_{5} +  \text{h.c.}\,,
\end{equation}
\item Model 5 (adding $T_1+Q_1$):
\begin{align}
    -\mathcal{L}_{5}^{\rm int} =& \lambda^u_{Q_1}\bar{Q}_{1L}\tilde{\phi}\, u_{R} + \lambda^d_{Q_1} \bar{Q}_{1L} \phi\, d_{R} +\frac{\lambda_{T_1}}{2}\,\bar{T}_{1R}^I\phi^\dagger \sigma^I\, q_{L} +\frac{\lambda_{T_1 Q_1}}{2}\,\bar{T}_{1}^I\phi^\dagger \sigma^I\, Q_{1}+\text{h.c.}\,, 
\end{align}
\item Model 6 (adding $T_1+Q_5$):
\begin{equation}
     -\mathcal{L}_{6}^{\rm int} = \lambda_{Q_5} \bar{Q}_{5L} \tilde{\phi}\, d_{R} +\frac{\lambda_{T_1}}{2}\,\bar{T}_{1R}^I\phi^\dagger \sigma^I\, q_{L} + \frac{\lambda_{T_1 Q_5}}{2}\,\bar{T}_{1}^I\tilde{\phi}^\dagger \sigma^I\, Q_{5}  +\text{h.c.}\,,
\end{equation}
\item Model 7 (adding $T_2+Q_1$):
\begin{align}
    -\mathcal{L}_{7}^{\rm int} &=  \lambda^u_{Q_1}\bar{Q}_{1L}\tilde{\phi} u_{R} + \lambda^d_{Q_1} \bar{Q}_{1L} \phi d_{R} + \frac{\lambda_{T_2}}{2} \bar{T}_{2R}^I\tilde{\phi}^\dagger \sigma^Iq_{L}+\frac{\lambda_{T_2Q_1}}{2}\,\bar{T}_{2}^I\tilde{\phi}^\dagger \sigma^I Q_{1} +\text{h.c.}\,,
\end{align}
\item Model 8 (adding $T_2+Q_7$):
\begin{equation}
    -\mathcal{L}_{8}^{\rm int} = \lambda_{Q_7} \bar{Q}_{7L}\phi\, u_{R} + \frac{\lambda_{T_2}}{2}\, \bar{T}_{2R}^I\tilde{\phi}^\dagger \sigma^Iq_{L} + \frac{\lambda_{T_2Q_7}}{2}\,\bar{T}_{2}^I\phi^\dagger \sigma^I Q_{7} +\text{h.c.}\,.
    \label{eq:Lag_M8}
\end{equation}
\end{itemize}
\begin{table}[t]
    \centering
    \renewcommand{\arraystretch}{1.6}
    \setlength{\tabcolsep}{12pt}
    \begin{tabular}{|c|c||c|c|}
        \hline
        $[\cO_{u\phi}]^{rp}$ & $\bar q_L^r \tilde{\phi} u_R^p\,\phi^\dagger\phi$ & $[\cO_{d\phi}]^{rp}$ & $\bar q_L^r {\phi} d_R^p\,\phi^\dagger\phi$ \\
        $[\cO_{\phi u}]^{rp}$ & $(i \phi^\dagger \overleftrightarrow{D}_\mu \phi)(\bar{u}_R^r \gamma^\mu u_R^p)$ & $[\cO_{\phi d}]^{rp}$ & $(i \phi^\dagger \overleftrightarrow{D}_\mu \phi)(\bar{d}_R^r \gamma^\mu d_R^p)$ \\
        $[\cO_{\phi q}^{(1)}]^{rp}$ & $(i\phi^\dagger \overleftrightarrow{D}_\mu \phi)(\bar{q}_L^r \gamma^\mu q_L^p)$ & $[\cO_{\phi q}^{(3)}]^{rp}$ & $(i\phi^\dagger \overleftrightarrow{D}_\mu^I \phi)(\bar{q}_L^r \gamma^\mu \sigma^I q_L^p)$ \\
        $[\cO_{\phi ud}]^{rp}$ & $(i\tilde\phi^\dagger {D}_\mu \phi)(\bar{u}_L^r \gamma^\mu d_L^p)$ & & \\ \hline
    \end{tabular}
    \caption{Dimension-6 SMEFT operators generated by Models 1 -- 8 at tree-level.}
    \label{tab:tree-level_ops}
\end{table}

In each model, two types of Yukawa-like interactions appear: either two VLQs interact with the Higgs boson or one VLQ interacts with the Higgs and a SM quark. In the latter case, the couplings (for example $\lambda_U$, $\lambda_{Q_1}^u$, and $\lambda_{Q_1}^d$ in Model I) are three-vectors in flavour space, and carry a flavour index of the SM quark field. Taking $M_\Psi\gg v$, the VLQs can be integrated out from the low-energy dynamics resulting in the tree-level contribution to the operators shown in Tab.~\ref{tab:tree-level_ops}. We collect the matching to the corresponding Wilson coefficients for all models in Tab.~\ref{tab:treelevelmodelItoVIII}.
Note that we have neglected the contributions proportional to the marginal Yukawa couplings of $\mathcal{L}_{\rm SM}$. To explain why such contributions are expected to be subleading, let us consider the result of the full tree-level matching to $\cO_{u\phi}$ in Model 1 with the new states coupled to the first-generation quarks only. We obtain
\begin{equation}
    [\cC_{u\phi}]_{11} = \frac{[y_u^*]_{11} |[\lambda_U]_1|^2}{2M_U^2} + \frac{[y_u^*]_{11} |[\lambda_{Q_1}^u]_1|^2}{2M_{Q_1}^2} - \frac{\lambda_{UQ_1} [\lambda_U^*]_1 [\lambda_{Q_1}^u]_1}{M_U\,M_{Q_1}}\,,
    \label{eq:WCuphi11}
\end{equation}
such that the up-quark mass reads
\begin{equation}
    m_u = \frac{v}{\sqrt{2}}\left([y_u]_{11}-\frac{v^2}{2}[\cC_{u\phi}]_{11}\right).
    \label{eq:up_mass}
\end{equation}
In order to reproduce the physical up-quark mass, there needs to exist some amount of tuning between the marginal Yukawa coupling in $\mathcal{L}_{\rm SM}$ and the NP contribution encoded in $\cC_{u\phi}$. For that reason, we expect $[y_u]_{11}\simeq v^2 [\cC_{u\phi}]_{11}/2$, such that the first two terms in Eq.~\eqref{eq:WCuphi11} effectively act as $v^4/\Lambda^4$, where $\Lambda$ is the placeholder for vector-like quark masses $M_U$ and $M_{Q_1}$. Therefore, in Tab.~\ref{tab:treelevelmodelItoVIII}, we report the contributions proportional to the new physics couplings only and write $\propto y_{u,d}$ when the leading contribution is achieved through the insertion of the pure SM coupling. 

The amount of tuning necessary to reproduce the correct quark masses depends on the total enhancement we want to achieve in the effective Higgs-quark Yukawa couplings. As we will see later after studying all applicable constraints, the largest enhancement can be achieved for up-quark Yukawa couplings with $\kappa_u\approx 1000$, translating to the largest tuning in this setup to be 1 per mille between $[y_u]_{11}$ and $v^2 [\cC_{u\phi}]_{11}/2$. This tuning is stable under radiative corrections and is not larger than the one in the SM CKM sector which has no a priori reason
to be close to the identity~\cite{deGouvea:2003xe}.

For natural values of the new physics couplings in Eqs.~\eqref{eq:Lag_M1}-\eqref{eq:Lag_M8}, i.e.~$\lambda\sim \cO(1)$, the enhancement in the first-generation quark Yukawa coupling of the order $\kappa_{u,d}\sim\cO(10^3)$ (or $\kappa_{c,s}\sim\cO(10)$ for the second-generation) can be obtained with TeV-scale vector-like quarks. However, as can be seen from Tab.~\ref{tab:treelevelmodelItoVIII}, the vector-like quarks with generic couplings in all models lead to flavour-changing $Z$-couplings at tree-level
\begin{align}
\mathcal{L}_{\rm NP}^{Z} \supset -\sqrt{g_2^2+g_1^2}\, Z^{\mu} \left( \delta g_{L\,ij}^{Zf}\,\bar f^i_L \gamma_{\mu} f_L^j
+\delta g_{R\,ij}^{Zf}\,\bar f^i_R \gamma_{\mu}f_R^j \right)\,,
\label{eq:Z_BSM}
\end{align}
where $i,j$ are the flavour indices of the corresponding quark fields $f=\{u,d\}$, $g_2$ and $g_1$ are the $SU(2)$ and $U(1)_Y$ gauge couplings, respectively, and 
\begin{align}
\delta g_{L\,ij}^{Zu}=&\,-\frac{v^2}{2}V_{ik} \left(
[\cC_{\phi q}^{(1)}]_{kl}-[\cC_{\phi q}^{(3)}]_{kl}
\right) V_{lj}^{\dagger}\,,\label{eq:delta_g_ZuL}\\
\delta g_{R\,ij}^{Zu}=&\,-\frac{v^2}{2}[\cC_{\phi u}]_{ij}\,,\label{eq:delta_g_ZuR}\\
\delta g_{L\,ij}^{Zd}=&-\frac{v^2}{2}\left([\cC_{\phi q}^{(1)}]_{ij}+[\cC_{\phi q}^{(3)}]_{ij}\right)\,,\label{eq:delta_g_ZdL}\\
\delta g_{R\, ij}^{Zd}=&\,-\frac{v^2}{2}[\cC_{\phi d}]_{ij}\,,\label{eq:delta_g_ZdR}
\end{align}
with $V$ being the CKM matrix, and we assumed alignment to the down-quark mass basis, $q_L^i = (V_{ik}u_L^k,\,d_L^i)^T$.

Therefore, coupling vector-like quarks to more than one light generation leads to flavour-changing neutral currents at tree level suppressed by $v^2/\Lambda^2$ only, pushing the scale of new particles $\Lambda$ to a multi-TeV range and diminishing the contributions to the effective Yukawa couplings. Hence, in order to have a realistic prospect of observing vector-like quarks which lead to a dramatic enhancement of light quark Yukawa couplings, these new states must couple to only \emph{one generation at a time}. We follow this flavour non-universal assumption in this work.

Furthermore, it is clear from Eqs.~\eqref{eq:delta_g_ZuL}--\eqref{eq:delta_g_ZdR} that some flavour violation is induced by the presence of the CKM matrix. This is only problematic for models that induce both $\delta g_L^{Zu}$ and $\delta g_L^{Zd}$ simultaneously, as one can always choose either up- or down-type quarks to be mass eigenstates. Concretely, in Models 1--4, we find that the contribution to $\cC_{\phi q}^{(1)}$ and $\cC_{\phi q}^{(3)}$ are equal up to a sign, resulting in these models affecting $Z$-couplings to either up or down left-handed quarks, but not both. In contrast, for Models 5--8, which involve vector-like states that are triplets under the $SU(2)$, contributions to $\cC_{\phi q}^{(1)}$ and $\cC_{\phi q}^{(3)}$ are also different in magnitude, which means that both $\delta g_L^{Zu}$ and $\delta g_L^{Zd}$ are generated. Indeed, in these models, there is an irreducible flavour violation due to the CKM matrix which results in a higher new physics scale and less contribution to the effective Yukawa couplings. We will discuss this in more detail in Sec.~\ref{sec:flavour_physics}.

\begin{landscape}
\begin{table}[t]
{\renewcommand{\arraystretch}{1.5}
    \begin{tabular}{|c|c|c|c|c| c|}
    \hline
    Class                             &    Coefficient                                  &   Model 1 & Model 2 & Model 3 & Model 4\\
    \hline
    \multirow{2}{*}{$\psi^2\phi^3$}   &$\left[\cC_{u\phi}\right]^{rp}$        & 
 $-\frac{\lambda_{UQ_1}}{M_UM_{Q_1}}\left[\lambda_U^*\right]^r\left[\lambda^u_{Q_1}\right]^p$  & $\propto y_u$  &
$ - \frac{\lambda_{UQ_7}}{M_U M_{Q_7}} \left[\lambda_U^*\right]^r \left[\lambda_{Q_7}\right]^p $
&
0
\\
                                   &$\left[\cC_{d\phi}\right]^{rp}$  &$\propto y_d$ 
                                   &  $  -\frac{\lambda_{DQ_1}}{M_{Q_1}M_D} \left[\lambda^*_{D}\right]^{r} \left[\lambda^d_{Q_1}\right]^{p}$ & 0 &
                                   $- \frac{\lambda_{DQ_5}}{M_D M_{Q_5}} \left[\lambda^*_D\right]^r \left[\lambda_{Q_5}\right]^p   $ 
                                   \\   
    \hline
	\multirow{5}{*}{$\psi^2\phi^2 D$} &$\left[\cC^{(1)}_{\phi q}\right]^{rp}$	& $\frac{1}{4M_U^2}\left[\lambda_U^*\right]^r\left[\lambda_U\right]^p$ & $-\frac{1}{4M_D^2} \left[\lambda_D^*\right]^r \left[\lambda_D\right]^p$ & $ \frac{1}{4M^2_U}\left[\lambda_U^*\right]^{r}\left[\lambda_U\right]^p $ & $   -\frac{1}{4M^2_D} \left[\lambda_D\right]^{p} \left[\lambda_D^*\right]^r$	\\
	                                  &$\left[\cC^{(3)}_{\phi q}\right]^{rp}$	& $-\frac{1}{4M_U^2}\left[\lambda_U^*\right]^r\left[\lambda_U\right]^p$	& $-\frac{1}{4M_D^2} \left[\lambda_D^*\right]^r \left[\lambda_D\right]^p$ & $ -\frac{1}{4M^2_U}\left[\lambda_U^*\right]^{r}\left[\lambda_U\right]^p    $  & $-\frac{1}{4M^2_D} \left[\lambda_D\right]^{p} \left[\lambda_D^*\right]^r$	\\
                                      &$\left[\cC_{\phi u}\right]^{rp}$		& $-\frac{1}{2M_{Q_1}^2}\left[\lambda^{u*}_{Q_1}\right]^{r}\left[\lambda^u_{Q_1}\right]^{p}	$ &
  $-\frac{1}{2M_{Q_1}^2} \left[\lambda^{u*}_{Q_1}\right]^{r} \left[\lambda^u_{Q_1}\right]^{p}$ &   $\frac{1}{2M^2_{Q_7}} \left[\lambda^{*}_{Q_7}\right]^r \left[\lambda_{Q_7}\right]^p$ & 0                                                                     \\	
	                                  &$\left[\cC_{\phi d}\right]^{rp}$		& $\frac{1}{2M_{Q_1}^2}\left[\lambda^{d*}_{Q_1}\right]^{r}\left[\lambda^d_{Q_1}\right]^{p}$ &
                                   $\frac{1}{2M_{Q_1}^2} \left[\lambda^{d*}_{Q_1}\right]^{r} \left[\lambda^d_{Q_1}\right]^{p}$ & 0 &  $-\frac{1}{2M_{Q_5}^2} \left[\lambda_{Q_5}^{*}\right]^r \left[\lambda_{Q_5}\right]^{p}$ 
                                   \\
	                                  &$\left[\cC_{\phi ud}\right]^{rp}$		& $\frac{1}{M^2_{Q_1}}\left[\lambda^{u*}_{Q_1}\right]^{r}\left[\lambda^d_{Q_1}\right]^p$ &	 $\frac{1}{M^2_{Q_1}} \left[\lambda^{u*}_{Q_1}\right]^{r} \left[\lambda^d_{Q_1}\right]^p$ &0 &0 \\	\hline\hline 

    Class                             &    Coefficient                                  &   Model 5 & Model 6 & Model 7 & Model 8\\
    \hline
    \multirow{2}{*}{$\psi^2\phi^3$}   &$\left[\cC_{u\phi}\right]^{rp}$        & 
$ - \frac{\lambda_{T_1 Q_1} }{2 M_{T_1} M_{Q_1}} \left[\lambda_{T_1}\right]^{*r} \left[\lambda^u_{Q_1}\right]^p$ &
$\propto y_u$
& $ - \frac{\lambda_{T_2Q_1} }{4M_{Q_1}M_{T_2}}\left[\lambda_{T_2} \right]^{*r} \left[\lambda^u_{Q_1}\right]^p $  
    & $  \frac{\lambda_{T_2Q_7} }{4M_{Q_7}M_{T_2}} \left[\lambda_{T_2}\right]^{*r} \left[\lambda_{Q_7}\right]^{p}   $ 
\\
                                   &$\left[\cC_{d\phi}\right]^{rp}$  & $ - \frac{ \lambda_{T_1 Q_1}}{4M_{T_1} M_{Q_1}} \left[\lambda_{T_1}\right]^{*r} \left[\lambda^d_{Q_1}\right]^p$
                                   & $  \frac{\lambda_{T_1Q_5}}{4M_{T_1} M_{Q_5}}  \left[\lambda_{T_1}^*\right]^r \left[\lambda_{Q_5}\right]^p$  & $ - \frac{\lambda_{T_2Q_1}}{2 M_{Q_1}M_{T_2}} \left[\lambda_{T_2} \right]^{*r} \left[\lambda^d_{Q_1}\right]^p $  & $\propto y_d$
                                   
                                   \\   
    \hline
	\multirow{5}{*}{$\psi^2\phi^2 D$} &$\left[\cC^{(1)}_{\phi q}\right]^{rp}$	& $-\frac{3}{16M^2_{T_1}}\left[\lambda_{T_1}\right]^{*r}\left[\lambda_{T_1}\right]^p$ & $-\frac{3}{16M^2_{T_1}}\left[\lambda_{T_1}\right]^{*r}\left[\lambda_{T_1}\right]^p$ & $\frac{3}{16M^2_{T_2}} \left[ \lambda_{T_2} \right]^p \left[\lambda_{T_2}\right]^{*r}$ & 	$\frac{3}{16M^2_{T_2}} \left[ \lambda_{T_2} \right]^p \left[\lambda_{T_2}\right]^{*r}$\\
	                                  &$\left[\cC^{(3)}_{\phi q}\right]^{rp}$	& $\frac{1}{16M^2_{T_1}}\left[\lambda_{T_1}\right]^{*r}\left[\lambda_{T_1}\right]^p$	& $\frac{1}{16M^2_{T_1}}\left[\lambda_{T_1}\right]^{*r}\left[\lambda_{T_1}\right]^p$ & $\frac{1}{16M^2_{T_2}} \left[ \lambda_{T_2} \right]^p \left[\lambda_{T_2}\right]^{*r}$  & $\frac{1}{16M^2_{T_2}} \left[ \lambda_{T_2} \right]^p \left[\lambda_{T_2}\right]^{*r}$ 	\\
                                      &$\left[\cC_{\phi u}\right]^{rp}$		& $-\frac{1}{2M_{Q_1}^2}\left[\lambda^{u}_{Q_1}\right]^{*r}\left[\lambda^u_{Q_1}\right]^{p}$ &0
 & $-\frac{1}{2M_{Q_1}^2}\left[\lambda^{u}_{Q_1}\right]^{*r}\left[\lambda^u_{Q_1}\right]^{p}$  & $\frac{1}{2M^2_{Q_7}} \left[\lambda_{Q_7}\right]^{*r} \left[\lambda_{Q_7}\right]^p$                                                                      \\	
	                                  &$\left[\cC_{\phi d}\right]^{rp}$		& $\frac{1}{2M_{Q_1}^2}\left[\lambda^{d}_{Q_1}\right]^{*r}\left[\lambda^d_{Q_1}\right]^{p}$ & $-\frac{1}{2M_{Q_5}^2} \left[\lambda_{Q_5}^{*}\right]^r \left[\lambda_{Q_5}\right]^{p}$
                                    & $\frac{1}{2M_{Q_1}^2}\left[\lambda^{d}_{Q_1}\right]^{*r}\left[\lambda^d_{Q_1}\right]^{p}$ & 0  
                                   \\
	                                  &$\left[\cC_{\phi ud}\right]^{rp}$		& $\frac{1}{M_{Q_1}^2}\left[\lambda^{u}_{Q_1}\right]^{*r}\left[\lambda^d_{Q_1}\right]^{p}$ &	0  & $\frac{1}{M_{Q_1}^2}\left[\lambda^{u}_{Q_1}\right]^{*r}\left[\lambda^d_{Q_1}\right]^{p}$ &0 \\ \hline	
    \end{tabular}
}    
    \caption{Tree-Level matching for Model 1 -- 8.}
    \label{tab:treelevelmodelItoVIII}
\end{table}
\end{landscape}

\begin{table}[t]
    \centering
    \renewcommand{\arraystretch}{1.6}
    \setlength{\tabcolsep}{12pt}
    \begin{tabular}{|c|c||c|c||c|c|}
        \hline
        $\cO_{\phi\Box}$ & $(\phi^\dagger\phi)\Box(\phi^\dagger\phi)$ & $\cO_{\phi G}$ & $\phi^\dagger \phi G_{\mu\nu}^A G^{\mu\nu A}$ & $\cO_{\phi B}$ & $\phi^\dagger \phi B_{\mu\nu} B^{\mu\nu}$\\ \hline
        $\cO_{\phi D}$ & $|\phi^\dagger D_\mu \phi|^2$ & $\cO_{\phi W}$ & $\phi^\dagger \phi W_{\mu\nu}^I W^{\mu\nu I}$ & $\cO_{\phi WB}$ & $\phi^\dagger \sigma^I \phi W_{\mu\nu}^I B^{\mu\nu}$\\ \hline
    \end{tabular}
    \caption{Relevant dimension-6 SMEFT operators generated by Models 1 -- 8 at loop-level.}
    \label{tab:loop-level_ops}
\end{table}

Finally, matching the models at the one-loop level generates additional operators that provide relevant constraints on the parameter space; such operators are listed in Tab.~\ref{tab:loop-level_ops}. 
We discuss the details of our phenomenological study in the following sections but note here that the matching at one-loop has been performed using both \texttt{Matchete}~\cite{Fuentes-Martin:2022jrf} and \texttt{SOLD}~\cite{Guedes:2023azv}. For the moment, the \texttt{SOLD} package is not able to perform the matching to $\mathcal{O}_{\phi \Box}$ and $\mathcal{O}_{\phi D}$, so we have explicitly cross-checked the \texttt{Matchete} output against our analytical computation finding agreement. We present our matching computation and the results in the limit of equal VLQ masses in App.~\ref{app:CHD}.
\section{Constraints}
\label{sec:constraints}
The presence of vector-like quarks at a high scale affects various low-energy processes. In the following sections, we identify the most relevant constraints on the parameter space of the models that we consider. We discuss in detail the effects in flavour, electroweak, and Higgs physics, together with direct searches for the new states. 
\subsection{Flavour physics}
\label{sec:flavour_physics}

Vector-like quarks with couplings to the SM fermions affect flavour transitions in a nontrivial way. For instance, when their Yukawa couplings are flavour-generic, they induce flavour-changing neutral currents that push their mass scale into the multi-TeV regime. Hence, in order to identify the viable models at the TeV scale that enhance the effective light quark Yukawa couplings by several orders of magnitude, we are directed to regard only the models in which vector-like quarks couple to one light generation at a time. In this section, we are interested in quantifying the consequences of such an assumption. 

For Models 1--4, integrating out the vector-like quarks results in modifying the $Z$ boson couplings to either up- or down-type left-handed quarks. In these cases, we can always choose the basis such that there are no flavour-changing amplitudes. For example, let us consider Model 1 with vector-like quarks coupled to the first generation. In this case, we align the left-handed quarks to the up-quark mass basis, $q_L^i=(u_L^i,\, V_{ij}^* d_L^j)^T$, such that the $Z$-coupling modifications read
\begin{alignat}{2}
&\delta g_{L\,11}^{Zu}=-\frac{v^2}{4M_U^2}|[\lambda_U]_1|^2\,,\qquad &&\delta g_{L\,11}^{Zd}=0\,,\label{eq:delta_g_ZL_M1}\\
&\delta g_{R\,11}^{Zu}=\,\frac{v^2}{4M_{Q_1}^2}|[\lambda_{Q_1}^u]_1|^2\,,\qquad &&\delta g_{R\, 11}^{Zd}=-\frac{v^2}{4M_{Q_1}^2}|[\lambda_{Q_1}^d]_1|^2\,,\label{eq:delta_g_ZR_M1}
\end{alignat}
with all other $\delta g_{L\,ij}^{Zf}$ and $\delta g_{R\,ij}^{Zf}$ vanishing. In contrast, for Model 2, we align to the down-quark mass basis, $q_L^i=(V_{ik}u_L^k,\, d_L^i)^T$, and the corresponding modifications to the $Z$-couplings read
\begin{alignat}{2}
&\delta g_{L\,11}^{Zu}=0\,,\qquad &&\delta g_{L\,11}^{Zd}=\frac{v^2}{4M_D^2}|[\lambda_D]_1|^2\,,\label{eq:delta_g_ZL_M2}\\
&\delta g_{R\,11}^{Zu}=\,\frac{v^2}{4M_{Q_1}^2}|[\lambda_{Q_1}^u]_1|^2\,,\qquad &&\delta g_{R\, 11}^{Zd}=-\frac{v^2}{4M_{Q_1}^2}|[\lambda_{Q_1}^d]_1|^2\,,\label{eq:delta_g_ZR_M2}
\end{alignat}
with all other $\delta g_{L\,ij}^{Zf}=\delta g_{R\,ij}^{Zf}=0$. Therefore, for Models 1 and 2 with vector-like quarks coupled to the first-generation quarks, the flavour-changing neutral currents remain unaffected and we avoid the corresponding constraints. The same conclusion remains for Models 3 and 4, as well as if vector-like quarks are coupled only to the second-generation quarks. 

\begin{table}[t]
    \centering
    \renewcommand{\arraystretch}{1.4}
    \begin{tabular}{|c|c|c|c|}
    \hline
        Particle & $K^+\to \pi^+\nu\bar\nu$ & $K_L\to\mu^+\mu^-$ & $D\to\mu^+\mu^-$ \\
        \hline
        $U\sim (\mathbf{3},\mathbf{1})_{2/3}$ & -- & -- & $7.8\times{\sqrt{|[\lambda_U]_1 [\lambda_U]_2|}}$ \\
        \hline
        $D\sim (\mathbf{3},\mathbf{1})_{-1/3}$ & $\!\!\!\!126\times{\sqrt{|[\lambda_D]_1 [\lambda_D]_2|}}$ & $\!\!\!\!100\times{\sqrt{{\rm Re}\left([\lambda_D^*]_1 [\lambda_D]_2\right)}}$ & -- \\
        \hline
        $Q_1\sim (\mathbf{3},\mathbf{2})_{1/6}$ & $126\times{\sqrt{|[\lambda_{Q_1}^d]_1 [\lambda_{Q_1}^d]_2|}}$ & $100\times{\sqrt{{\rm Re}\left([\lambda_{Q_1}^{d*}]_1 [\lambda_{Q_1}^d]_2\right)}}$ & $7.8\times{\sqrt{{\rm Re}\left([\lambda_{Q_1}^{u*}]_1 [\lambda_{Q_1}^u]_2\right)}}$ \\
        \hline
        $Q_5\sim (\mathbf{3},\mathbf{2})_{-5/6}$ & $126\times{\sqrt{|[\lambda_{Q_5}]_1 [\lambda_{Q_5}]_2|}}$ & $100\times{\sqrt{{\rm Re}\left([\lambda_{Q_5}^*]_1 [\lambda_{Q_5}]_2\right)}}$ & --\\
        \hline
        $Q_7\sim (\mathbf{3},\mathbf{2})_{7/6}$ & -- & -- & $7.8\times{\sqrt{{\rm Re}\left([\lambda_{Q_7}^*]_1 [\lambda_{Q_7}]_2\right)}}$\\
        \hline
        $T_1\sim (\mathbf{3},\mathbf{3})_{-1/3}$ & $63\times{\sqrt{|[\lambda_{T_1}]_1 [\lambda_{T_1}]_2|}}$ & $35\times{\sqrt{{\rm Re}\left([\lambda_{T_1}^{*}]_1 [\lambda_{T_1}]_2\right)}}$ & $3.9\times{\sqrt{{\rm Re}\left([\lambda_{T_1}^{*}]_1 [\lambda_{T_1}]_2\right)}}$\\
        \hline
        $T_2\sim (\mathbf{3},\mathbf{3})_{2/3}$ & $89\times{\sqrt{|[\lambda_{T_2}]_1 [\lambda_{T_2}]_2|}}$ & $50\times{\sqrt{{\rm Re}\left([\lambda_{T_2}^{*}]_1 [\lambda_{T_2}]_2\right)}}$ & $2.8\times{\sqrt{{\rm Re}\left([\lambda_{T_2}^{*}]_1 [\lambda_{T_2}]_2\right)}}$ \\
        \hline
    \end{tabular}
    \caption{Constraints from $\Delta F =1$ processes on the particle mass measured in TeV. We assume that the new physics contribution interferes constructively with the SM.}
    \label{tab:DeltaF=1}
\end{table}

Still, it is worth considering what happens if the new states simultaneously couple to both light generations. In this case, there are flavour-changing neutral currents mediated by the $Z$-boson through $\delta g_{L\,12}^{Zf}\ne 0$ and $\delta g_{R\,12}^{Zf}\ne0$.
A comprehensive analysis of such effects has been performed in Ref.~\cite{Ishiwata:2015cga} and we summarise the leading constraints from $\Delta F=1$ processes ($K^+\to\pi^+\nu\bar\nu$, $K_L\to\mu^+\mu^-$, $D\to\mu^+\mu^-$) in Tab.~\ref{tab:DeltaF=1}, and $\Delta F=2$ processes ($K-\bar K$ and $D-\bar D$ mixing) in Tab.~\ref{tab:DeltaF=2}. A difference compared to Ref.~\cite{Ishiwata:2015cga} is that in the models we study, two vector-like states often contribute to the same amplitude allowing for the possibility of cancellation among them. However, we follow a natural assumption that such tunings are absent. Moreover, the results reported in Ref.~\cite{Ishiwata:2015cga} are slightly different due to updates and progress in experimental measurements. Specifically, we use the recently updated measurement of ${\rm BR}(K^+\to \pi^+\nu\bar\nu)= (13.0^{+3.3}_{-2.9})\times 10^{-11}$ by NA62~\cite{NA62:2021zjw,2024KtoPinunu,Buras:2022wpw}, and ${\rm BR}(D^0\to \mu^+\mu^-)<3.1\times 10^{-9}$ by LHCb~\cite{LHCb:2022jaa}. For meson mixing, we use the one-loop amplitudes mediated by the vector-like quarks computed in~\cite{Ishiwata:2015cga} and the latest SM global fits performed by the UTfit collaboration to extract the constraints from $\Delta F=2$ transitions~\cite{Bona:2024bue}. As can be seen, the mass scale of vector-like quarks is around $\mathcal{O}(50-100)$ TeV for generic $\mathcal{O}(1)$ couplings. Thus, the contribution to the effective Yukawa couplings from these states in a generic flavour setup is practically negligible, and it is largest for the up-quark Yukawa in Model 3 where the lowest scale for both vector-like states is $M_U=M_{Q_7}=47$ TeV:
\begin{equation}
    |\kappa_u - 1|= \frac{v^3}{\sqrt{2}\, m_u\, M_U\, M_{Q_7}} \simeq  2.1\,,
    \label{eq:kappa_12on}
\end{equation}
where we also set the couplings to one. The $\kappa_u$ value in Eq.~\eqref{eq:kappa_12on} is virtually unobservable even at future colliders (see Sec.~\ref{sec:future}), thus motivating our assumption to study flavour structures with vector-like quarks coupled to only one family.

\begin{table}[t]
    \centering
    \renewcommand{\arraystretch}{1.4}
    \begin{tabular}{|c|c|c|}
    \hline
        Particle & $K-\bar K$ & $D-\bar D$ \\
        \hline
        $U\sim (\mathbf{3},\mathbf{1})_{2/3}$ & -- & $47\times\sqrt{{\rm Re}\left[\left([\lambda_U^*]_1 [\lambda_U]_2\right)^2\right]}$ \\
        \hline
        $D\sim (\mathbf{3},\mathbf{1})_{-1/3}$ & $\!\!\!\!\!31\times\sqrt{{\rm Re}\left[\left([\lambda_D^*]_1 [\lambda_D]_2\right)^2\right]}$ & -- \\
        \hline
        $Q_1\sim (\mathbf{3},\mathbf{2})_{1/6}$ & $44\times\sqrt{{\rm Re}\left[\left([\lambda_{Q_1}^{d*}]_1 [\lambda_{Q_1}^{d}]_2\right)^2\right]}$ & $67\times\sqrt{{\rm Re}\left[\left([\lambda_{Q_1}^{u*}]_1 [\lambda_{Q_1}^{u}]_2\right)^2\right]}$ \\
        \hline
        $Q_5\sim (\mathbf{3},\mathbf{2})_{-5/6}$ & $31\times\sqrt{{\rm Re}\left[\left([\lambda_{Q_5}]_1 [\lambda_{Q_5}]_2\right)^2\right]}$ & --\\
        \hline
        $Q_7\sim (\mathbf{3},\mathbf{2})_{7/6}$ & -- & $47\times\sqrt{{\rm Re}\left[\left([\lambda_{Q_7}]_1 [\lambda_{Q_7}]_2\right)^2\right]}$ \\
        \hline
        $T_1\sim (\mathbf{3},\mathbf{3})_{-1/3}$ & $9\times\sqrt{{\rm Re}\left[\left([\lambda_{T_1}]_1 [\lambda_{T_1}]_2\right)^2\right]}$ & $13\times\sqrt{{\rm Re}\left[\left([\lambda_{T_1}]_1 [\lambda_{T_1}]_2\right)^2\right]}$ \\
        \hline
        $T_2\sim (\mathbf{3},\mathbf{3})_{2/3}$ & $9\times\sqrt{{\rm Re}\left[\left([\lambda_{T_1}]_1 [\lambda_{T_2}]_2\right)^2\right]}$ & $13\times\sqrt{{\rm Re}\left[\left([\lambda_{T_2}]_1 [\lambda_{T_2}]_2\right)^2\right]}$ \\
        \hline
    \end{tabular}
    \caption{Constraints from the real part of $\Delta F =2$ processes on the particle mass measured in TeV. The imaginary parts of $\Delta F =2$ amplitudes lead to constraints that are a factor of $\mathcal{O}(18)$ better for $K-\bar{K}$ mixing, and a factor of $\mathcal{O}(5)$ for $D-\bar{D}$ mixing.}
    \label{tab:DeltaF=2}
\end{table}

Nevertheless, even in this case, there are important constraints on the vector-like masses imposed by the CKM unitarity violation. For that purpose, let us analyze the couplings of the $W$-boson to the left-handed quark mass eigenstates
\begin{equation}
    \mathcal{L}_{\rm eff}\supset -\frac{g_2}{\sqrt{2}} W^{+}_\mu\,{\bar u}_L^{\,i} \gamma^\mu\left(V_{ij}+v^2 V_{ik}\,[\cC_{\phi q}^{(3)}]_{kj}\right) d_L^{\,j}\,,
    \label{eq:Weff}
\end{equation}
where $V$ would be the SM CKM matrix in the new physics decoupling limit, $VV^\dagger = \mathbb{1}$. Due to the mixing with the vector-like quarks, a $3\times 3$ matrix encoding interactions of the quarks with the $W$-boson
\begin{equation}
    Y = V+v^2\, V\,\cC_{\phi q}^{(3)}\,,
    \label{eq:Y_matrix}
\end{equation}
is no longer unitary, and the deviations from the unitarity can be encoded by defining 
\begin{equation}
    S = YY^\dagger = \mathbb{1} + v^2\, V\left(\cC_{\phi q}^{(3)}+\cC_{\phi q}^{(3)\dagger}\right) V^\dagger + \mathcal{O}(v^4/\Lambda^4)\,.
\end{equation}
In the models that we consider, the vector-like quarks couple to either the first- or second-generation quarks, and the constraints arise from the unitarity violation in the first and second CKM row, respectively~\cite{ParticleDataGroup:2024cfk}
\begin{align}
    S_{11} &= 0.9984\pm0.0007\,,\\
    S_{22} &= 1.001 \pm 0.012\,.
\end{align}
Due to the recent reduction of the value of $|V_{ud}|$, there
is a 2.3$\sigma$ tension with unitarity in the SM in the first row that the vector-like quarks considered in this work could even help to address. However, we constrain the vector-like contribution by demanding it to be smaller than the maximal difference between the $2\sigma$ range of the measurement and the SM prediction. This results in the following constraints 
\begin{align}
    2 v^2 \left|[\cC_{\phi q}^{(3)}]_{11}\right|<0.003\,,\quad\quad 2 v^2 \left|[\cC_{\phi q}^{(3)}]_{22}\right|<0.025\,,
\end{align}
imposing the requirements on the masses of vector-like quarks coupled to the first-generation
\begin{align}
    M_{\Psi} > 3.2\times|[\lambda_\Psi]_1|\,\,\,{\rm TeV}\,,\quad\quad &\Psi = U,D\,,\label{eq:Mpsi_UD_1st}\\
    M_{\Psi} > 1.6\times|[\lambda_\Psi]_1|\,\,\,{\rm TeV}\,,\quad\quad &\Psi = T_1,T_2\,,\label{eq:Mpsi_T1T2_1st}
\end{align}
and
\begin{align}
    M_{\Psi} > 1.1\times|[\lambda_\Psi]_2|\,\,\,{\rm TeV}\,,\quad\quad &\Psi = U,D\,,\label{eq:Mpsi_UD_2nd}\\
    M_{\Psi} > 0.6\times|[\lambda_\Psi]_2|\,\,\,{\rm TeV}\,,\quad\quad &\Psi = T_1,T_2\,,\label{eq:Mpsi_T1T2_2nd}
\end{align}
for models with vector-like quarks coupled to the second generation. The vector-like states $Q_1, Q_5$, and $Q_7$ remain unconstrained as they do not contribute to $\cC_{\phi q}^{(3)}$.

In addition to the constraints established by the CKM unitarity, the models with vector-like states that are triplets under the $SU(2)$ group get restrictions from flavour-changing neutral current transitions. This occurs due to the simultaneous modifications of the $Z$-boson couplings to left-handed up- and down-quarks. To illustrate this point, let us write the $Z$-boson interactions in the presence of the vector-like triplet $T_1\sim(\mathbf{3},\mathbf{3})_{-1/3}$ coupled to the first-quark generation in the basis aligned to the down-quark mass basis
    \begin{align}
\mathcal{L}_{\rm NP}^{Z} \supset -\frac{\sqrt{g_2^2+g_1^2}\,v^2}{8 M_{T_1}^2} |[\lambda_{T_1}]_1|^2\left(V_{i1} V_{1j}^\dagger \,\bar u^i_L \gamma_{\mu} u_L^j
+\frac{1}{2}\,\bar d^1_L \gamma_{\mu} d_L^1 \right) Z^{\mu} \,,
\label{eq:Z_BSM_triplets_down}
\end{align}
or in the basis aligned to the up-quark mass basis
    \begin{align}
\mathcal{L}_{\rm NP}^{Z} \supset -\frac{\sqrt{g_2^2+g_1^2}\,v^2}{8 M_{T_1}^2} |[\lambda_{T_1}]_1|^2\left(\bar u^1_L \gamma_{\mu} u_L^1
+\frac{1}{2}\,V_{i1}^\dagger V_{1j}\,\bar d^i_L \gamma_{\mu} d_L^j \right) Z^{\mu} \,,
\label{eq:Z_BSM_triplets_up}
\end{align}
such that flavour-changing neutral currents are induced in the up- or down-sector, respectively. After the electroweak symmetry-breaking, the couplings in Eqs.~\eqref{eq:Z_BSM_triplets_down} and~\eqref{eq:Z_BSM_triplets_up} match onto the operators in the Low-Energy Effective Field Theory (LEFT)~\cite{Jenkins:2017jig} which describe the flavour violating transitions. In particular, the strongest constraints come from
neutral meson oscillations described by
\begin{align}
    \mathcal{L}_{\rm meson} \supset - C_K^1\,(\bar d_L\gamma_\mu s_L)^2 - C_D^1\,(\bar u_L\gamma_\mu c_L)^2\,,
    \label{eq:LEFT_CK1_CD1}
\end{align}
bounding the scale of $M_{T_1}$ and $M_{T_2}$ from $K-\bar K$ and $D-\bar D$ mixing. For the state $T_1$ coupled to the first generation, we obtain
\begin{alignat}{2}
    C_K^1 &= \frac{g_Z^2 v^4}{512 m_Z^2 M_{T_1}^4} (V_{11}^\dagger V_{12})^2 |[\lambda_{T_1}]_1|^4\,,\quad&&\text{up-quark mass basis}\,,\\
    C_D^1 &= \frac{g_Z^2 v^4}{128 m_Z^2 M_{T_1}^4} (V_{11}^\dagger V_{12})^2 |[\lambda_{T_1}]_1|^4\,,\quad&&\text{down-quark mass basis}\,.
\end{alignat}
Assuming only the contributions to the real part of the mixing amplitude, we employ the constraints from the UTfit collaboration~\cite{Bona:2024bue} on ${\rm Re}(C_K^1)$ and ${\rm Re}(C_D^1)$ to obtain 
\begin{alignat}{2}
    M_{T_1} &> 4.2 \times |[\lambda_{T_1}]_1|\,\,\,{\rm TeV}\,,\quad&&\text{down-quark mass basis}\,,\\
    M_{T_1} &> 2.3\times |[\lambda_{T_1}]_1|\,\,\,{\rm TeV}\,,\quad&&\text{up-quark mass basis}\,.
\end{alignat}
Similarly, for the state $T_2$, we have 
\begin{alignat}{2}
    M_{T_2} &> 2.9 \times |[\lambda_{T_2}]_1|\,\,\,{\rm TeV}\,,\quad&&\text{down-quark mass basis}\,,\\
    M_{T_2} &> 3.3\times |[\lambda_{T_2}]_1|\,\,\,{\rm TeV}\,,\quad&&\text{up-quark mass basis}\,,
\end{alignat}
and the same holds for the vector-like triplets coupled to the second generation only.

\begin{table}[t]
    \centering
    \renewcommand{\arraystretch}{1.4}
    \begin{tabular}{|c|c|c|}
    \hline
        Particle & CKM unitarity & $\Delta F=2$ \\
        \hline
        $U\sim (\mathbf{3},\mathbf{1})_{2/3}$ & $3.2\times |[\lambda_U]_{1}|$ & -- \\
        \hline
        $D\sim (\mathbf{3},\mathbf{1})_{-1/3}$ & $3.2\times |[\lambda_D]_{1}|$ & -- \\
        \hline
        $T_1\sim (\mathbf{3},\mathbf{3})_{-1/3}$ & $1.6\times |[\lambda_{T_1}]_{1}|$ & $2.3\times |[\lambda_{T_1}]_{i=1,2}|$ \\
        \hline
        $T_2\sim (\mathbf{3},\mathbf{3})_{2/3}$ & $1.6\times |[\lambda_{T_2}]_{1}|$ &  $2.9\times |[\lambda_{T_2}]_{i=1,2}|$\\
        \hline
    \end{tabular}
    \caption{Flavour constraints on the particle mass measured in TeV in the scenarios when the particles couple to only one light quark generation at a time $i=\{1,2\}$. For the vector-like triplets $T_1$ and $T_2$, we work in a basis that gives less stringent $\Delta F=2$ constraints. The states $Q_1$, $Q_5$, and $Q_7$ stay unconstrained from the flavour point of view when coupled to one quark generation only. We do not show the constraints on the second-generation couplings from the CKM unitarity violation as these remain loose and do not play a role in the global fit in Sec.~\ref{sec:results}. 
    }
    \label{tab:flav_summ}
\end{table}

Summarising, flavour physics constraints point towards flavour non-universal structure where vector-like quarks dominantly couple to only one light generation. Such scenarios can arise in a framework with a gauged flavor non-universal symmetry at the scale $\Lambda$, under which the first-generation SM and vector-like quarks are charged. At this scale, new Yukawa interactions break the global symmetry down to $U(2)^3\times U(1)_{B_1}$ in the quark sector, which continues to forbid flavor-changing neutral currents. However, this symmetry cannot be exact and is necessarily broken by the SM Yukawa interactions. Since flavor violation follows the same pattern as in the SM—being absent at tree level and mediated by the CKM matrix—we have estimated its impact on the bounds derived in this work within such a Minimal Flavour Violating (MFV)-like scenario~\cite {DAmbrosio:2002vsn}. We find that the strongest constraints in Tab.~\ref{tab:DeltaF=1}, particularly from $K^+\to \pi^+\nu\bar\nu$, are sufficiently relaxed so as not to significantly affect our results. Therefore, this single-family-dominant coupling ansatz enables substantial modifications to light quark Yukawa couplings while remaining compatible with stringent constraints from flavor-changing neutral currents.

Nonetheless, even such models with strong flavour protection lead to violations of the CKM unitarity which ultimately result in non-trivial constraints on the specific vector-like states given in Eqs.~\eqref{eq:Mpsi_UD_1st}-\eqref{eq:Mpsi_T1T2_2nd}. Furthermore, the models with vector-like states that are triplets under the $SU(2)$ induce flavour-changing neutral currents either in the down- or up-quark sector proportional to the CKM matrix elements. For the rest of the paper, we choose to work in the basis which results in the less stringent constraints. For instance, we assume the up-quark mass basis when quoting the results for the models that contain $T_1$ and take $M_{T_1}>2.3\times|[\lambda_{T_1}]_i|$ TeV, while we assume the down-quark mass basis when quoting the results for the models that contain $T_2$ and take $M_{T_2}>2.9\times|[\lambda_{T_2}]_i|$ TeV. We summarise the constraints from flavour physics that we will use in the rest of the paper in Tab.~\ref{tab:flav_summ}. The vector-like states that are doublets under the $SU(2)$ gauge group remain unconstrained from the flavour point of view when coupled to one quark generation only. However, there are important direct searches performed at LHC that constrain the masses of these states. We discuss this next.

\subsection{Direct searches \label{sec:directsearch}}

A direct search for pair production of VLQs with subsequent decay into $W^{\pm}q$, where $q$ generically stands for light quarks, has been performed by ATLAS in Ref.~\cite{ATLAS:2024zlo}. 
We use this search to set a lower bound on the VLQ masses that we collectively denote by $\Lambda$. 
In general, the VLQs in the scenario that we consider decay into $W^{\pm}q$, $hq$ and $Zq$. The ATLAS search considers only the $Wq$ final states for the produced pairs' decays. In the ATLAS study, two separate assumptions on the decay branching ratios were made leading to the following bounds on the VLQ masses:
\begin{itemize}
    \item Assuming $\text{BR}(\psi\rightarrow Wq)=1$, a lower bound on the mass was set to $1530$ GeV at $95\%$ CL. We argue that when studying the decays of the VLQs with non-SM charges (i.e.~states with an electric charge $Q=- 4/3$ and $Q=+ 5/3$), this assumption always holds, while for the VLQs with SM charges, one has to assume that some of the VLQ couplings are negligible with respect to the others;
    \item Ref.~\cite{ATLAS:2024zlo} also considers the scenario $\text{BR}(\psi\rightarrow Wq:Zq:hq)=0.5:0.25:0.25$. 
    In this case, the lower bound on the mass was set to $1150$ GeV. This scenario can be recovered in the models we study when analysing the decays of the VLQs with SM charges, but only by making appropriate assumptions on the Yukawa couplings of the VLQs with the SM quarks.
\end{itemize}
For branching ratios different from the indicated values the bounds have to be accordingly rescaled. In particular, if the VLQs have sizeable BRs into $hq$ or $Zq$ the limits presented by ATLAS become weaker. We will now check which values of branching ratios can be concretely realised in the considered models.

\begin{figure}[t!]
    \centering
    \begin{subfigure}[t]{0.4\textwidth}
    \includegraphics[width=\linewidth]{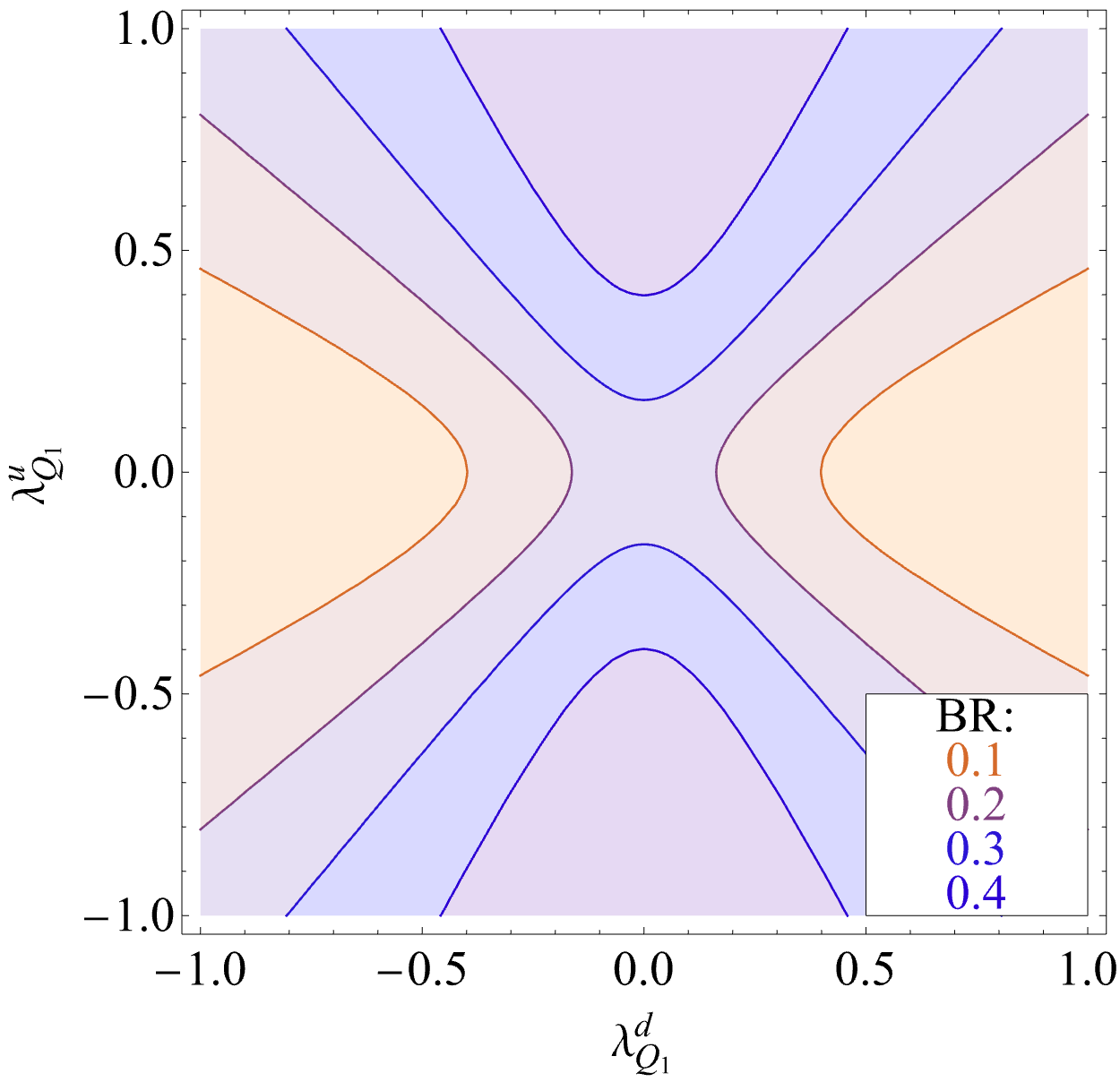}
    \caption{$T$ or $U$ decay into $hu$ or $Zu$.}\label{fig:model1neutralbrT23U}
    \end{subfigure}
    \hspace{1cm}
    \begin{subfigure}[t]{0.4\textwidth}
    \includegraphics[width=\linewidth]{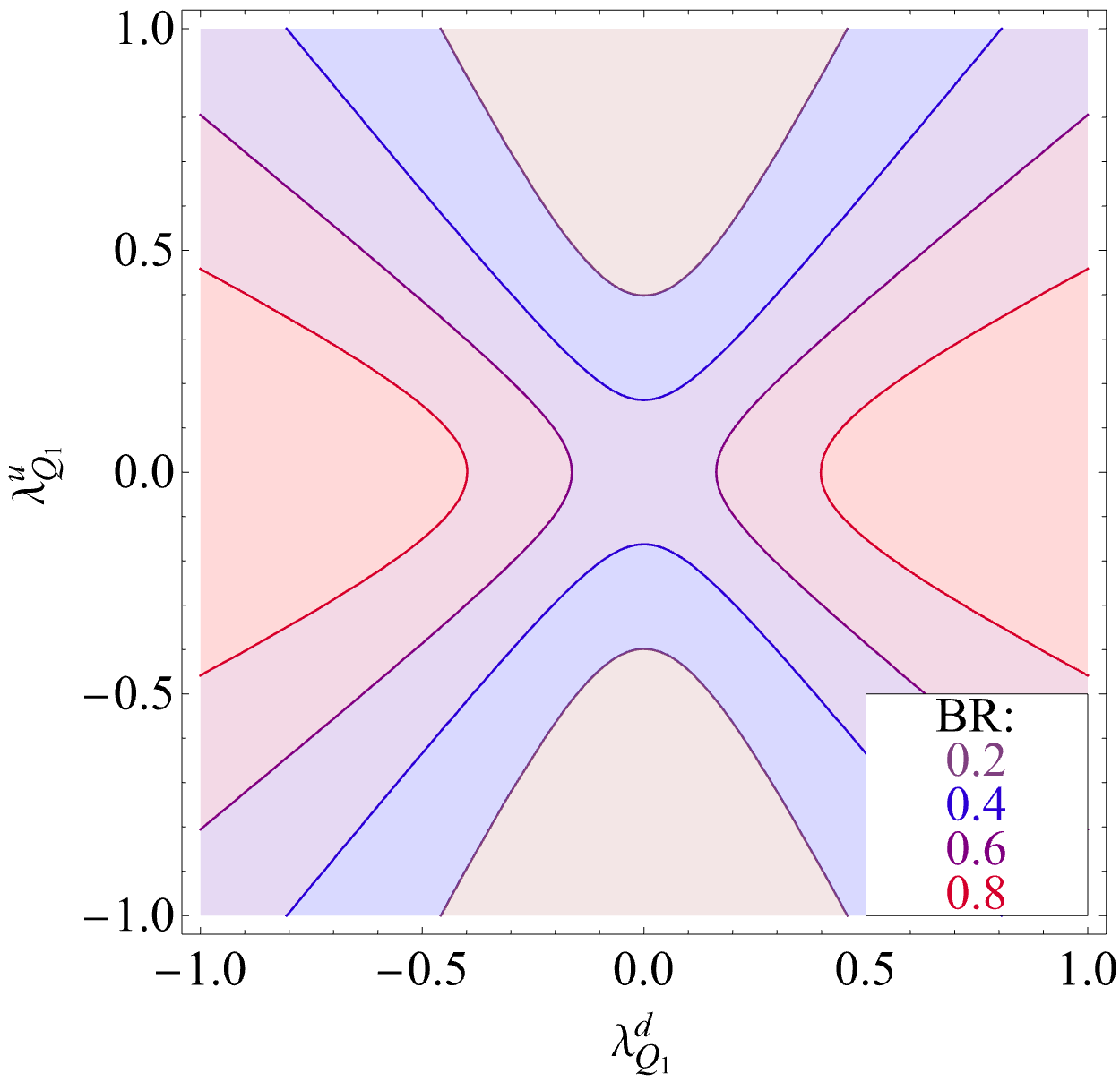}
    \caption{$T$ or $U$ decay into $W^+d$.}\label{fig:model1chargedbrT23U}
    \end{subfigure}\\
    \vspace{10pt}
    \begin{subfigure}[t]{0.4\textwidth}
    \includegraphics[width=\linewidth]{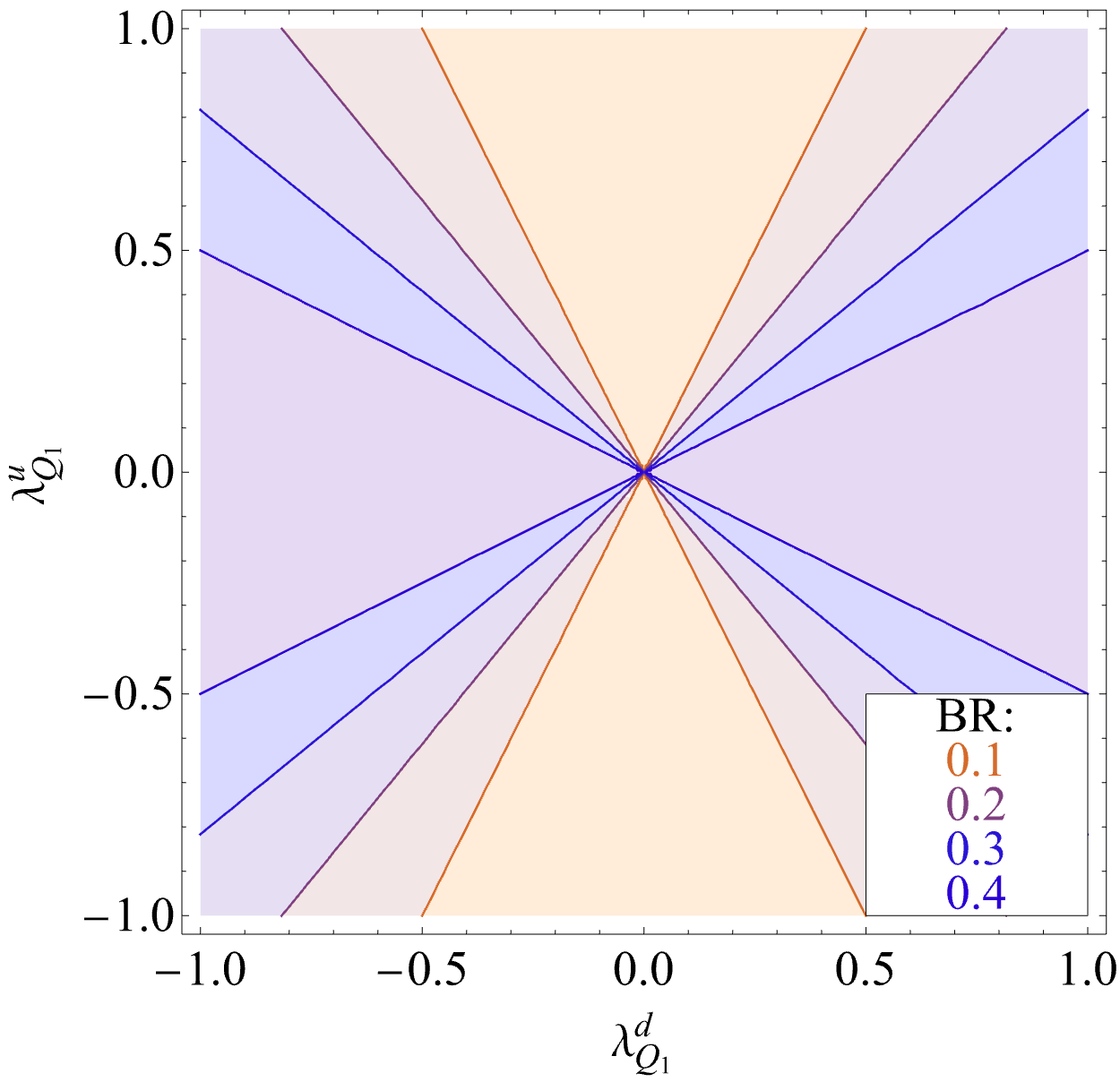}
    \caption{$B$ decay into $hd$ or $Zd$.}\label{fig:model1neutralbrB}
    \end{subfigure}
    \hspace{1cm}
    \begin{subfigure}[t]{0.4\textwidth}
    \includegraphics[width=\linewidth]{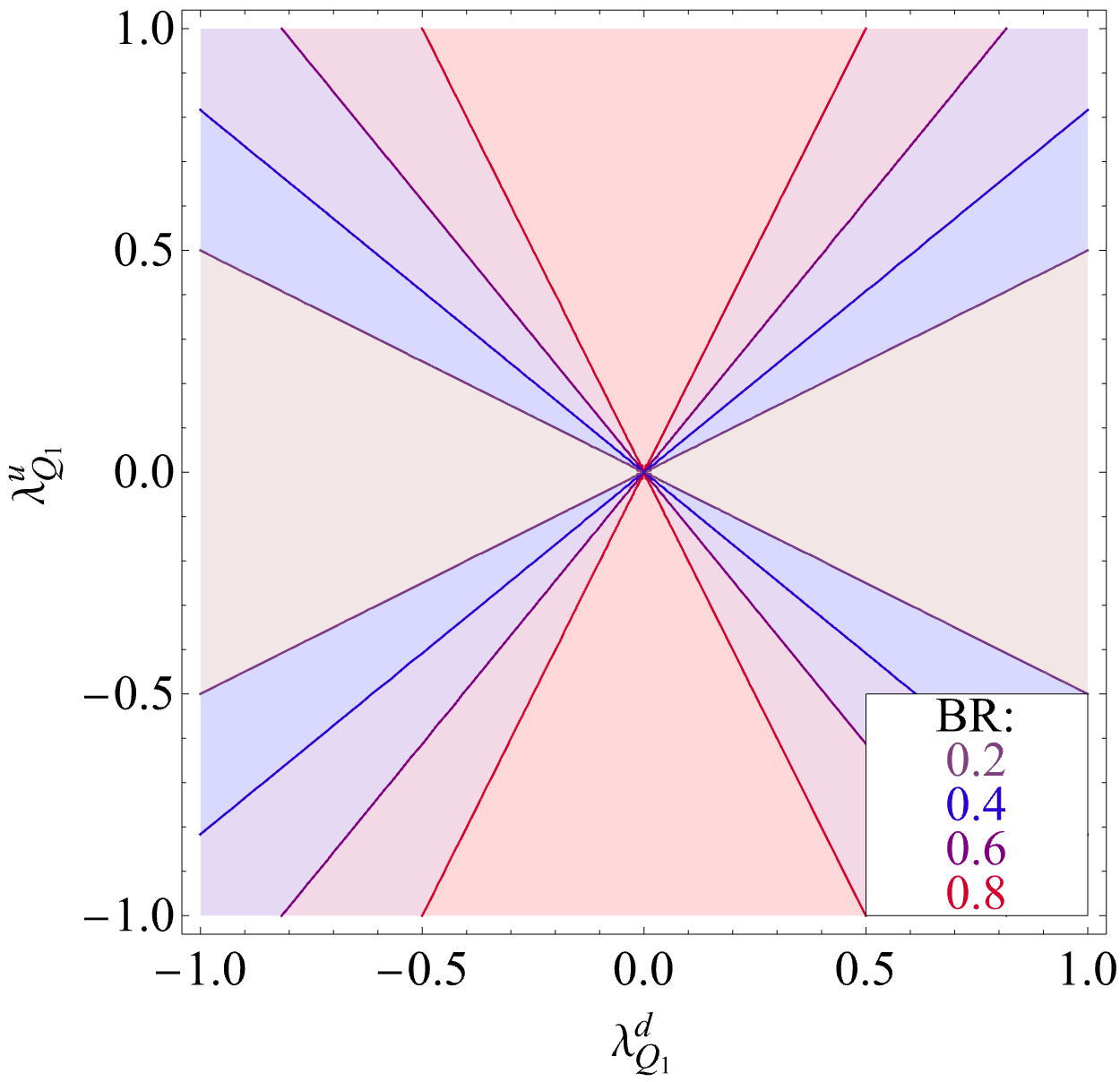}
    \caption{$B$ decay into $W^- u$.}\label{fig:model1chargedbrB}
    \end{subfigure}
    \caption{Branching ratios for the decays of the VLQs introduced in Model 1. The plots are shown as a function of the couplings $\lambda^u_{Q_1}$ and $\lambda_{Q_1}^d$. For the first two plots, the third coupling is set to the best-fit value of the combined electroweak + Higgs fit, $\lambda_U=-0.23$ (see Sec.~\ref{sec:results}).}
    \label{fig:branchingratiosmodel1}
\end{figure}

Let us start from Model 1, where the mass eigenstates after the electroweak symmetry-breaking are $B$ and $T$, mainly originating from the doublet $Q_1=(T,B)^T$, and an additional up-type partner $U$ associated to the $SU(2)$ singlet. The states $T$ and $U$ can decay into $hu$, $Zu$ and $W^+ d$, whereas the final states states for $B$ are $hd$, $Zd$ and $W^-u$, assuming first-generation couplings for the VLQs. The discussion can be applied equivalently to the second generation since the light quarks remain untagged. In the limit that $\Lambda \gg v$, so that a perturbative treatment in the expansion parameter $v/\Lambda$ can be performed, the leading terms for the branching ratios for $T$ and $U$ (Fig.~\ref{fig:model1neutralbrT23U} and \ref{fig:model1chargedbrT23U}) are the same and read
\begin{align}
    &\text{BR}\left( T/U \rightarrow hu \right) = \text{BR}\left( T/U\rightarrow Zu \right) = \frac{1}{2} \frac{\lambda_U^2 + \lambda^{u2}_{Q_1}}{\lambda^{u\,2}_{Q_1} + \lambda^{d\,2}_{Q_1} +2\lambda_U^2 } \overset{\lambda^{u2}_{Q_1}\rightarrow \lambda^{d2}_{Q_1}}{\longrightarrow} \frac{1}{4}  \,,\label{eq:brTUneutralmodel1}\\
    &\text{BR}\left( T/U \rightarrow W^+ d \right) = \frac{\lambda_U^2 + \lambda^{d\,2}_{Q_1}}{\lambda^{u\,2}_{Q_1} + \lambda^{d\,2}_{Q_1} + 2\lambda_U^2} \overset{\lambda^{u2}_{Q_1}\rightarrow \lambda^{d2}_{Q_1}}{\longrightarrow} \frac{1}{2} \label{eq:brTUchargedmodel1}\,.
\end{align}
As for the branching ratios for the decays of $B$ (Fig.~\ref{fig:model1neutralbrB} and \ref{fig:model1chargedbrB}), they are
\begin{align}
    & \text{BR}\left(B \rightarrow hd\right) = \text{BR}\left(B \rightarrow Zd\right) = \frac{1}{2} \frac{\lambda^{d\,2}_{Q_1}}{\lambda^{u\,2}_{Q_1} + \lambda^{d\,2}_{Q_1}} \overset{\lambda^{u2}_{Q_1}\rightarrow \lambda^{d2}_{Q_1}}{\longrightarrow} \frac{1}{4} \,,\label{eq:brBneutralmodel1}\\
    & \text{BR} \left( B \rightarrow W^- u \right) = \frac{\lambda_{Q_1}^{u\,2}}{\lambda_{Q_1}^{u\,2} + \lambda_{Q_1}^{d\,2}} \overset{\lambda^{u2}_{Q_1}\rightarrow \lambda^{d2}_{Q_1}}{\longrightarrow} \frac{1}{2}\,. \label{eq:brBchargedmodel1}
\end{align}
As the equations show, the case of $\text{BR}(\psi\rightarrow Wq:Zq:hq)=0.5:0.25:0.25$ can be recovered if the state $Q_1$ has the same coupling to both the up and down type right-handed SM quarks. Instead, $\text{BR}(\psi\rightarrow Wq)=1$ is recoverable only in the limit in which one of the $Q_1$ couplings is much larger than the others. Specifically, one needs large $\lambda^d_{Q_1}$ for decays of $T$ and $U$, and a large $\lambda^u_{Q_1}$ for $B$. This means that as long as the couplings are all of the same order the limit from direct searches for Model 1 is lower than $1.53$ TeV.

The study of the branching ratios can be generalised to all the models being considered. For example, in Model 2 with particle content consisting in the doublet $Q_1=(T,B)^T$ and the singlet $D$, the branching ratios for the three available decay modes are related to those of Model 1 by the exchange of the particles with positive electric charge with the negative ones and by replacing the couplings $\lambda_U\mapsto \lambda_D$, $\lambda^u_{Q_1}\mapsto \lambda^d_{Q_1}$, and $\lambda^d_{Q_1} \mapsto \lambda^u_{Q_1}$. The branching ratio plots for the particles in Model 2 can be obtained from those of Model 1 in Fig.~\ref{fig:branchingratiosmodel1} by performing the same substitutions. Also in this case, the ATLAS assumption  $\text{BR}(\psi\rightarrow Wq:Zq:hq)=0.5:0.25:0.25$ is recovered in the limit that $Q_1$ couplings $\lambda^d_{Q_1}$ and $\lambda^u_{Q_1}$ are the same. The generalisation for other models is discussed in App.~\ref{app:GeneralisedBR}. 

\indent Finally, we consider the decays of $X$ (with electric charge $Q=5/3$) and $Y$ (with electric charge $Q=-4/3$) which stem from the vector-like states $Q_5$, $Q_7$, $T_1$, and $T_2$ after electroweak symmetry-breaking. For these particles, there is only one final state involving a SM quark
\begin{equation}
    X\rightarrow W^+u\,, \quad Y \rightarrow W^- d\,,
\end{equation}
as the decay into a final state involving a neutral boson would require the presence of light quarks with a charge $Q=5/3$ or $Q=-4/3$. In this case, the bounds obtained by ATLAS are $M_{X/Y}>1.53\text{ TeV}$. Such exotic particles are introduced in all models except for Models 1 and 2. 
To ease the comparison between the different models, we set the new physics scale generically to $\Lambda = 1.6$ TeV as imposed by direct searches. We consider this a conservative option as the electroweak symmetry-breaking necessarily induces mass splitting between the various vector-like states resulting in different collider signatures than those studied by ATLAS. 
For instance, it is natural to expect sequential decays like $X\to W^+ T(\to Zu)$ from $Q_7=(X,T)^T$ with multi-particle final states decreasing ${\rm BR}(X\to W^+ u)$. 
Therefore, in addition to flavour constraints in Tab.~\ref{tab:flav_summ} which affect the states $U,D,T_1$, and $T_2$, direct searches allow us to restrict the masses of vector-like quarks that are doublets under the $SU(2)$ gauge group, $M_\Psi>1.6$ TeV, with $\Psi=\{Q_1,Q_5,Q_7\}$. We use this as underlying restrictions on the model parameters in the following discussion of electroweak and Higgs processes. 

\subsection{Electroweak precision observables \label{sec:EWPTs}}
\begin{table}[t]
    \centering
    \renewcommand{\arraystretch}{1.5}
    \begin{tabular}{|c|c||c|c|}
        \hline
          Observable & Definition & Observable & Definition \\\hline
          $\Gamma_Z$ & $\sum_f \Gamma(Z\to f\bar f)$ & $R_{uc}$ & $\frac{\Gamma(Z\to u\bar u) + \Gamma(Z\to c\bar c)}{2\sum_q \Gamma(Z\to q\bar q)}$ \\ \hline
          $\sigma_{\rm had}$ & $\frac{12\pi}{m_Z}\frac{\Gamma(Z\to e^+ e^-)\Gamma(Z\to q\bar q)}{\Gamma_Z^2}$ & $m_W$ & $m_W$ \\ \hline
          $R_f$ & $\frac{\Gamma(Z\to f\bar f)}{\sum_q \Gamma(Z\to q\bar q)}$ & $\Gamma_W$ & $\sum_{f_1, f_2} \Gamma(W\to f_1 f_2)$ \\ \hline
          $A_f$ & $\frac{\Gamma(Z\to f_L \bar f_L)-\Gamma(Z\to f_R \bar f_R)}{\Gamma(Z\to f\bar f)}$ & ${\rm BR}(W \to \ell\nu)$ & $\frac{\Gamma(W \to \ell\nu)}{\Gamma_W}$  \\ \hline
          $A_{\rm FB}^{0,\ell}$ & $\frac{3}{4} A_e A_\ell$  &$R_{W_c}$ & $\frac{\Gamma(W\to cs)}{\Gamma(W\to ud) + \Gamma(W\to cs)}$    \\ \hline
          $A_c^{\rm FB}$ & $\frac{3}{4} A_e A_c$  &  $A_b^{\rm FB}$  & $\frac{3}{4} A_e A_b$\\ \hline
    \end{tabular}
    \caption{Observables and their definitions used to construct the EW fit.}
    \label{tab:ewpos}
\end{table}

To better understand the effect of vector-like quarks on electroweak physics, we construct the likelihood to perform the electroweak fit. For this purpose, we use the electroweak observables (EWPOs) measured at the $Z$-pole, $O_{\rm EWPO}^Z =\{\Gamma_Z,\,\sigma_{\rm had},\, R_f,\, A_f,\, A_{\rm FB}^{0,\ell},\, A_q^{\rm FB},\, R_{uc}\}$, and $W$-pole, $O_{\rm EWPO}^W =\{m_W$, $\Gamma_W$, ${\rm BR}(W\to \ell \nu)$, $R_{W_c}\}$, with $f=\{$$e$, $\mu$, $\tau$, $(s)$, $c$, $b\}$\footnote{\,There is no measurement of $R_s$.}, $\ell=\{e,\,\mu,\,\tau\}$, and $q=\{c,\,b\}$. The observables are defined in Tab.~\ref{tab:ewpos}, while the corresponding measurements and the SM predictions have been extracted from \cite{ALEPH:2005ab,ALEPH:2013dgf,Janot:2019oyi,dEnterria:2020cgt,SLD:2000jop,
ParticleDataGroup:2020ssz,CDF:2005bdv,LHCb:2016zpq,ATLAS:2016nqi,D0:1999bqi,ATLAS:2020xea} and are collected in~\cite{Breso-Pla:2021qoe}. Using this information, we define the $\chi_{\rm EWPO}^2$-function as
\begin{equation}
\chi_{\rm EWPO}^2 = \sum_{ij}[O_{i,{\rm exp}}-O_{i,{\rm th}}] (\sigma^{-2})_{ij}[O_{j,{\rm exp}}-O_{j,{\rm th}}] \,,
\label{eq:chi2_EWPO}
\end{equation}
with $\sigma^{-2}$ being the inverse of the covariance matrix~\cite{ALEPH:2005ab,ALEPH:2013dgf}.

To describe new physics effects, we work in the $\{\alpha_{EM},m_Z,G_F\}$ input scheme, where the relevant effective Lagrangian describing dynamics of the electroweak gauge bosons reads
\begin{align}
\mathcal{L}_{\rm eff} &\supset - g_2 \left[\left(W^{+\mu} j^{-}_{\mu}+{\rm h.c.}\right)+ Z^\mu j_\mu^Z\right]+\frac{g_2^2 v^2}{4}(1+\delta m_W)^2 W^{+\mu} W^-_{\mu}+\frac{g_2^2 v^2}{8 c_W^2}Z^{\mu} Z_{\mu}\,,\\
j^{-}_{\mu} &= \frac{1}{\sqrt{2}}\left[\bar u_L^i \gamma_{\mu} \left(V_{ij}+\delta g_{ij}^{Wq}\right)d_L^j + \bar \nu_L^i \gamma_{\mu} \left(\delta_{ij}+\delta g_{ij}^{W \ell }\right)e_L^j\right]\,,\\
j_\mu^Z &= \,\frac{1}{c_W} \left[\bar f^i_L \gamma_{\mu}
\left( g_L^{Zf}\delta_{ij}+ \delta g_{L\,ij}^{Zf}\right) f_L^j
+\bar f^i_R \gamma_{\mu}
\left( g_R^{Zf} \delta_{ij}+ \delta g_{R\,ij}^{Zf}\right) f_R^j
\right]\,,
\label{eq:EW_eff_Lag}
\end{align}
where
\begin{align}
g_L^{Zf}=\,T_f^3-s_W^2 Q_f\,,\quad\quad g_R^{Zf}= -s_W^2 Q_f\,,
\end{align}
with $T^3_f$ and $Q_f$ being the third component of weak isospin and the electric charge of the fermion $f$, respectively, and $c_W$ ($s_W$) is the cosine (sine) of the Weinberg angle. 

\begin{figure}[t]
\centering
\begin{subfigure}[b]{0.48\linewidth}
\centering
    \begin{tikzpicture}
        \begin{feynman}
            \vertex (phi1) at (0,0) {\(\phi\)};
            \vertex[right = 4 em of phi1] (A);
            \vertex[right = 4.5 em of A] (B);
            \vertex[right = 3.75em of B] (W1) {\(W^I_\mu\)};
            \vertex[below = 4.5 em of phi1] (phi3) {\(\phi^*\)};
            \vertex[below = 4.5 em of A] (D);
            \vertex[below = 4.5 em of B] (C);
            \vertex[below = 4.5 em of W1] (B1) {\(B_\nu\)};
            \diagram*{
                (phi1) -- [scalar] (A);
                (W1) -- [boson] (B);
                (phi3) -- [scalar] (D);
                (B1) -- [boson] (C);
                (A) -- [fermion, edge label  = \(Q_1\)] (B) -- [fermion, edge label = \(Q_1\)] (C) -- [fermion, edge label = \(Q_1\)] (D) -- [fermion, edge label = \(U\)] (A);
            };
        \end{feynman}
    \end{tikzpicture}    
\end{subfigure}
\begin{subfigure}[b]{0.48\linewidth}
\centering      
    \begin{tikzpicture}
        \begin{feynman}
            \vertex (phi1) at (0,0) {\(\phi\)};
            \vertex[right = 4em of phi1] (A);
            \vertex[right = 4.5 em of A] (B);
            \vertex[right = 3.75em of B] (phi4) {\(\phi^*\)};
            \vertex[below = 4.5 em of phi1] (phi3) {\(\phi^*\)};
            \vertex[below = 4.5 em of A] (D);
            \vertex[below = 4.5 em of B] (C);
            \vertex[below = 4.5 em of phi4] (phi2) {\(\phi\)};
            \diagram*{
                (phi1) -- [scalar] (A);
                (phi4) -- [scalar] (B);
                (phi2) -- [scalar] (C);
                (phi3) -- [scalar] (D);
                (A) -- [fermion, edge label  = \(Q_1\)] (B) -- [fermion, edge label = \(U\)] (C) -- [fermion, edge label = \(Q_1\)] (D) -- [fermion, edge label = \(U\)] (A);
            };
        \end{feynman}
    \end{tikzpicture}
\end{subfigure}
\caption{Representative box diagrams in Model 1 involving the VLQ-VLQ-Higgs coupling which match onto the SMEFT operators $\cO_{\phi WB}$ and $\cO_{\phi D}$.}
\label{fig:CHD_CHWB}
\end{figure}
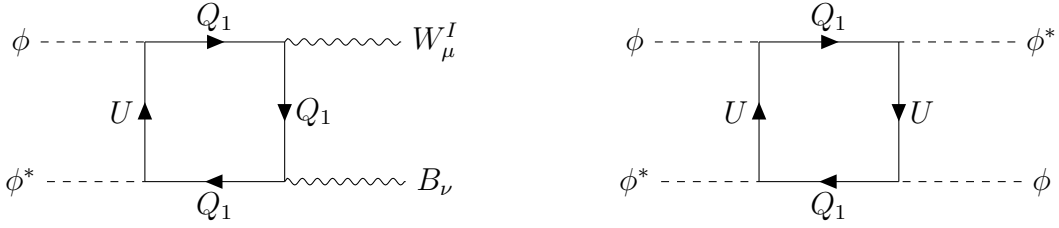

Therefore, we capture the NP effects through the shifts in the gauge couplings and the $W$-boson mass. A part of the results for $\delta g^Z$ have been shown in Eqs.~\eqref{eq:delta_g_ZuL}--\eqref{eq:delta_g_ZdR}, while the rest of the expressions are collected in App.~\ref{app:EWPO} for completeness. 
In addition to the tree-level contributions to the operators in Tab.~\ref{tab:treelevelmodelItoVIII} which modify the electroweak gauge couplings and always depend on two new physics couplings at most, there is an additional universal contribution arising at one loop
\begin{align}
\delta g_{L\,ii}^{Zf}=
-\frac{v^2}{4}\left(T_f^3+\frac{g_1^2}{g_2^2-g_1^2}Q_f\right)\cC_{\phi D}-v^2\frac{g_2g_1}{g_2^2-g_1^2}Q_f\,\cC_{\phi WB}\,.
\label{eq:universal_shift}
\end{align}
Two representative diagrams involved in the evaluation of such universal contribution for the specific example of Model 1 are shown in Fig.~\ref{fig:CHD_CHWB}. Importantly, these contributions make the electroweak fit sensitive to the Yukawa coupling between two vector-like quarks and the Higgs boson which enters the effective light quark Yukawa. Consequently, we can probe all three couplings that determine the contributions to $\cC_{u\phi}$ and $\cC_{d\phi}$ in Tab.~\ref{tab:treelevelmodelItoVIII}, constraining the possible size of the light quark Yukawa couplings through the electroweak precision measurements. 

We show the results obtained when confronting Models 1--8 with $\chi^2_{\rm EWPO}$ in Sec.~\ref{sec:results}. 

\subsection{Higgs physics \label{sec:Higgs}}
Beyond electroweak observables, the processes involving the Higgs boson get modified. The main effect of new vector-like quarks in Higgs physics is the generation of effective Higgs couplings to gluons and photons at one-loop level, as well as a universal shift to all the couplings via the operators $\mathcal{O}_{\phi D}$ and $\mathcal{O}_{\phi\Box}$.
Furthermore, strongly enhanced light quark Yukawa couplings open up new production channels where the Higgs boson can directly couple to the partons. In addition, the decays to light quarks will get enhanced. Since they remain untagged, this is reflected in all the SM branching ratios (BRs) being reduced. 
\par
Typically, the possibility of an enhanced light quark Yukawa coupling is not considered in the available global fits. Thus, we will first check -- not considering that the models also enhance the light quark Yukawa couplings, hence not accounting for the suppressed BRs and the opening up of the new production modes -- how a global fit constrains the parameter space taking into account the VLQ matching to the other SMEFT operators. For this purpose we can make use of the results of the global fit performed in \cite{ATLAS:2024lyh}\footnote{ Based on the measurements performed in \cite{ATLAS:2020wny, ATLAS:2022fnp, ATLAS:2023hyd, ATLAS:2023pwa}.} in which simplified template cross-sections and fiducial differential cross-sections in different decay channels are interpreted in terms of SMEFT. As already mentioned, this parameterisation does not consider the enhanced light quark Yukawa case. Moreover, the fit assumes that operators that include the light quark generations are flavour universal. Those results can hence be used assuming that the second generation contributions are suppressed with respect to the first generation by the small charm/strange parton distribution functions. We implement the results in a Gaussian approximation. 
\par 
In the second fit, we consider the modifications due to the new production modes and modified BRs using the data provided in \cite{CMS:2022dwd}.
We perform the fit using the signal strengths
\begin{equation}
\mu_i=\frac{\sigma \cdot \text{BR}_{i}}{\sigma^{\rm SM} \cdot \text{BR}_{i}^{\rm SM}} \label{eq:signalstrenght}
\end{equation}
with $i=W^+W^-, \, ZZ, \, b\bar{b}, \, \tau^+ \tau^-, \mu^+ \mu^-, \,\gamma\gamma$. The values for the Standard Model BRs and the signal strength measurements reported by \cite{CMS:2022dwd} are summarised in Tab. \ref{tab:higgdecaychannels}. The Higgs production cross-section $\sigma$ is obtained as follows
\begin{equation}
\sigma= \left[48.68 + 2.83\!\cdot\! 10^{4}\, v^2 \cC_{\phi G}+(8.52\,  \kappa_u^2 +2.71 \, \kappa_d^2)\! \cdot\! 10^{-6}+2.53\!\cdot 10^{-3} \kappa_s^2 + 0.25\, \kappa_c^2\right]\text{pb}\,, \label{eq:Higgs_productionxs}
\end{equation}
where $\kappa_u$, $\kappa_d$, $\kappa_c$, and $\kappa_s$ are defined with respect to the reference mass values $m_u = 2.2\text{ MeV}$, $m_d = 4.7 
\text{ MeV}$, $m_c = 1.27\text{ GeV}$, and $m_s = 95
\text{ MeV}$. 
The terms proportional to $\kappa_{u,d}$ are from Higgs production from the initial parton quarks, which becomes comparable to gluon fusion for values of $\kappa$ of $\mathcal{O}(1000)$, instead the first two terms are the SM Higgs pair production in gluon fusion cross section and the second term the contribution to gluon fusion from the operator $\mathcal{O}_{\phi G}$, assuming the dominance of gluon fusion with respect to other Higgs production modes. We have computed the numerical values for the gluon fusion production cross sections in terms of $\cC_{\phi G}$ using \texttt{higlu}~\cite{higlu} and including a $K$-factor to scale to the $\text{N}^3\text{LO}$ cross-section from~\cite{NNNLO, Mistlberger:2018etf}. For the quark fusion to a Higgs boson, we used the results of~\cite{Alasfar:2019pmn, Balzani:2023jas} scaling again by a $K-$factor to NLO.
We note that the dependence on $\kappa_{u,d}$ is taken into account quadratically. Even though in the SMEFT language this corresponds formally to a $\mathcal{O}(1/\Lambda^4)$ term, it is essential to include it as the $\mathcal{O}(1/\Lambda^2)$ terms are suppressed by the small quark masses. 
\begin{table}[t]
\centering
\renewcommand{\arraystretch}{1.5}
\begin{tabular}{|c|c|c|}
\hline
Decay Channel $i$ & BR$^{\text{SM}}_i$ &   $\mu^\text{exp}_i$ \\
\hline
$h\rightarrow WW$ &       $22.00\%$              &          $0.97\pm0.09$                 \\\hline
$h\rightarrow ZZ$ &            $2.71\%$         &         $0.97\pm 0.12$                 \\\hline
$h\rightarrow b\bar{b}$ &       $57.63\%$              &      $1.05\pm0.22$       \\\hline
$h\rightarrow \tau^+ \tau^-$ &    $6.21\%$                 &      $0.85\pm 0.10$                     \\\hline
$h\rightarrow \mu^+\mu^-$ &      $0.0216\%$               &        $1.21\pm 0.44$                    \\\hline
$h\rightarrow \gamma\gamma$ &    $0.227\%$                 &        $1.13\pm0.09$                   \\
\hline 
\end{tabular}
\caption{Values for the Standard Model branching ratios for each of the decay channels, along with the channels' experimental signal strength with errors reported in the Gaussian approximation, used in the construction of the Higgs fit based on Ref.~\cite{CMS:2022dwd}.}
\label{tab:higgdecaychannels}
\end{table}

The Higgs branching ratios were computed using the formulae provided in \cite{Brivio:2019myy} for the individual and the total width. To the total width, we added the light quark Yukawa contributions by appropriate rescaling the $h\to b\bar{b}$ width. In addition, we added the SM contributions to the $h\to \gamma \gamma$ and $h\to gg$ rates rescaled by $\cC_{\phi,kin}$ in Eq.~\eqref{eq:C_kin}, since \cite{Brivio:2019myy} takes into account only the LO contributions. The explicit expressions are reported in App.~\ref{app:Hfit}.

We perform a $\chi_{\rm Higgs}^2$ fit using the correlation matrix provided in the auxiliary material of Ref.~\cite{CMS:2022dwd}.

\section{Results}
\label{sec:results}
In this section, we analyze the simplified models introduced in Sec.~\ref{sec:EFT_models} in the context of the combined constraints from direct searches, flavour, electroweak and Higgs physics. We present the results for the viable parameter space and show which Yukawa coupling modifications can be achieved. We keep the discussion detailed for Model 1 remarking that the analogous points hold for other models for which we show only the final results. We recall that we couple the vector-like quarks to one generation only so we start the discussion with the first generation and discuss the second generation afterwards. In the following, we set a universal mass scale for the vector-like quarks (for all models) to be $\Lambda=1.6\text{ TeV}$ to avoid any bounds from direct searches discussed in Sec.~\ref{sec:directsearch}. 

\subsection{First generation}
\begin{figure}[t]
\centering
\includegraphics[width=0.45\textwidth]{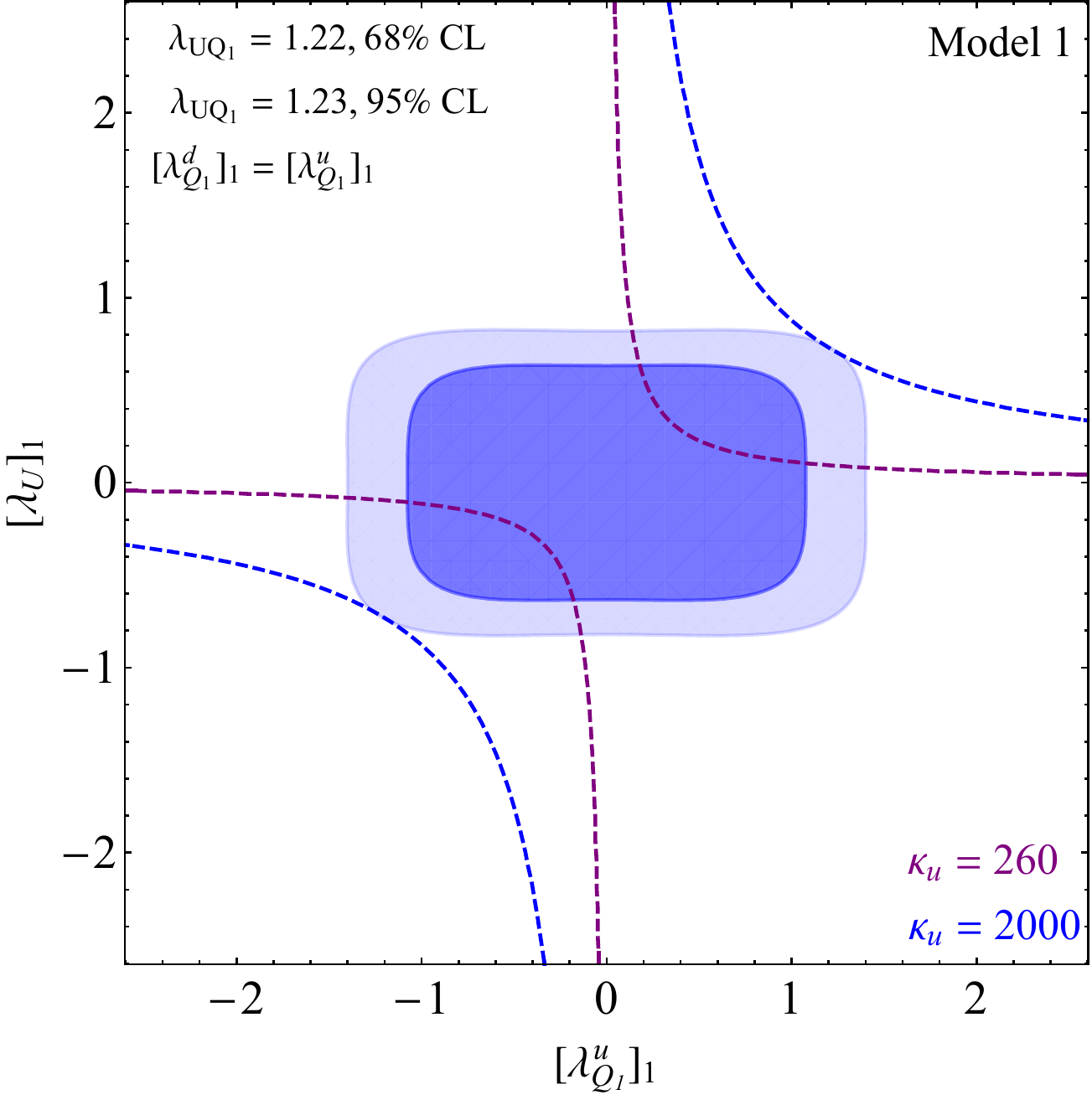} \quad
\includegraphics[width=0.45\textwidth]{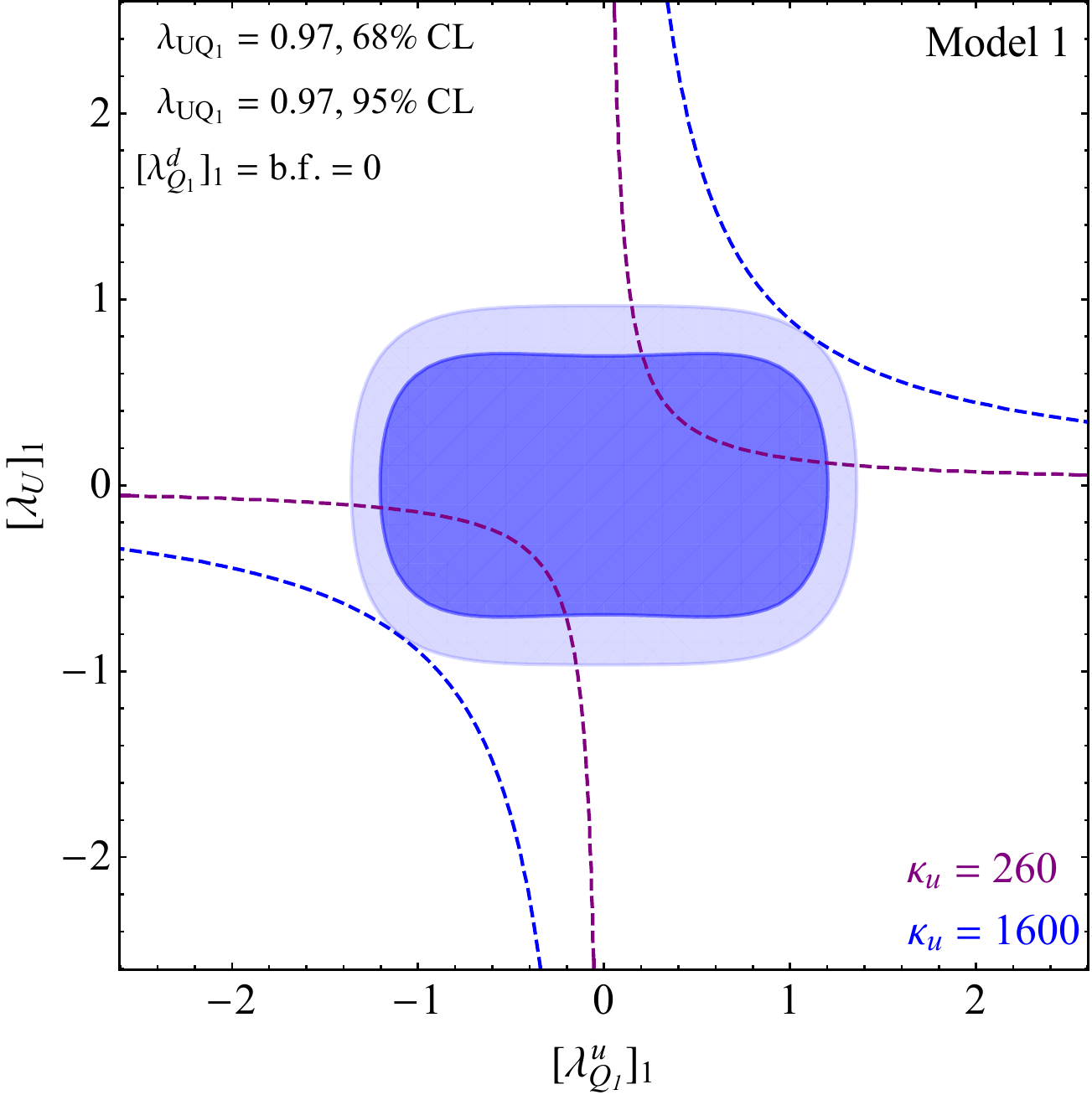}
\caption{Electroweak fit for Model 1. The lighter (darker) color shows the 95\% (68\%) CL interval. The dashed lines indicate fixed values of $\kappa_u$ in the parameter plane. The masses of the VLQs have been set to a universal value $\Lambda=1.6\text{ TeV}$. \textbf{Left:} Fit setting $\lambda_{Q_1}^d=\lambda_{Q_1}^u$. \textbf{Right:}  Fit setting $\lambda_{Q_1}^d$ to its best-fit (b.f.) value. \label{fig:EWPTfitsMod1}}
\end{figure}

\begin{figure}[ht!]
    \centering
    \includegraphics[width=0.95\linewidth]{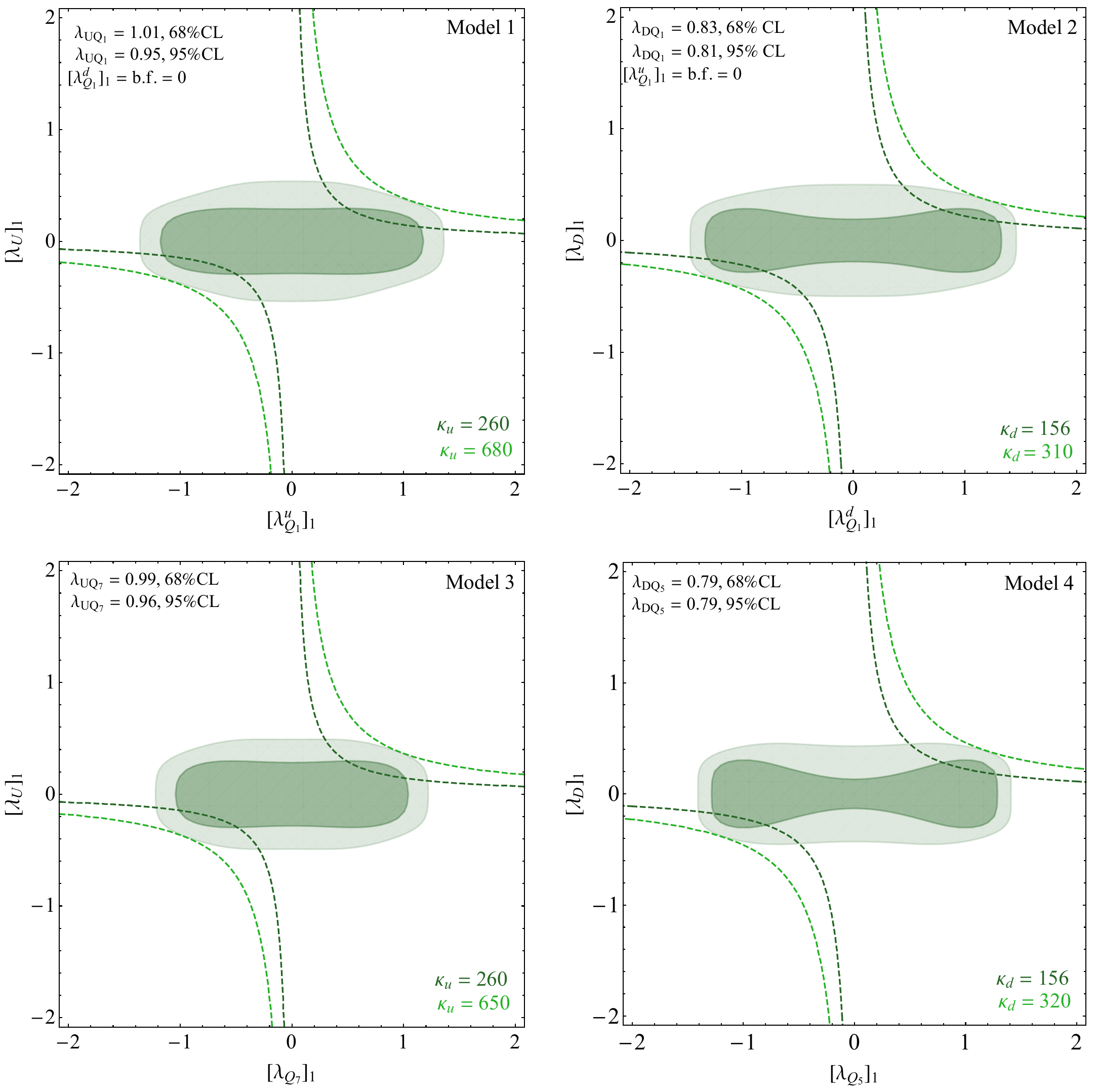}
    \caption{Allowed parameter range from a fit to EWPT and Higgs data for Models 1--4 with VLQs coupled to the first-generation quarks. The lighter (darker) color shows the 95\% (68\%) CL interval. The dashed lines show fixed values of $\kappa_q$ in the parameter plane.}
    \label{fig:1st_global_M1-M4}
\end{figure}

\begin{figure}[t!]
    \centering
    \includegraphics[width=0.95\linewidth]{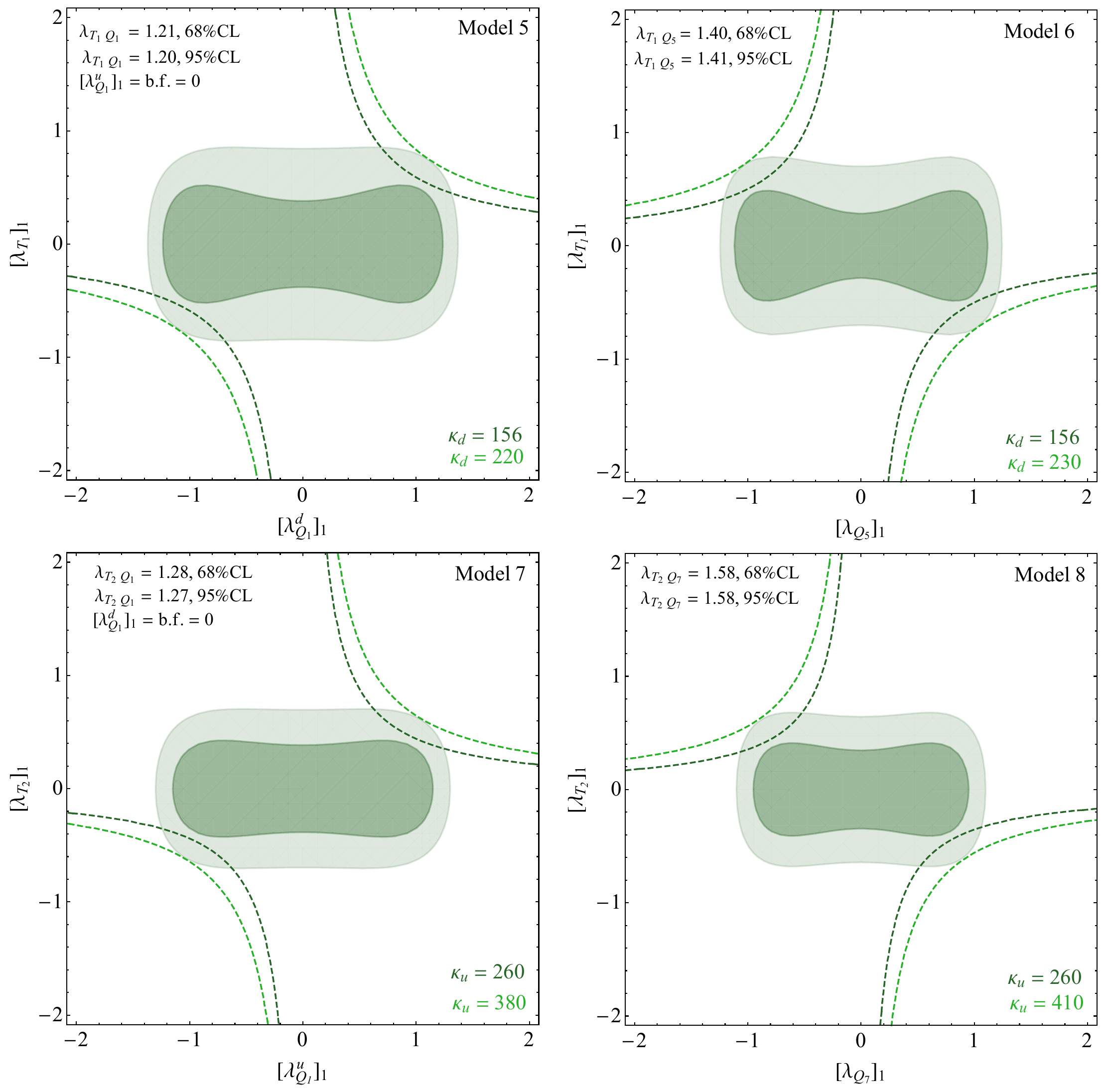}
    \caption{Allowed parameter range from a fit to EWPT and Higgs data for Models 5--8 with VLQs coupled to the first-generation quarks. The lighter (darker) color shows the 95\% (68\%) CL interval. The dashed lines show fixed values of $\kappa_q$ in the parameter plane.}
    \label{fig:1st_global_M5-M8}
\end{figure}

In Fig.~\ref{fig:EWPTfitsMod1}, we show the allowed parameter space in the plane of $\lambda_U$ and $\lambda_{Q_1}^u$ from electroweak precision data only. The left plot shows the case in which $\lambda_{Q_1}^d=\lambda_{Q_1}^u$ while the right plot shows the case where $\lambda_{Q_1}^d$ is set to its best-fit value obtained by minimising $\chi^2_{\rm EWPO}$ in Eq.~\eqref{eq:chi2_EWPO}. The fourth coupling $\lambda_{UQ_1}$ is set to maximise $\kappa_u$ under the condition 
\begin{equation}
    \chi^2_{\rm EWPO} <  \chi^2_{\rm EWPO}\Big|_{\rm min}\!\! + \chi^2(3,95\%)\,,
    \label{eq:lamdaQQ}
\end{equation}
where $\chi^2(3,95\%)$ is the value of the chi-squared distribution with three degrees of freedom and $p$-value of 0.05 corresponding to the $95\%$ CL interval. The dashed lines show fixed values of $\kappa_u$ in the $\lambda_U-\lambda_{Q_1}^u$ plane, where one of the lines corresponds to the projected sensitivity for $\kappa_u=260$ from~\cite{Balzani:2023jas}, while the other line lies tangent to the 95\% CL region. We note that the bounds derived from electroweak precision tests (EWPTs) do not stem directly from the light quark Yukawa modifications but come from the other operators that these models introduce, as discussed in Sec.~\ref{sec:EWPTs}. As can be seen from Fig.~\ref{fig:EWPTfitsMod1}, the choice for the $Q_1$ coupling to $d_R^1$, $[\lambda_{Q_1}^d]_1$, makes a slight difference in the final result for the possible value of $\kappa_u$ one can achieve. The $SU(2)_R$-symmetric choice $[\lambda_{Q_1}^d]_1 = [\lambda_{Q_1}^u]_1$ allows for a slightly larger $\kappa_u=2000$ compared to $\kappa_u=1600$ when $[\lambda_{Q_1}^d]_1$ is set to its best-fit value (in this case $[\lambda_{Q_1}^d]_1={\rm b.f.} =0$). This is mainly due to the resulting value of $\lambda_{UQ_1}$ from Eq.~\eqref{eq:lamdaQQ} when $[\lambda_{Q_1}^d]_1$ is included in $\chi^2_{\rm EWPO}$ or not. As we make no prior assumptions on the value of the couplings that appear in the individual models but do not influence the $\kappa_q$ values, we always keep them in the corresponding likelihood and set them to their best-fit values. This is done when presenting the final results for all models in Figs.~\ref{fig:1st_global_M1-M4}--\ref{fig:2nd_global_M5-M8}. However, we note that taking the UV-inspired choice, like the $SU(2)_R$-symmetric one with $[\lambda_{Q_1}^d]_1 = [\lambda_{Q_1}^u]_1$, does not make an influential difference in our conclusions.

\begin{figure}[t]
\centering
\includegraphics[width=0.45\textwidth]{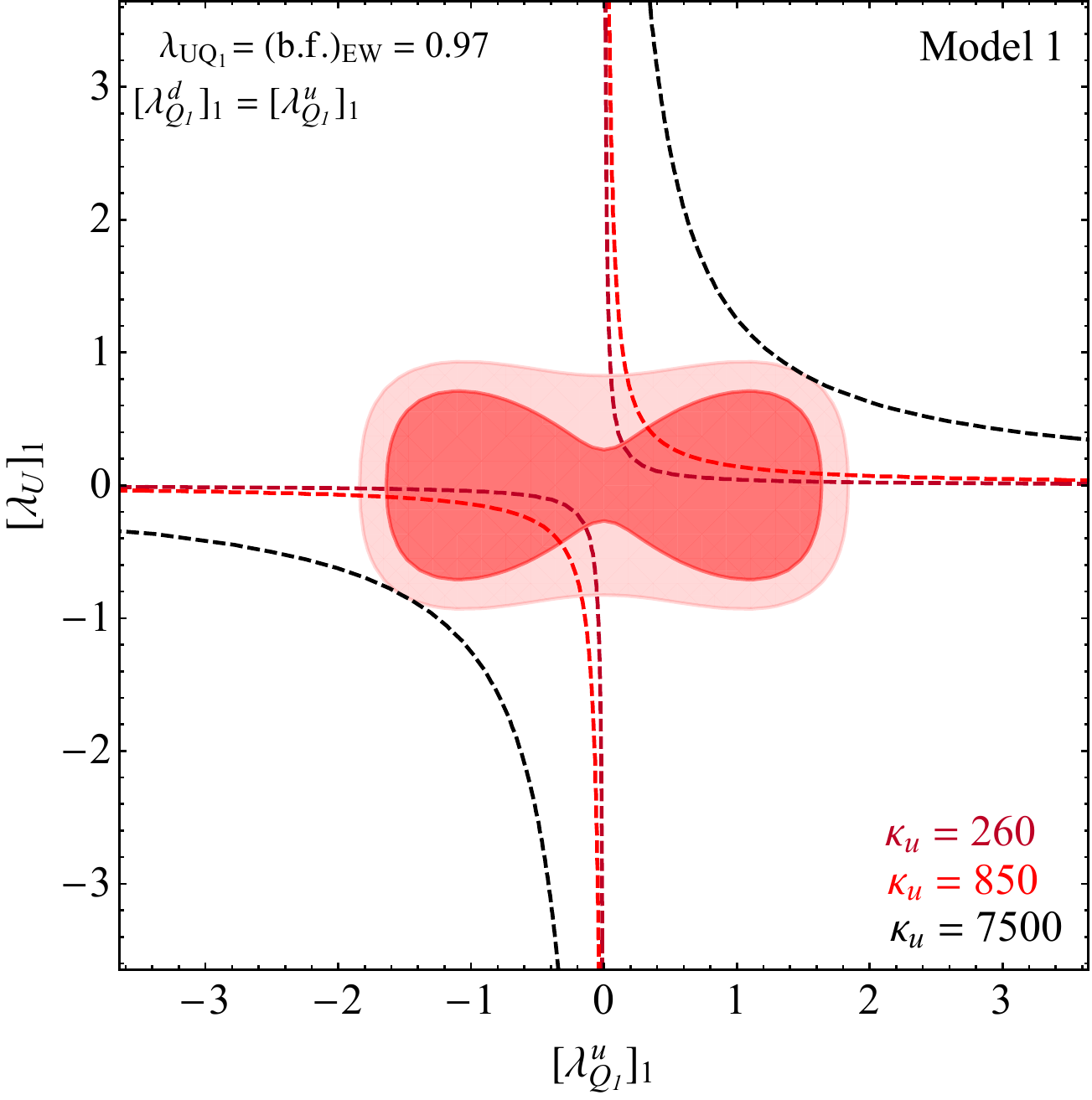}\quad
\includegraphics[width=0.45\textwidth]{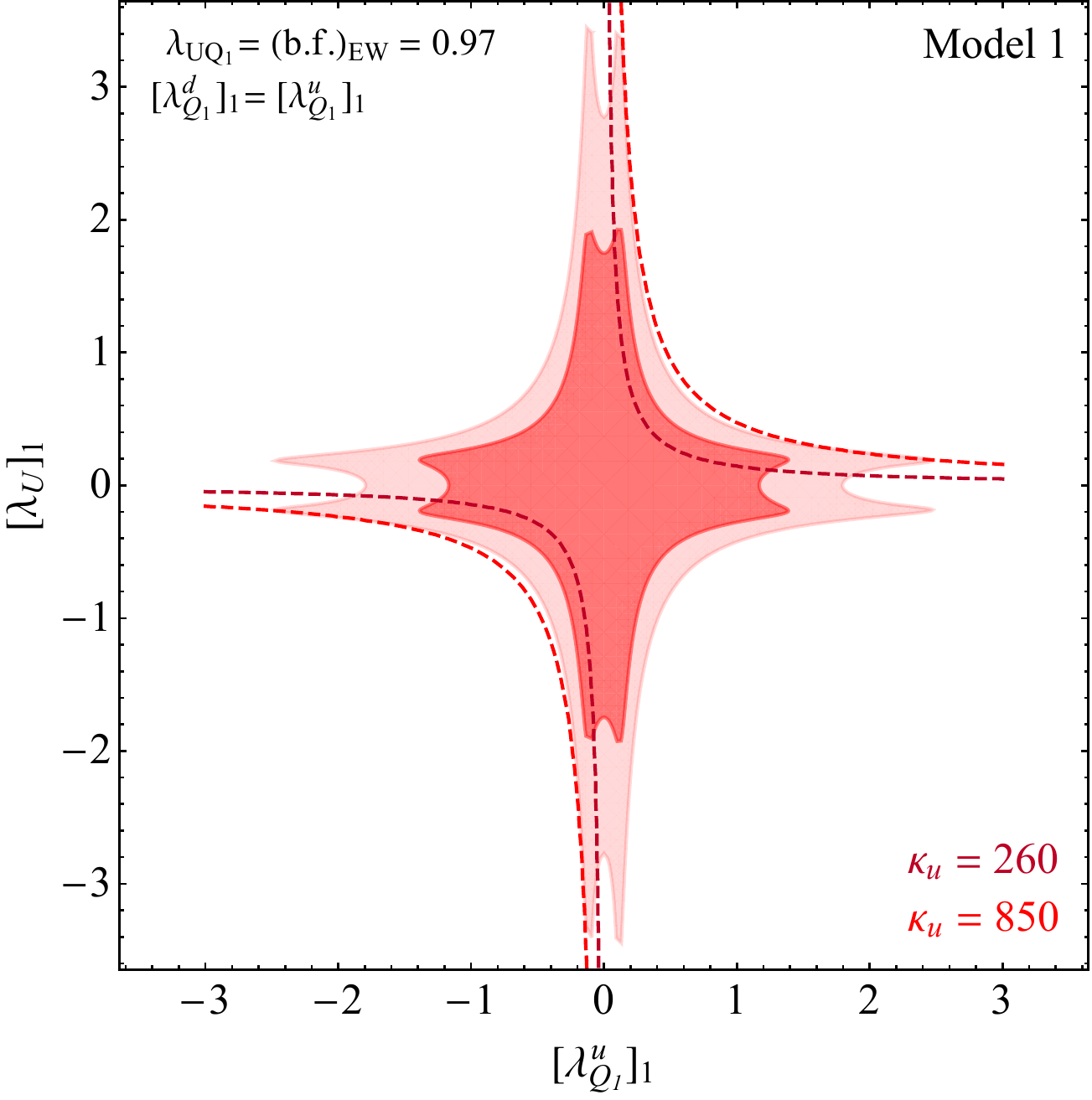} 
\caption{ Allowed parameter range from Higgs data for Model 1. The lighter (darker) color shows the 95\% CL (68\% CL) interval. The dashed lines show fixed values of $\kappa_u$ in the parameter plane.  \textbf{Left:} ATLAS global fit without the modifications due to enhanced $\kappa_q$~\cite{ATLAS:2024lyh}. \textbf{Right:} Fit on the total rates of CMS~\cite{CMS:2022dwd} including also the modifications due to enhanced $\kappa_q$ of the first generation. In both plots, $\lambda_{Q_1}^d =\lambda_{Q_1}^u$ while the value of $\lambda_{UQ_1}$ is set to its best-fit value from the EW fit.
\label{fig:HiggsfitsMod1}}
\end{figure}

Furthermore, the vector-like quarks affect the Higgs processes and Fig.~\ref{fig:HiggsfitsMod1} shows two different fits to Higgs data for the case where $\lambda_{UQ_1}$ is set to the value extracted from the EW fit using Eq.~\eqref{eq:lamdaQQ}. This is done because the electroweak fit shows great sensitivity to $\lambda_{UQ_1}$ through $\cC_{\phi D}\propto |\lambda_{UQ_1}|^4$ affecting universally all light quark couplings to the electroweak gauge bosons. Therefore, the EW fit constrains $\lambda_{UQ_1}$ in a precise range in contrast to the Higgs fit which allows it to vary considerably. Moreover, we set $[\lambda_{Q_1}^d]_1 = [\lambda_{Q_1}^u]_1$ as the best-fit value for $[\lambda_{Q_1}^d]_1$ obtained by minimising $\chi_{\rm Higgs}^2$ turns out unnaturally large ($[\lambda_{Q_1}^d]_1^{\rm b.f.} > 3$).
In the plot on the left of Fig.~\ref{fig:HiggsfitsMod1} we use the global fit by ATLAS of Ref.~\cite{ATLAS:2024lyh} which takes into account differential data, but not the possibility of enhanced light quark Yukawa couplings, hence no new production processes with respect to the SM or decay modes ones are considered. Consequently, we can only bound the operators generated by the model that influence Higgs physics without light quark Yukawa couplings, namely the loop-induced $\mathcal{O}_{\phi G}$, $\mathcal{O}_{\phi B}$, $\mathcal{O}_{\phi W}$, and $\mathcal{O}_{\phi WB}$. In the plot on the right-hand side, instead, we fit to total signal strengths but take into account all the new effects due to the enhancement of the light quark Yukawa couplings as described in Sec.~\ref{sec:Higgs}. Indeed, the fact that the two plots look vastly different, emphasises that already with the Higgs data at hand, the light quark Yukawa enhancements can be constrained and that this should be taken into account when performing global fits. 

The constraints from flavour physics discussed in Sec.~\ref{sec:flavour_physics} are implemented by adding to the total likelihood a term 
\begin{equation}
    \chi_{\rm Flav}^2 = 3.84\, \left(\frac{3.2 \,|[\lambda_U]_1|}{1.6}\right)^2\,,
\end{equation}
which implements a constraint from the CKM unitarity violation for Model 1 in Tab.~\ref{tab:flav_summ}, $m_U\ge 3.2 \,|[\lambda_U]_1|$ TeV. Since we are using a universal mass scale for the vector-like quarks $\Lambda =1.6$ TeV to evade direct searches, it reflects in the limit $[\lambda_U]_1 \le 1.6/4.8$ at $95\%$ CL, and 3.84 signals one degree of freedom in the chi-squared distribution. For every other model subject to constraints from flavour physics, we use Tab.~\ref{tab:flav_summ} in the same way as presented here for Model 1. Effectively, flavour constraints always affect one of the couplings of vector-like quarks which are not doublets under $SU(2)$ gauge symmetry, reducing the allowed regions on the $y$-axis on the plots in Figs.~\ref{fig:1st_global_M1-M4}--\ref{fig:2nd_global_M5-M8}, compared to when only EW and Higgs fits are used.

\begin{figure}[t]
\centering
\begin{subfigure}[t]{0.48\textwidth}
\includegraphics[width=0.9\linewidth]{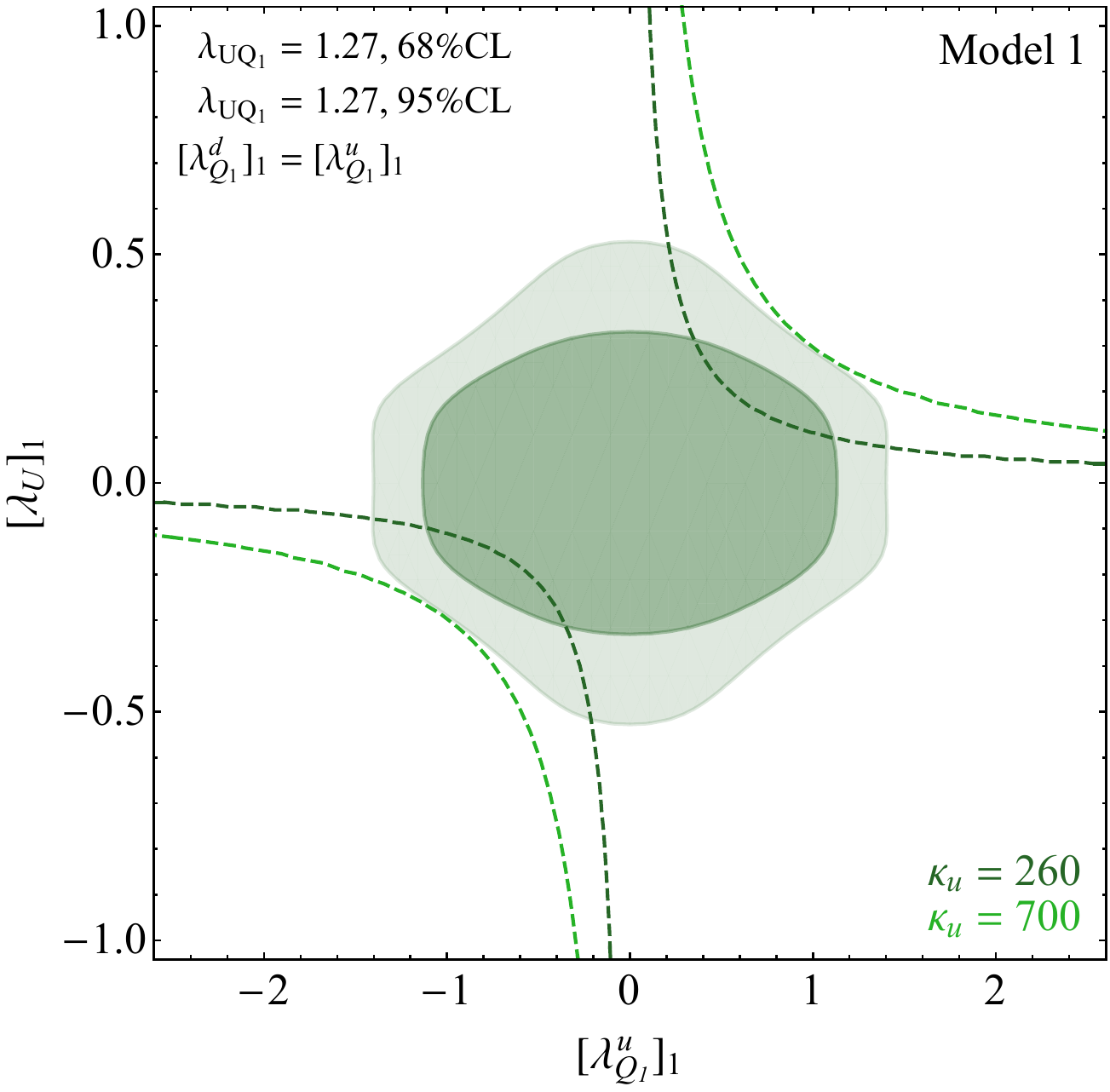}
\label{fig:combinedifitsMod1bestfit}
\end{subfigure}\quad
\begin{subfigure}[t]{0.48\textwidth}
\includegraphics[width=0.9\linewidth]{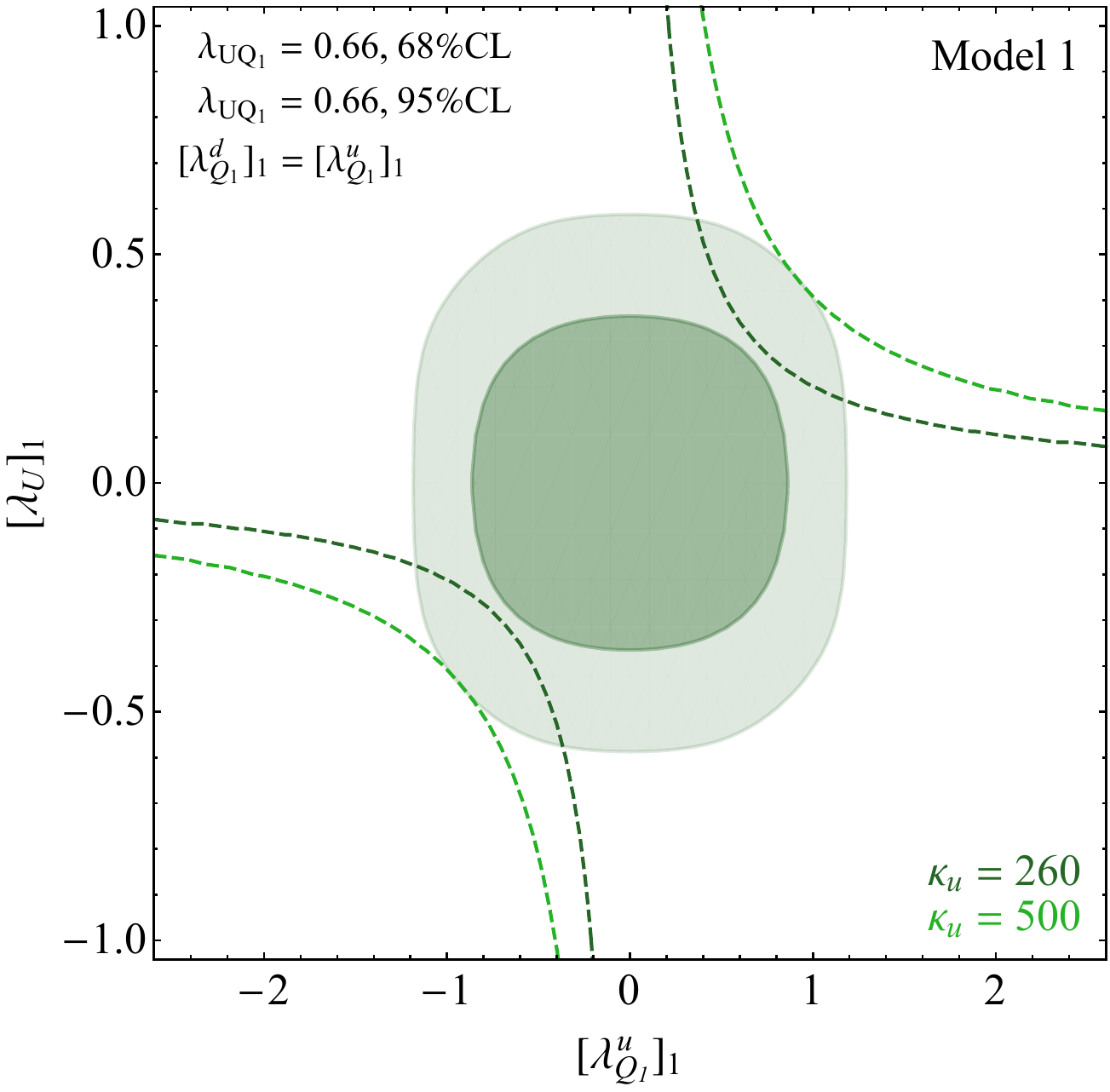}
\label{fig:combinedfitsMod1theobounds}
\end{subfigure}
\caption{Allowed parameter space from a combined fit to EWPT and Higgs data for Model 1. The lighter (darker) color shows the 95\% (68\%) CL interval. The dashed lines show fixed values of $\kappa_u$ in the parameter plane. In both plots, $[\lambda_{Q_1}^d]_1=[\lambda_{Q_1}^u]_1$ and $\Lambda=1.6$ TeV. \textbf{Left:} Setting $\lambda_{UQ_1}$ such that it maximises $\kappa_u$. \textbf{Right:} Using $|\lambda_{UQ_1}|\leq 0.66$ as theory constraint from Ref.~\cite{Adhikary:2024esf}. \label{fig:combinedfitsMod1}}
\end{figure}

In Fig.~\ref{fig:combinedfitsMod1}, we combine EWPT, Higgs, and flavour data for Model 1. In the case of Higgs physics, we used the fit to the full rates including the light quark Yukawa modifications. The plot on the left-hand side shows the result by choosing the value of the VLQ-VLQ-Higgs coupling $\lambda_{UQ_1}$ in such a way that it maximises $\kappa_u$ as in Eq.~\eqref{eq:lamdaQQ} with the likelihood
\begin{equation}
    \chi_{\rm TOT}^2
 = \chi^2_{\rm EWPO}+\chi^2_{\rm Higgs}+\chi^2_{\rm Flav}\,.
 \label{eq:chi2_TOT}
\end{equation} 
From the figure, we can infer that the excluded parameter space mostly corresponds to the one excluded by electroweak and flavour physics. At the same time, Higgs data establishes a direct connection with the light quark Yukawa couplings, cutting away the ``edges'' of the excluded parameter range of the EW data. In the plot on the right-hand side of Fig.~\ref{fig:combinedfitsMod1}, we employ the additional condition that $|\lambda_{UQ_1}|\le 0.66$, motivated by the fact that vector-like quarks modify the Higgs self-coupling via loop corrections and hence can render the vacuum unstable~\cite{Gopalakrishna:2018uxn, Adhikary:2024esf}. Requiring the Higgs potential to be bounded from below, or having the Higgs
self-coupling being positive up to the cut-off scale $\mu=100\text{ TeV}$, Ref.~\cite{Adhikary:2024esf} obtains a bound of $|\lambda_{UQ_1}|<0.66$ under the assumption that the VLQs do not couple to SM quarks and $\Lambda=1$ TeV. In our case, however, the beta function of the quartic coupling in the Higgs potential only obtains positive contributions from the couplings of the VLQs to the SM quarks, so those couplings at one-loop in the leading log approximation improve the situation. Moreover, a larger value of $\Lambda$ results in the larger theoretical limit on $|\lambda_{UQ_1}|$ approaching the values we used in our analysis. We do not discuss these theory bounds here any further\footnote{We also note that the results of Ref.~\cite{Adhikary:2024esf} are given only for what corresponds to Model 1. The other models are not discussed in the reference.} but note that while they might render the maximally allowed value of $\kappa_u$ somewhat smaller, the values of several hundred times the SM can still be achieved within Model 1.

To better understand the impact of different observables on the global fit in the models we study, let us define the \emph{pull} of an observable $O$ computed in a model M as follows
\begin{equation}
    P_{\rm M} = \frac{O-O_{\rm exp}}{\sigma_{\rm exp}}\,,
    \label{eq:pull}
\end{equation}
where $O_{\rm exp}$ is the experimental measurement with the corresponding uncertainty $\sigma_{\rm exp}$. We compute the observable $O$ in a model M by setting the model's parameters to their best-fit values obtained by minimising $\chi^2_{\rm TOT}$ in Eq.~\eqref{eq:chi2_TOT}. In Model 1, this procedure identifies four observables with the largest impact on the electroweak part of the fit: $m_W$, $A_e$, $R_\mu$, and $A_b^{\rm FB}$, with the relative pull with respect to the SM, $|P_1|-|P_{\rm SM}|$, shown in Tab.~\ref{tab:pulls_M1}. In addition, we show the pulls of different Higgs signal strengths. The absolute relative pulls of other EWPOs are below 0.1 and we do not report them.  
\begin{table}[t]
\centering
    \begin{tabular}{|c|c|c|c|}\hline
         Observable & $P_{\rm SM}$ & $P_{\rm 1}$ & ${|P_{\rm 1}|-|P_{\rm SM}|}$ \\\hline\hline
        $m_{W}$ &$-1.55$ & $-0.54$& $-1.01$\\
        $A_{e}$ &$-2.19$ & $-2.05$& $-0.14$ \\
        $R_\mu$ &$-1.50$ & $-1.61$& $0.11$ \\
        $A^{\rm FB}_b$ &$2.25$ & $2.38$& $0.13$\\\hline
        $\mu_{\tau\tau}$ &$1.50$ & $1.00$& $-0.50$\\
        $\mu_{WW}$ &$0.33$ & $-0.21$& $-0.12$\\
        $\mu_{ZZ}$ &$-0.26$ & $-0.17$& $-0.09$\\
        $\mu_{\mu\mu}$ &$-0.48$ & $-0.59$& $0.11$\\
        $\mu_{bb}$ &$-0.23$ & $-0.46$& $0.23$\\
        $\mu_{\gamma\gamma}$ &$-1.44$ & $-1.82$& $0.38$\\
        \hline
    \end{tabular}
    \caption{Pulls in Model 1 for the selected observables: EWPOs with the most influence on the global fit and all Higgs signal strengths.}
    \label{tab:pulls_M1}
\end{table}
The final column being negative (positive) signals that Model 1 reduces (increases) the tension with the experiment compared to the SM for a given observable. For experimental measurements, we use the results collected in~\cite{Breso-Pla:2021qoe,CMS:2022dwd} except for $m_W$ that we update with the recent ATLAS~\cite{ATLAS:2023fsi} and CMS~\cite{2024mWCMS} measurements. We perform LEP2, Tevatron, and LHC combination and obtain\,\footnote{We assume a pre-2022 CDF measurement for a Tevatron average of $m_W = 80.387 \pm 0.016$ GeV \cite{ParticleDataGroup:2024cfk}.}
\begin{equation}
    m_W = 80.3670 \pm 0.0071\,\,{\rm GeV}\,.
    \label{eq:mW_new}
\end{equation}
The results for the leading $|P_1|-|P_{\rm SM}|$ reflect the large indirect effect Model 1 introduces in shifting all fermion couplings universally through $\mathcal{O}_{\phi D}$ as given in Eq.~\eqref{eq:universal_shift}, as well as the $W$-boson mass   
\begin{equation}
\delta m_W=-\frac{v^2 g_2^2}{4(g_2^2-g_1^2)}\,\cC_{\phi D}
-\frac{v^2 g_2 g_1}{g_2^2-g_1^2}\,\cC_{\phi W B}\,.
\label{eq:WmassExpr}
\end{equation}

In addition to this indirect effect arising at one loop, there is a tree-level modification of the $Z$-boson couplings to the first-generation quarks through $\cO_{\phi q}^{(1,3)}$, $\cO_{\phi u}$, and $\cO_{\phi d}$, which directly feed into $\Gamma_Z$ that also affects $A_e$, $R_\mu$, and $A_b^{\rm FB}$. The electroweak part of the fit utilizes these new physics effects to address and reduce the SM tensions mainly in $m_W$, however, making simultaneously $A^{\rm FB}_b$ worse. Therefore, the competing effect between these observables eventually leads to the allowed parameter space shown in Fig.~\ref{fig:EWPTfitsMod1}. This is a recurring situation for other models as well, where $m_W$ is consistently made better by pairs of vector-like quarks while $A^{\rm FB}_b$ is made worse. 
Regarding the Higgs signal strengths, we discover that $\mu_{\tau\tau}$, $\mu_{WW}$, and $\mu_{ZZ}$ are slightly improved compared to the SM, while $\mu_{\mu\mu}$, $\mu_{bb}$, and $\mu_{\gamma\gamma}$ have a larger pull in Model 1 than in the SM. The detailed results for other models can be found in App.~\ref{sec:add_results}.

In Figs.~\ref{fig:1st_global_M1-M4}--\ref{fig:1st_global_M5-M8}, we show the results of the combined fit for all models with vector-like quarks coupled to the first generation. The dashed lines indicate fixed values of $\kappa_q$, with $q=u,d$. Like Model 1, Models 2, 5, and 7 introduce four new couplings, where one of them does not enter into the considered $\kappa_q$, and we set it to the best-fit value. In contrast, Models 3, 4, 6, and 8 have three parameters that directly determine $\kappa_q$. The coupling between the two representations of VLQs is always set to maximise $\kappa_{q}$ as in Eq.~\eqref{eq:lamdaQQ} with $\chi_{\rm EWPO}^2$ replaced by $\chi_{\rm TOT}^2$ in Eq.~\eqref{eq:chi2_TOT}. 

For Models 5 and 7, which generate both large $\kappa_d$ and $\kappa_u$, we show only curves for one of them. Specifically, in Model 5 we set $\luQone$ to the best-fit value and only show the lines for $\kappa_d$. However, by considering the explicit expressions for the coupling modifiers
\begin{align}
    \kappa^{}_u &= 1 + \frac{v^3}{2\sqrt{2}\, m_{u} M_{T_1} M_{Q_1}} \left[\luQone\right]_1 \left[\lambda_{T_1}\right]_1 \lambda_{T_1Q_1}\,, \label{eq:model5kugen1}\\
    \kappa^{}_{d} &= 1 + \frac{v^3}{4\sqrt{2}\, m_{d} M_{T_1} M_{Q_1}} \left[\ldQone\right]_1 \left[\lambda_{T_1}\right]_1 \lambda_{T_1Q_1}\,, \label{eq:model5kdgen1}
\end{align}
and the limit $\kappa_q\gg 1$
\begin{equation}
    \kappa_u = \frac{2 m_d}{m_u} \frac{[\lambda_{Q_1}^u]_1}{[\lambda_{Q_1}^d]_1} \kappa_d\,,
    \label{eq:kappas_M5}
\end{equation}
it is possible to learn about the maximum allowed value of $\kappa_u$ from the maximum value of $\kappa_d$ and the corresponding values of $[\lambda_{Q_1}^u]_1$ and $[\lambda_{Q_1}^d]_1$. In particular, in the $SU(2)_R$-symmetric limit, we have the maximal achievable $\kappa_u = 2 m_d/m_u \times 220 \simeq 940$. Alternatively, one can set $\ldQone$ to the best-fit value and leave $\luQone$ free to vary; in this case, the largest enhancement for the up quark coupling is found to be $\kappa_u=560$.\\
Likewise, in Model 7 we show the lines for $\kappa_u$ having set $\ldQone$ to the best fit value, but note that 
\begin{equation}
    \kappa_d = \frac{2 m_u}{m_d} \frac{[\lambda_{Q_1}^d]_1}{[\lambda_{Q_1}^u]_1} \kappa_u\,,
    \label{eq:kappas_M7}
\end{equation}
such that, in the limit $[\lambda_{Q_1}^d]_1=[\lambda_{Q_1}^u]_1$, the maximal achievable $\kappa_d = 2 m_u/m_d \times 380 \simeq 360$. Repeating the study setting instead $\luQone$ to the best-fit value leads to $\kappa_d=320$, once again consistent with the maximal achievable value in the $SU(2)_R$-symmetric limit.

From Figs.~\ref{fig:1st_global_M1-M4}--\ref{fig:1st_global_M5-M8}, it can be inferred that in all models it is possible to achieve $\kappa_q$ values of a few$\,\times\,\mathcal{O}(100)$. This clearly illustrates the potential of the light Yukawa probes mentioned in Introduction to explore interesting parameter space, motivating them in the EFT approach. This is qualitatively different from the model proposed in Ref.~\cite{Egana-Ugrinovic:2019dqu} which, while allowing for large light quark Yukawa coupling modifications, at the same time, involves rather light new Higgs bosons that can be discovered in searches for resonant Higgs pair production~\cite{Egana-Ugrinovic:2021uew}. 
\subsection{Second generation}

\begin{figure}[t!]
    \centering
    \includegraphics[width=0.47\linewidth]{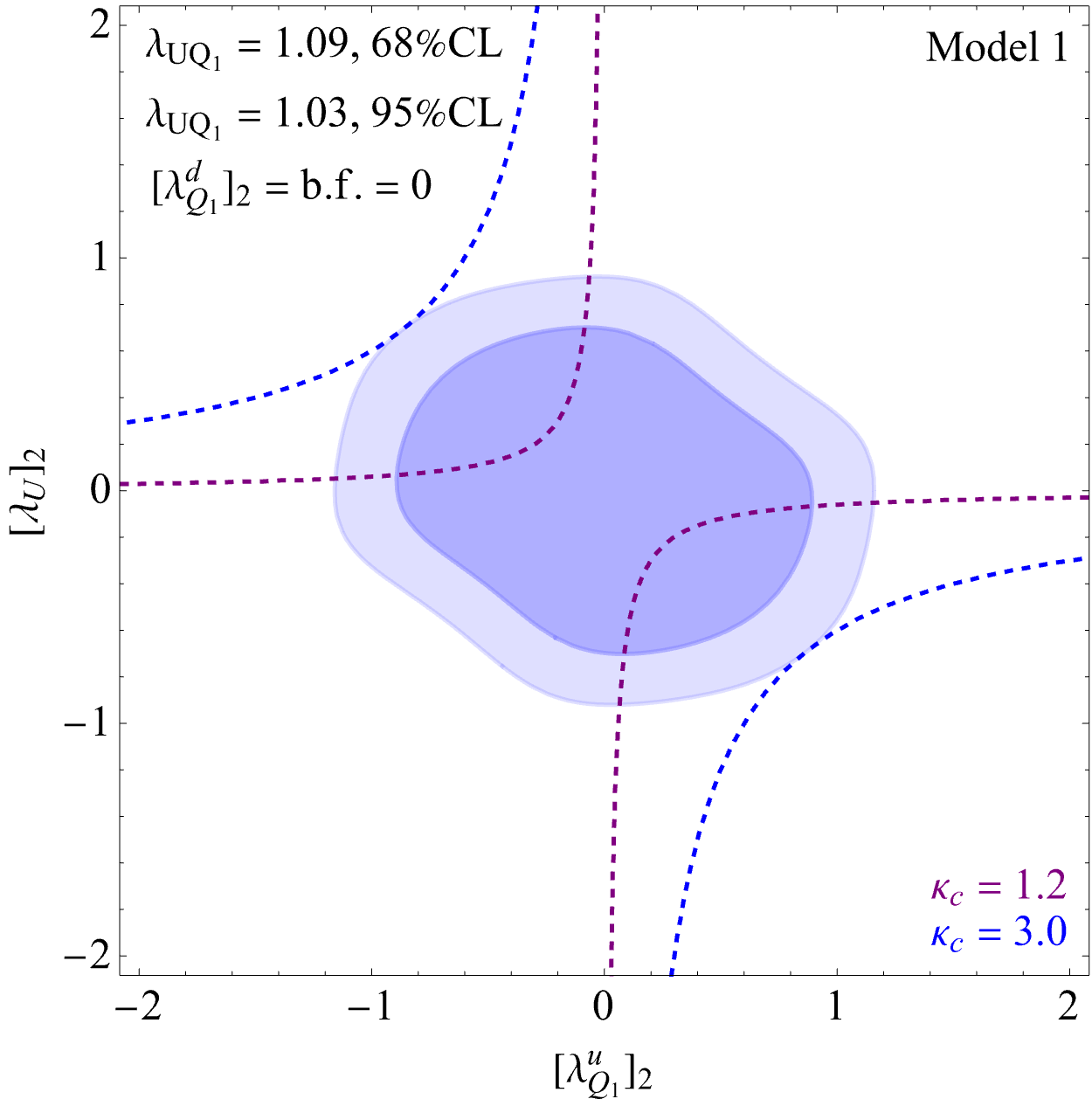} 
    \quad 
    \includegraphics[width=0.47\linewidth]{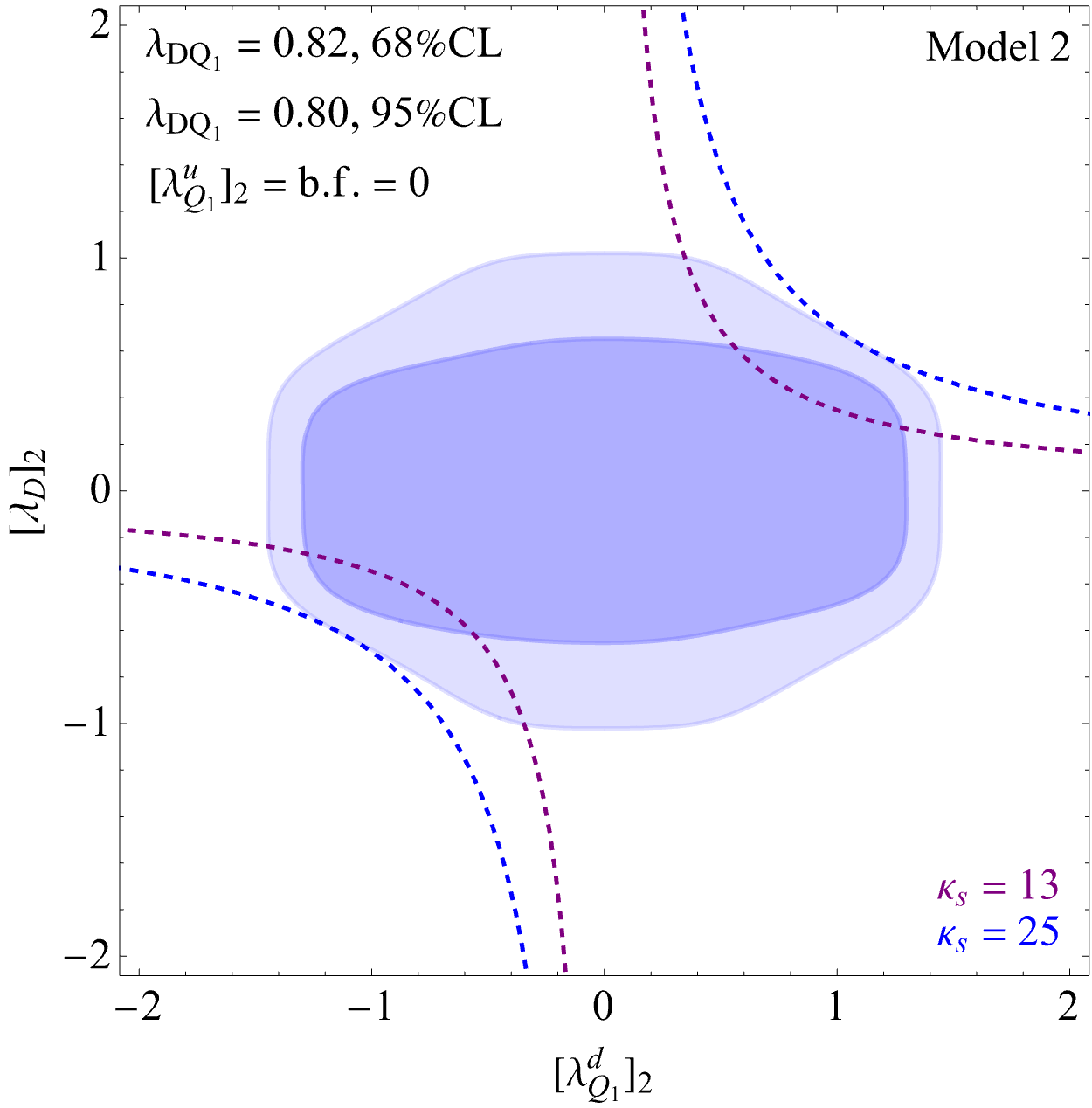}\\
    \includegraphics[width=0.47\linewidth]{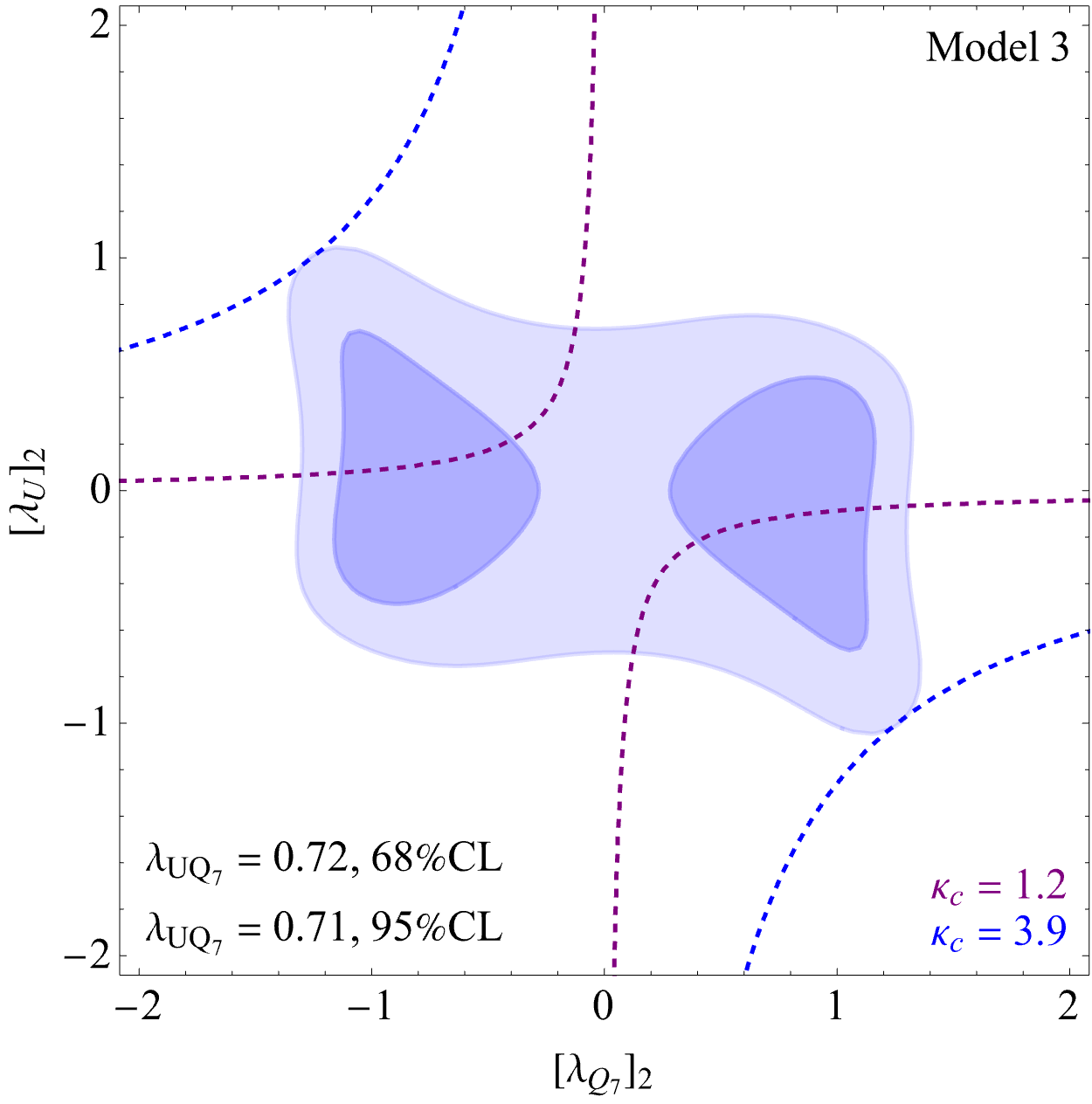} 
    \quad 
    \includegraphics[width=0.47\linewidth]{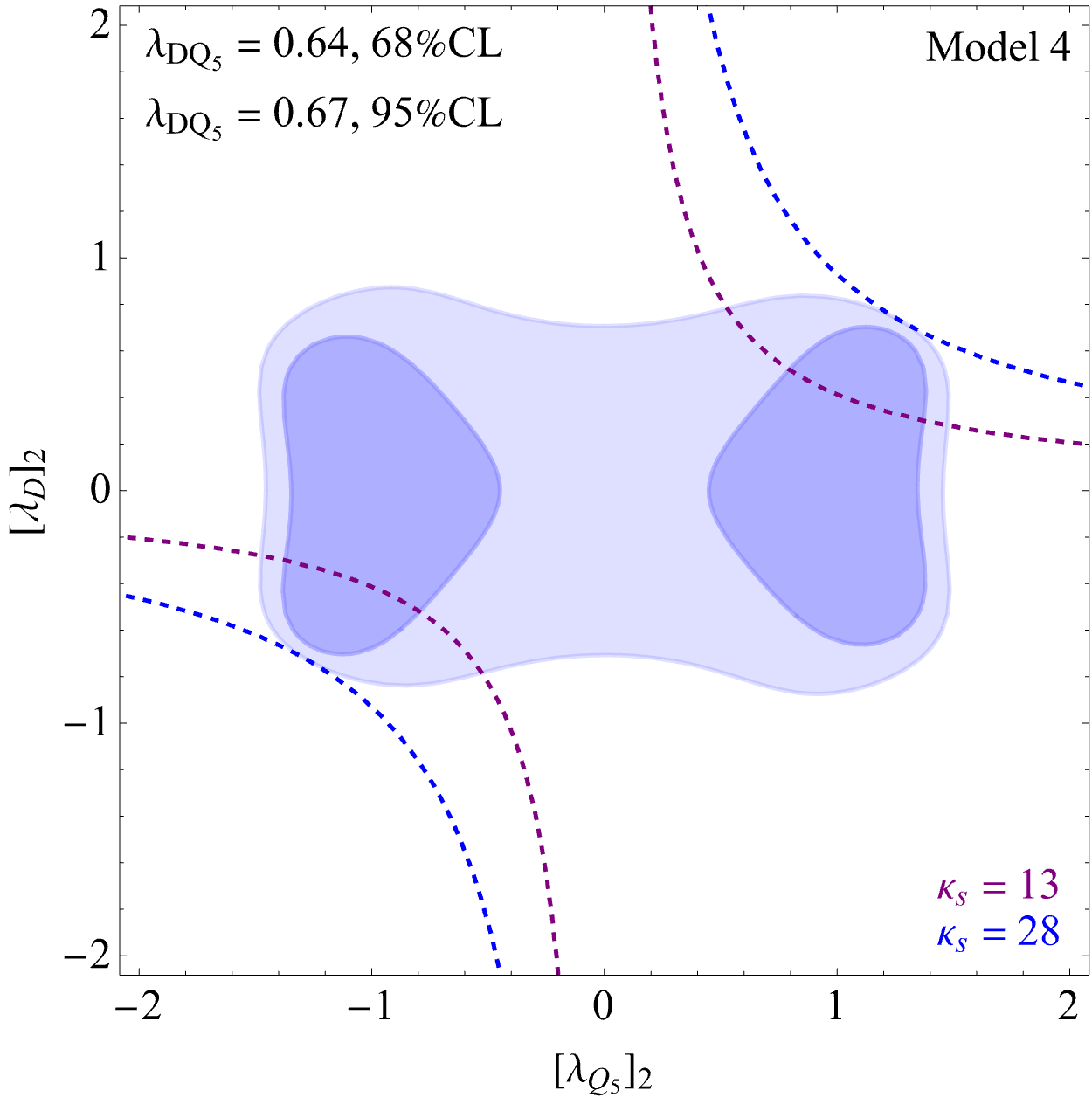}
    \caption{Allowed parameter range from a fit to EW, Higgs, and flavour data for Models 1--4 with VLQs coupled to the second-generation quarks. The lighter (darker) color shows the 95\% (68\%) CL interval. The dashed lines show fixed values of $\kappa_q$ in the parameter plane.}
    \label{fig:2nd_global_M1-M4}
\end{figure}

\begin{figure}[t!]
    \centering
    \includegraphics[width=0.47\linewidth]{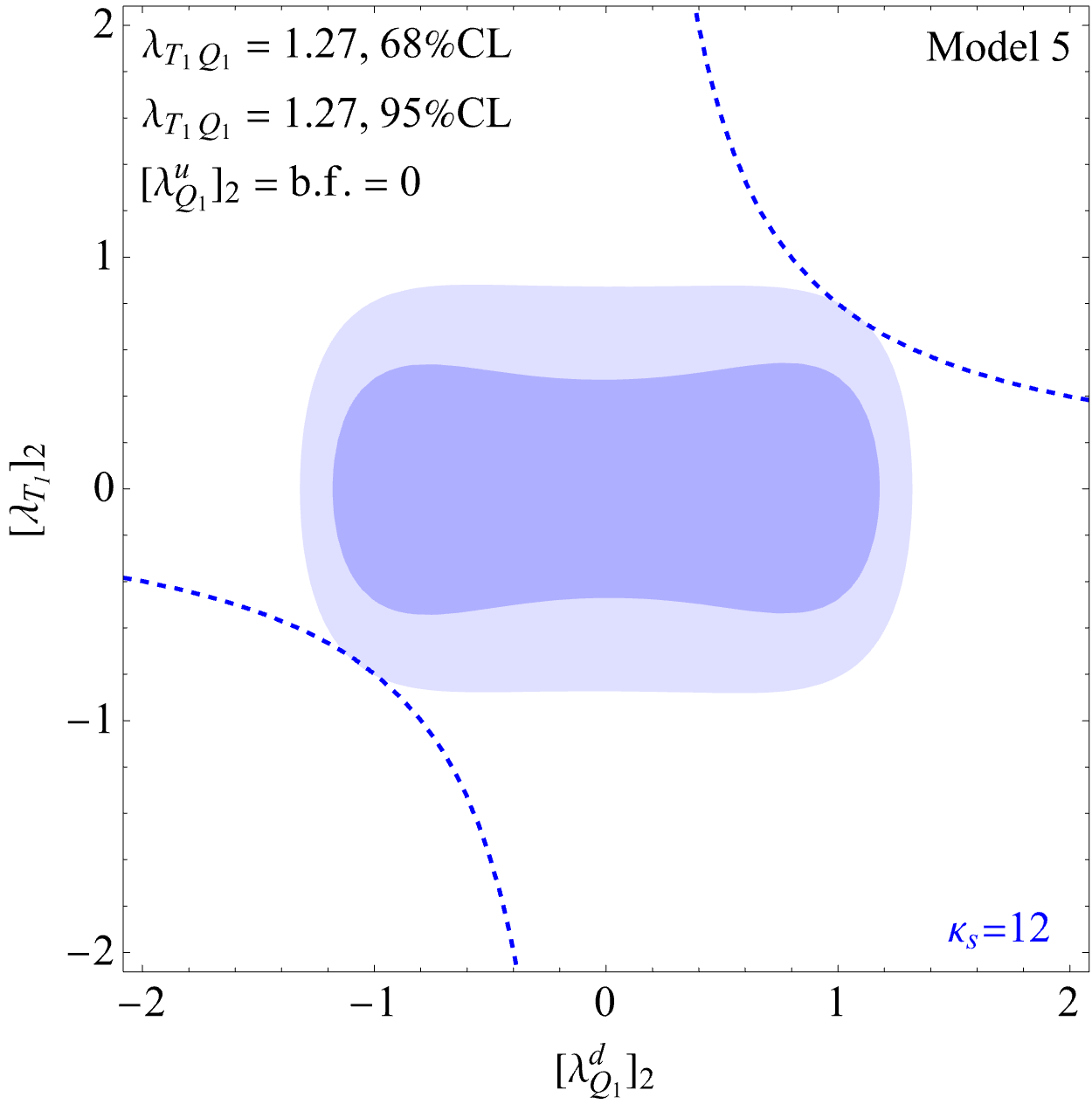} 
    \quad 
    \includegraphics[width=0.47\linewidth]{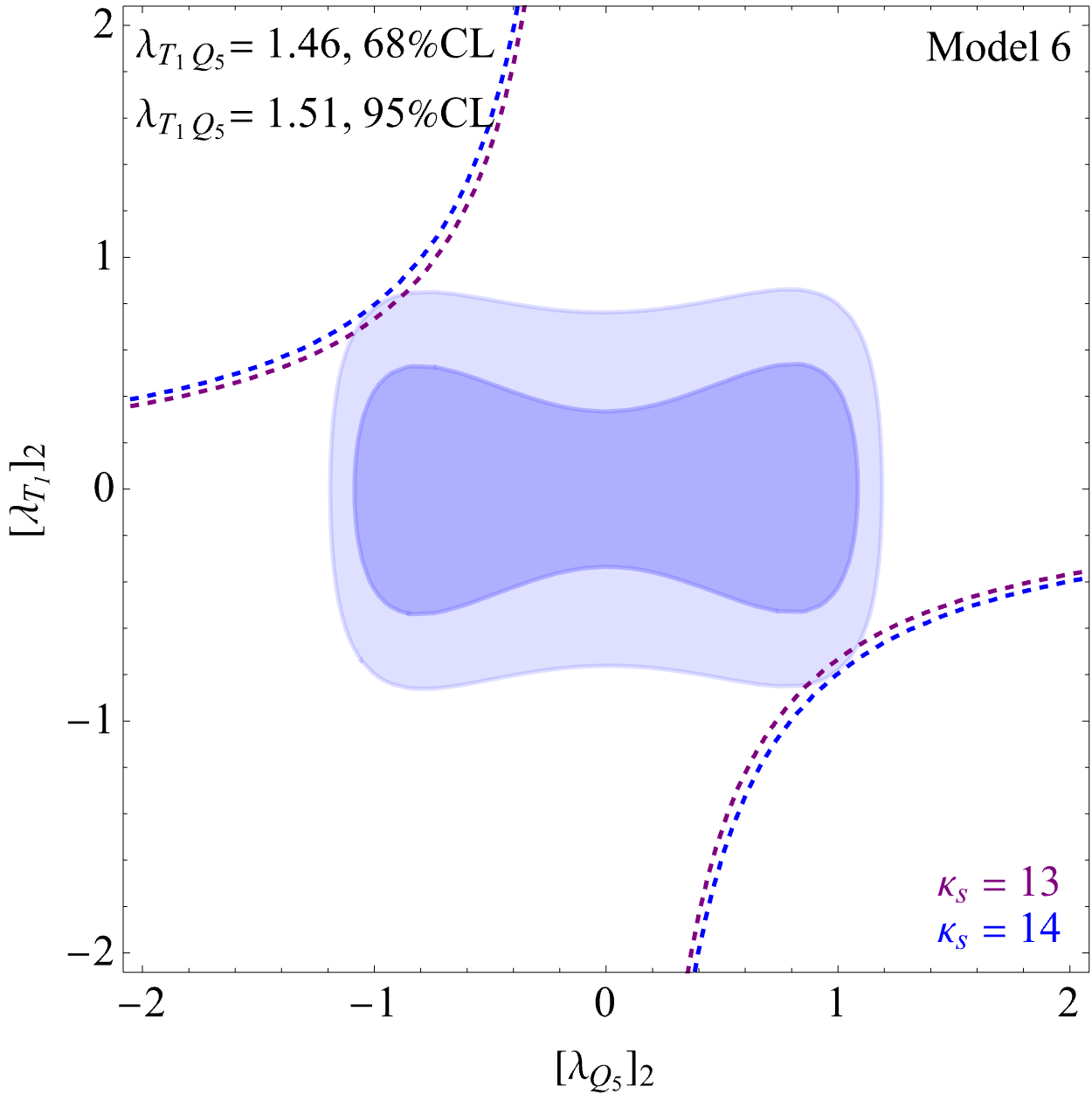}\\
    \includegraphics[width=0.47\linewidth]{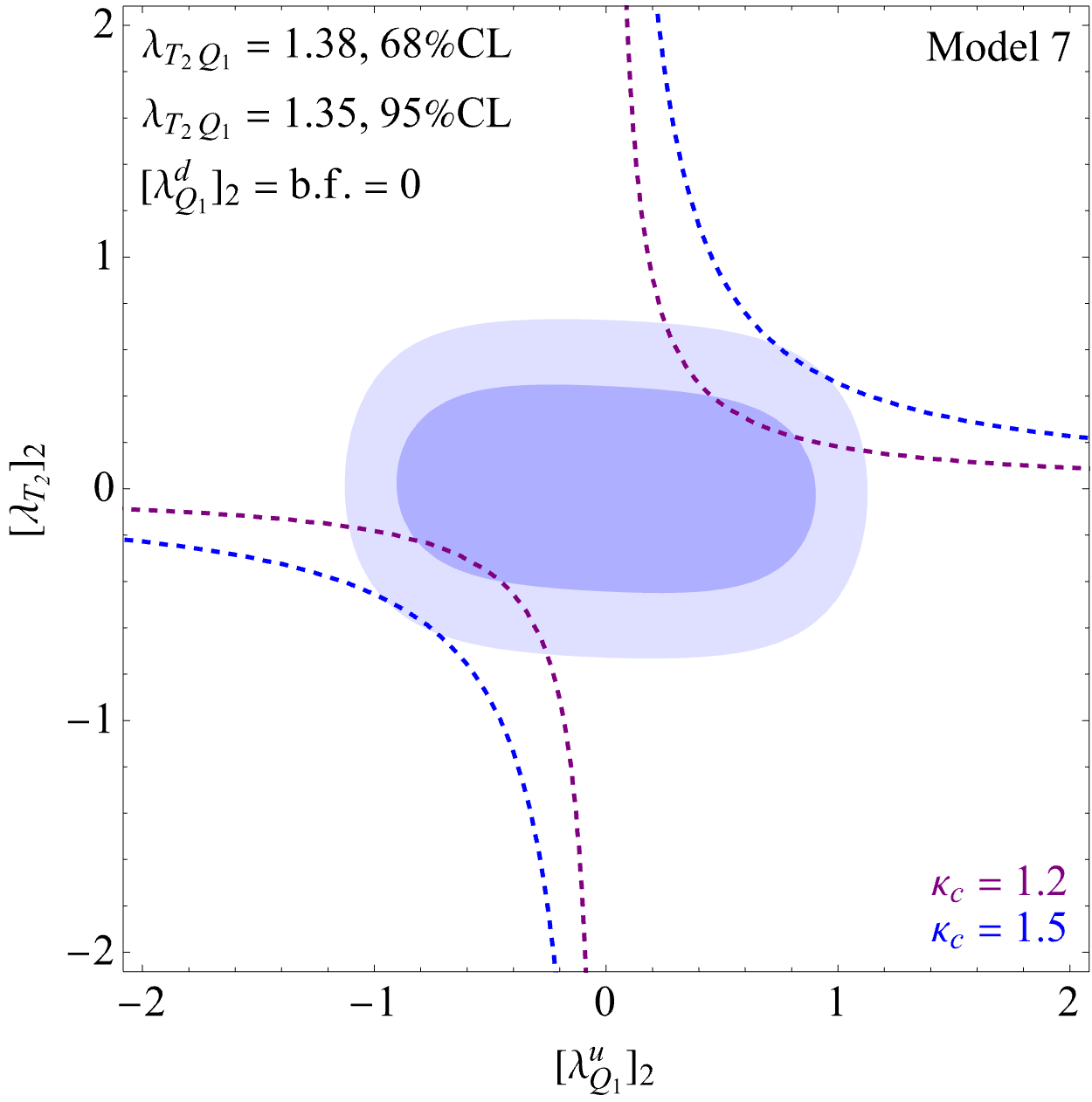} 
    \quad 
    \includegraphics[width=0.47\linewidth]{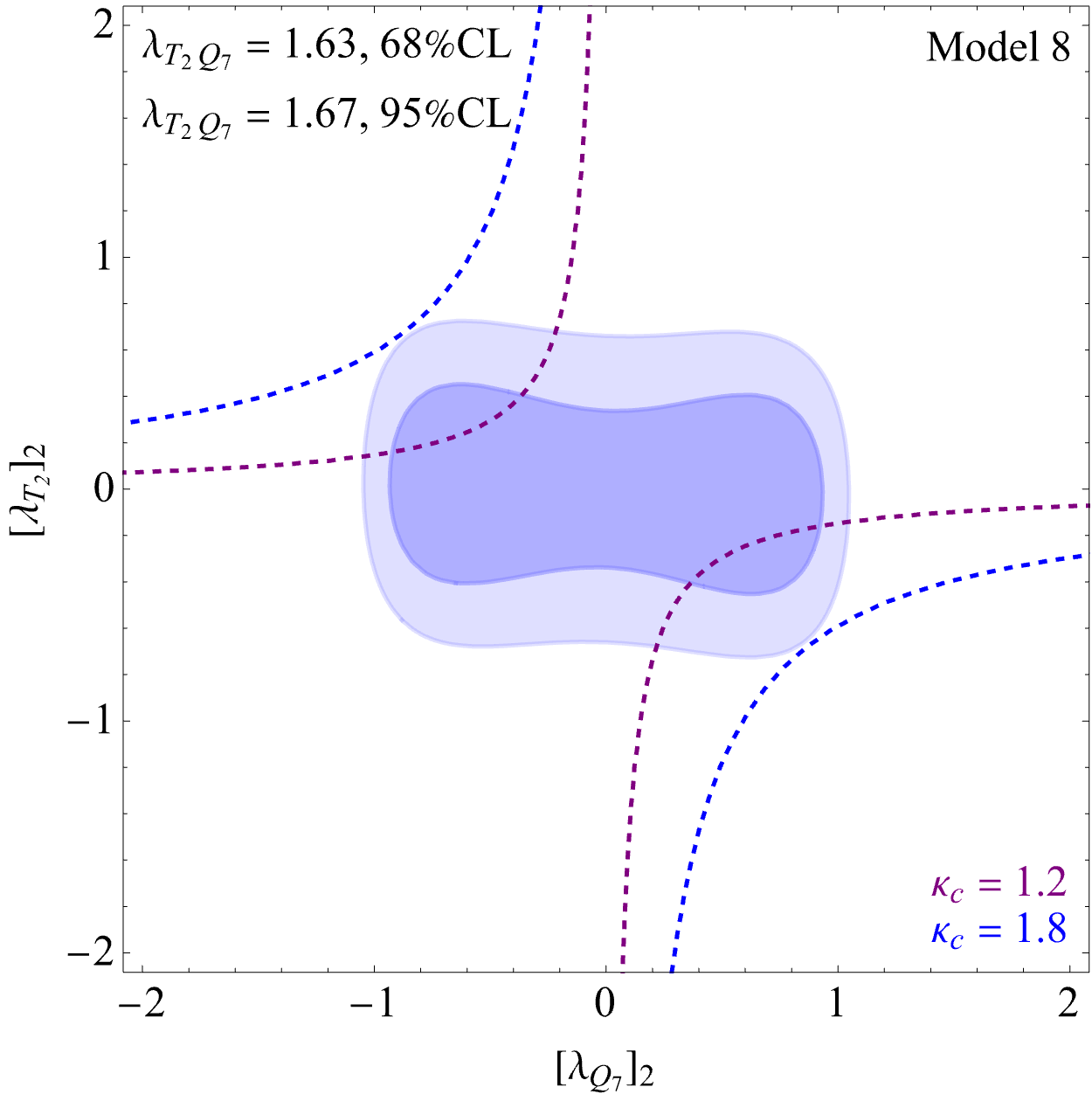}
    \caption{Allowed parameter range from a fit to EW, Higgs, and flavour data for Models 5--8 with VLQs coupled to the second-generation quarks. The lighter (darker) color shows the 95\% (68\%) CL interval. The dashed lines show fixed values of $\kappa_q$ in the parameter plane.}
    \label{fig:2nd_global_M5-M8}
\end{figure}

We now consider the models with vector-like quarks coupled only to the second generation. The results of a combined fit are shown in Figs.~\ref{fig:2nd_global_M1-M4}--\ref{fig:2nd_global_M5-M8}. We recall that the additional VLQ couplings to the SM quarks are set to their best-fit value if present and that the couplings among the different VLQ multiplets are set to maximise $\kappa_{c,s}$ in the same way as done for the models with first-generation couplings shown in Eq.~\eqref{eq:lamdaQQ} and with $\chi^2_{\rm EWPO}$ replaced by the total likelihood $\chi^2_{\rm TOT}$. 
Given these choices, the exclusion of the origin from the $68\%$ CL parameter space, which happens in Models 3 and 4 in Fig.~\ref{fig:2nd_global_M1-M4}, does not correspond to the exclusion of the SM. For the SM to be excluded, also the VLQ-VLQ-Higgs coupling would have to be set to zero. Again, as in the case of the first-generation couplings, the maximal coupling modifiers were considered to be the tangent lines to the outermost region of the $95\%$ CL parameter space.
All the models that are shown with the charm coupling modifier (Models 1, 3, 7, and 8)
present an allowed parameter space without the symmetry under the reflection about the $x$ and $y$ axes. 
This is particularly noticeable for Models 1 and 3 where the absence of flavour bounds makes the tilting of the allowed parameter space more pronounced. 
This shows the interplay of the different constraints on the allowed parameter space and motivates the importance of performing a global fit. Moreover, the fact that this effect is only visible when considering $\kappa_c = 1 \pm (v^3 \cC_{u\phi} )/({\sqrt{2} m_c})$ comes from the presence of $m_c$ in the denominator, such that a constant SM term in $\kappa_c$ can compete with the EFT contribution. Note that, for $\kappa_u = 1 \pm (v^3 \cC_{u\phi} )/({\sqrt{2} m_u})\simeq \pm(v^3 \cC_{u\phi} )/({\sqrt{2} m_u})$, the quadratic EFT term in $\kappa_u^2$ always dominates and we are never sensitive to the relative sign of the three couplings entering $\cC_{u\phi}$. The same holds when considering $\kappa_{d,s}$ and as $m_{d,s}$ are much lighter than $m_c$.

As the LEP measurements were not able to successfully distinguish between light generations, the models with vector-like quarks coupled to the second-generation quarks impact the electroweak physics in a similar way to what was already discussed in the previous section; the observables that mainly drive the fit still being $m_W$ and $A_{b}^{\rm FB}$ for most models. The detailed analysis of the most influential EWPOs and the effects of the Higgs signal strengths is given in App.~\ref{sec:add_results}. 
We did not consider in our plots the bound of $\kappa_c\in [-2.5,2.3]$ of Ref.~\cite{ATLAS:2022fnp} as it cannot be applied directly to our model as it assumes only $\kappa_c$ deviations. The results in Figs.~\ref{fig:2nd_global_M1-M4}--\ref{fig:2nd_global_M5-M8} show that VLQ models allow for deviations in the charm Yukawa couplings up to a few times its SM value, which means that the current dedicated searches just start to probe interesting parameter space. In particular, Model 3 can have $\kappa_c\leq 3.9$ which is close to the bound set in \cite{CMS:2022psv}. The models that instead generate couplings to the strange quark allow for values up to a few tens times the SM strange Yukawa coupling.  

Finally, we would like to comment on the recent study in Ref.~\cite{Nir:2024oor} in which various models that lead to enhanced charm Yukawa couplings are discussed. This study includes also Models 1, 3, 7, and 8 while it does not consider Model 5 which also leads to deviations in the charm Yukawa coupling.\footnote{From the Eqs. \eqref{eq:model5kugen1} and \eqref{eq:model5kdgen1}, it is possible to obtain the coupling modifiers for the second-generation quarks by substituting the first-generation indices with the second-generation ones.} While we discuss the models in more detail, for instance by taking into account also Higgs data, the authors clearly show that EWPTs provide an important constraint for the VLQ models they consider. Moreover, in addition to the results of Ref.~\cite{Nir:2024oor}, we perform the explicit one-loop matching and acquire the sensitivity to the VLQ-VLQ-Higgs coupling and constraint it through EWPTs and Higgs physics, thereby completing the information on all three couplings which determine the size of the effective light quark Yukawas. 

\section{Future projections}
\label{sec:future}

\begin{figure}[t]
    \centering
    \includegraphics[width=0.49\linewidth]{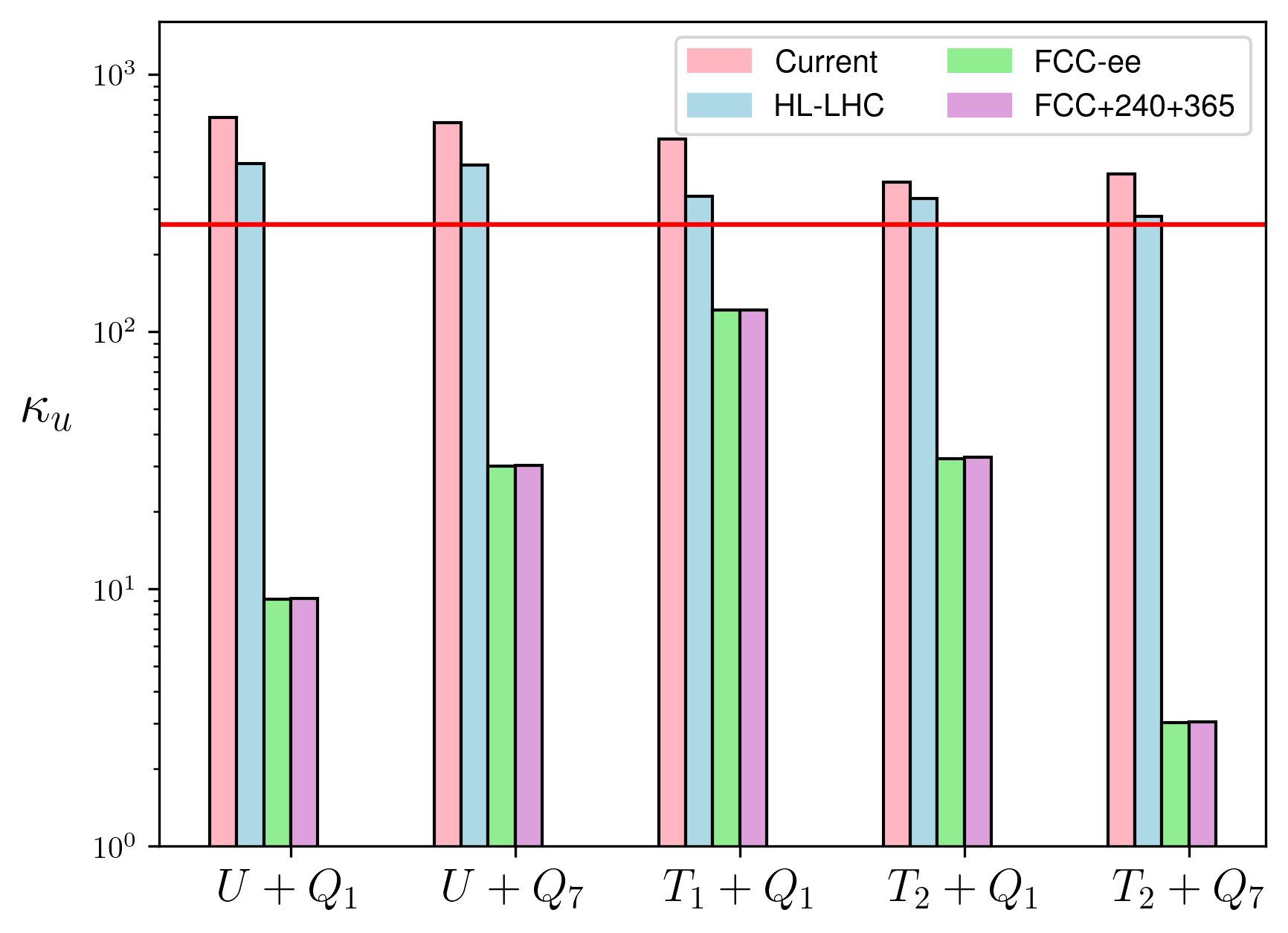}
    \hfill
    \includegraphics[width=0.49\linewidth]{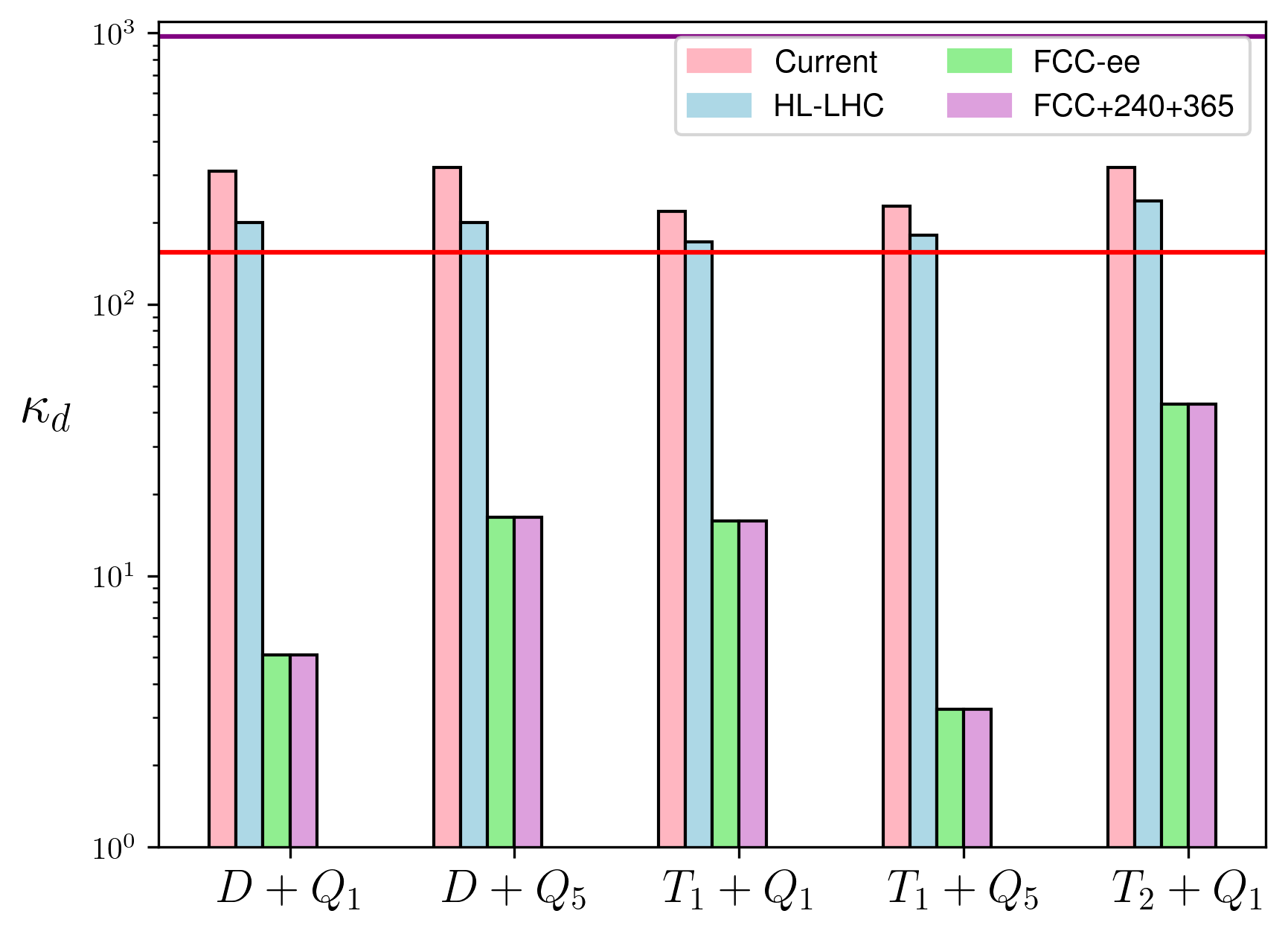}\\
    \includegraphics[width=0.49\linewidth]{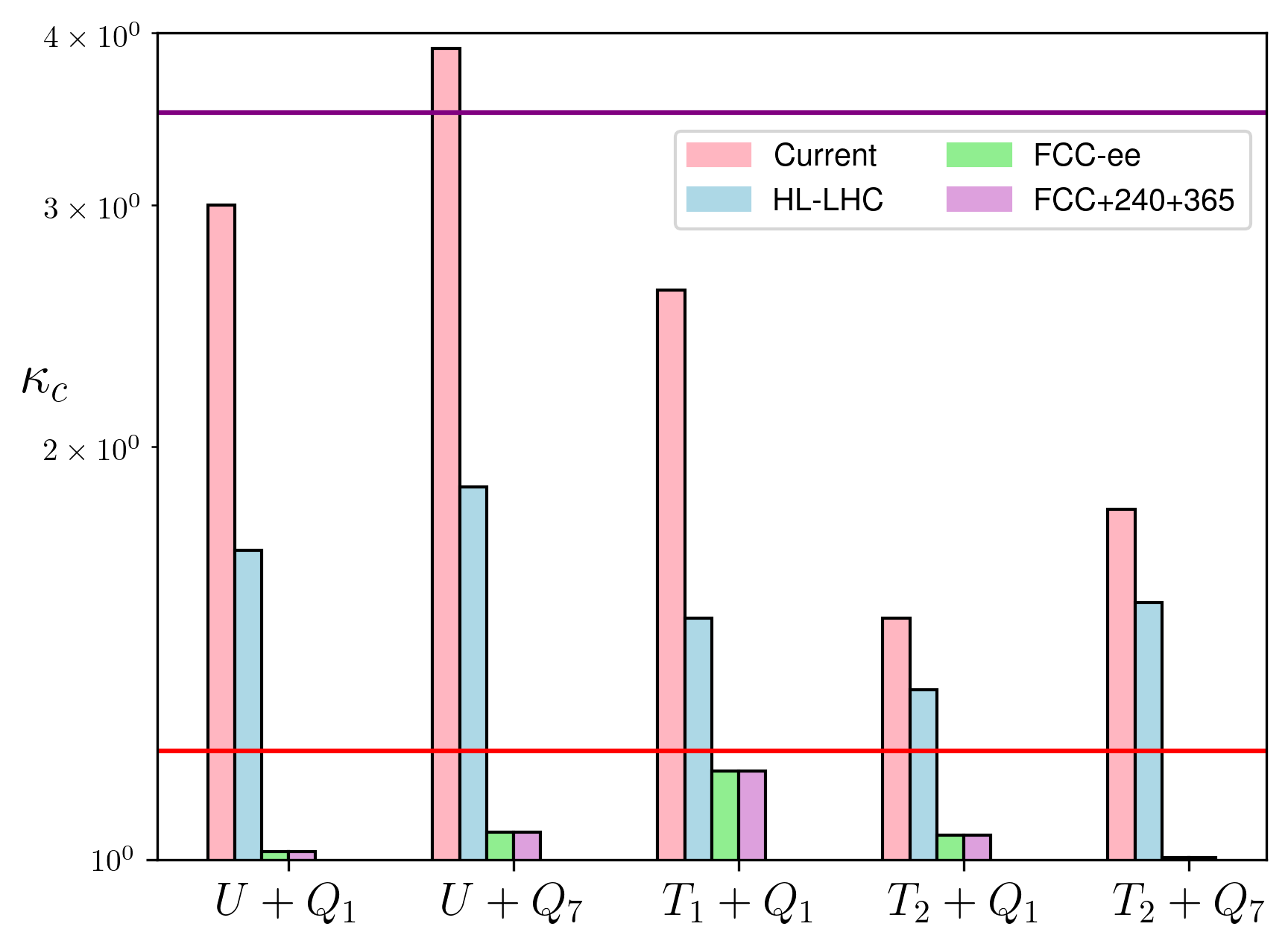}
    \hfill
    \includegraphics[width=0.49\linewidth]{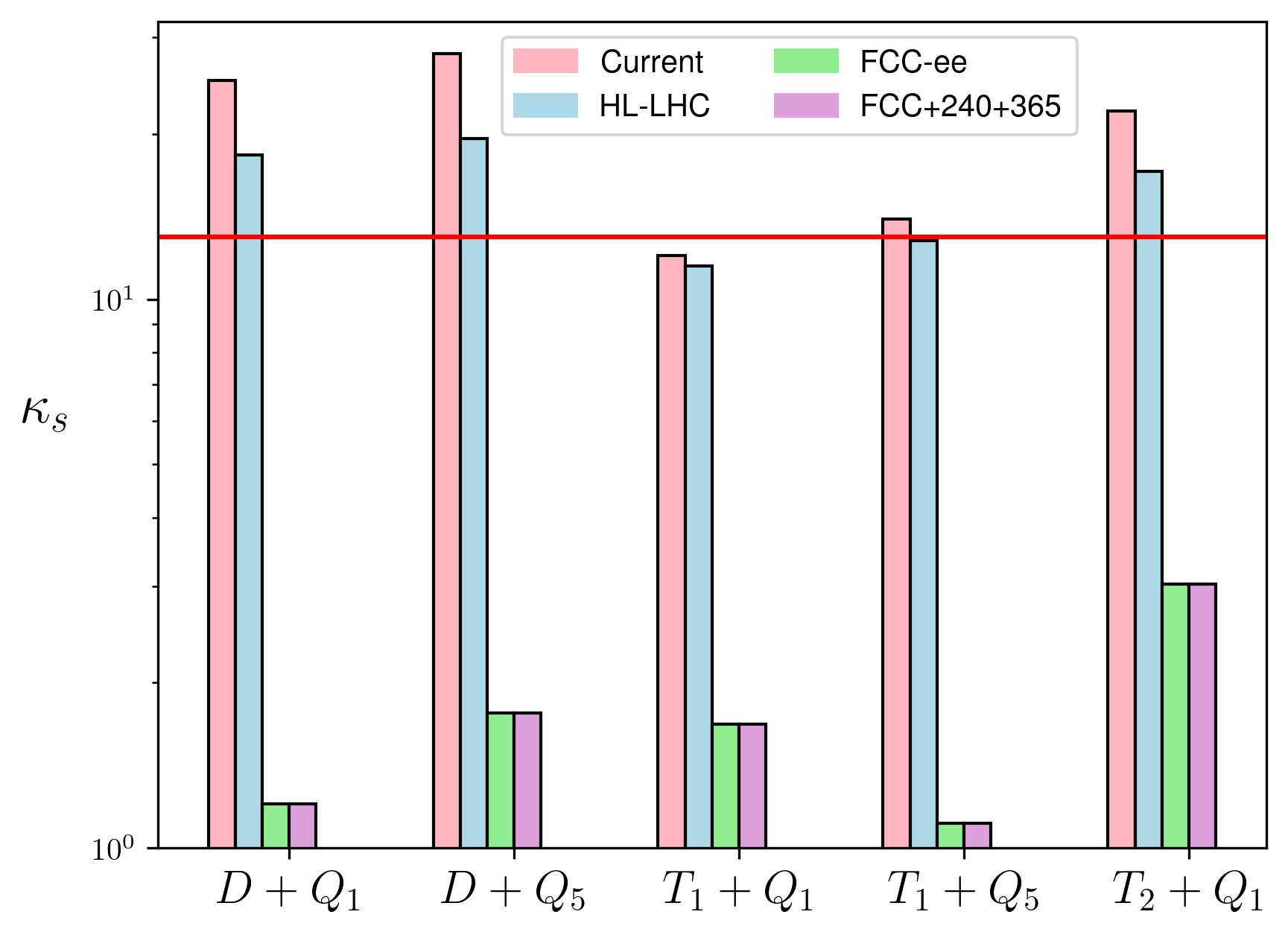}
    \caption{Overall largest allowed coupling enhancements $\kappa_q$ found with a 95\% CL. The different entries show results obtained by considering current experimental constraints or projections for future colliders (HL-LHC, and two FCC-ee scenarios). The red line corresponds to the HL-LHC projections, whereas the purple lines correspond to the upper bounds at a 95\% CL found by CMS~\cite{CMS:2025xkn}. }
    \label{fig:kappa_futurecolliders_comparison}
\end{figure}

Having so far accounted for the current experimental bounds on Higgs and electroweak observables, we now consider projections for future collider experiments: HL-LHC, FCC-ee Tera-Z and combined FCC-ee Tera-Z with the $\SI{240}{\giga\electronvolt}$+$\SI{365}{\giga\electronvolt}$ runs. In these cases, we fix the masses of the VLQs to be $\Lambda =\SI{2.4}{\tera\electronvolt}$ in accordance with Ref.~\cite{Freitas:2022cno} for VLQ searches at the HL-LHC. Our results for the largest allowed coupling enhancements for the different models considering various future experiments are summarised in Fig.~\ref{fig:kappa_futurecolliders_comparison}. 

\subsection{HL-LHC projections}
\begin{table}[t]
    \centering
    {\renewcommand{\arraystretch}{1.5}
    \begin{tabular}{|c|c|}
    \hline
    Decay channel $i$  &    $\delta\mu_i \left[\%\right]$ (S2) \\
    \hline
    $h\rightarrow WW$   & 2.5 \\
    \hline
    $h\rightarrow ZZ$  & 3.0 \\
    \hline
    $h\rightarrow b\bar{b}$  & 24.7 \\
    \hline
    $h\rightarrow \tau^+\tau^-$ & 4.6 \\
    \hline
    $h\rightarrow \mu^+\mu^-$ & 13.5 \\
    \hline
    $h\rightarrow \gamma\gamma$     & 2.8 \\  
    \hline    
    \end{tabular}
    }
    \caption{ The expected 68\% CL uncertainties on the gluon fusion Higgs production cross-section in the different decay modes at the HL-LHC (CMS)~\cite{Cepeda:2019klc}.}
    \label{tab:HLLHC_uncertainties}
\end{table}
In this first case we consider projections for Higgs physics observables at the HL-LHC. To this end, we introduce a new Higgs fit without modifying $\chi_{\rm EWPO}^2$ in Eq.~\eqref{eq:chi2_EWPO}, or the flavour constraints of Tab.~\ref{tab:flav_summ}.
Such Higgs fit is constructed using the uncertainties of the signal strengths reported in Tab.~\ref{tab:HLLHC_uncertainties} based on Ref.~\cite{Cepeda:2019klc} and setting the central values to one. The corresponding $\chi^2$-function is constructed as 
\begin{equation}
    \chi^2_{\rm Higgs} = \sum_{\alpha, \beta} \left(\mathcal{O}_{\alpha}^{\rm theo} - \mathcal{O}^{\rm exp}_\alpha\right)\,\sigma^{-2}_{\alpha\beta}\,\left(\mathcal{O}_\beta^{\rm theo} - \mathcal{O}^{\rm exp}_\beta\right)\,,
\label{eq:Higgschi2}\end{equation}
where the theoretical values correspond to the SMEFT predictions described in Sec.~\ref{sec:Higgs}, and the quantities in Tab.~\ref{tab:HLLHC_uncertainties} enter the inverse of the covariance matrix.

The largest allowed values for the coupling enhancements are the blue bars of Fig.~\ref{fig:kappa_futurecolliders_comparison}. Moreover, we include via horizontal red lines the projections for the coupling enhancements of Ref.~\cite{deBlas:2019rxi} that considers modification of the light quark Yukawa couplings only.
Finally, we note that the HL-LHC will most likely do better than what we show in our plots as we have considered total Higgs rates only. Indeed, differential Higgs measurements could offer another way to constrain light quark Yukawa couplings at the HL-LHC, as shown in Refs.~\cite{Balzani:2023jas, Soreq:2016rae, Bonner:2016sdg, Bishara:2016jga}.

\subsection{FCC-ee projections}

The results derived in Sec.~\ref{sec:results} demonstrate the importance of the electroweak and Higgs physics in constraining models that enhance the light quark Yukawa couplings. In that respect, in this section we study the potential of various runs at the future $e^+e^-$ circular collider FCC-ee.

\paragraph{Tera-Z run.}

Our starting point is Tab.~\ref{tab:FCCee_projections} which shows the current precision on EWPOs and the corresponding FCC-ee projections. For most of the observables we used the values reported in Ref.~\cite{Bernardi:2022hny} and, for the observables not listed there, we used the values of Ref.~\cite{DeBlas:2019qco}.
Notably, a projected uncertainty reduction in almost all observables is a factor of $\mathcal{O}(10)$, resulting in an outstanding indirect reach on the scale of new particles.\footnote{The excellent indirect reach of Tera-$Z$ for simple extensions of the SM was recently also shown in Ref.~\cite{Allwicher:2024sso,terHoeve:2025gey,Maura:2024zxz,Gargalionis:2024jaw}.} 

\begin{table}[t]
    \centering
    \begin{tabular}{|c|c|c||c|c|c|}
    \hline
    $(10^{-3})$            &  LEP/SLD & FCC-ee~\cite{Bernardi:2022hny}  & $(10^{-3})$ &  LEP/SLD    & FCC-ee~\cite{DeBlas:2019qco}\\
    \hline\hline
    $\Gamma_Z$ (GeV)       &  2.3 	  &  0.03	& $R_e$       &   50      & 6\\
    $\Gamma_W^*$ (GeV)     &  20.1 	  &  1	    & $R_\mu$     &   33      & 1\\
    $m_W^*$ (GeV)          &  7.1 	  &  0.4		& $R_\tau$    &   45     &  2\\
    $\sigma_{\rm had}$ (nb)&  37.0 	  &  4		& $A_\mu$     &   15     & 0.022 \\
    $A_{e}$ 			   &  4.9  	  &  0.02	& $A_\tau$    &   15     & 0.04 \\
    $A_{FB}^b$			   &  1.55	  &	 0.1	& $R_c$       &   3.0    & 0.26 \\
    $R_b$ 				   &  0.66 	  &  0.06       & $A_b$       &   20     & 3   \\
					    &		  &			& $A_c$       &   27     & 5\\
    \hline	
    \end{tabular}
    \caption{Current uncertainties in EWPOs at $10^{-3}$ level (e.g. $\delta \Gamma_Z = 0.0023$ GeV) measured at LEP and SLD compared to the FCC-ee projections from Ref.~\cite{Bernardi:2022hny} (left) and Ref.~\cite{DeBlas:2019qco} (right). *For $\Gamma_W$ we take the PDG average~\cite{ParticleDataGroup:2024cfk}, and for $m_W$ we take Eq.~\eqref{eq:mW_new}.}
    \label{tab:FCCee_projections}
\end{table}

\begin{figure}[t]
    \centering
    \includegraphics[width=0.47\linewidth]{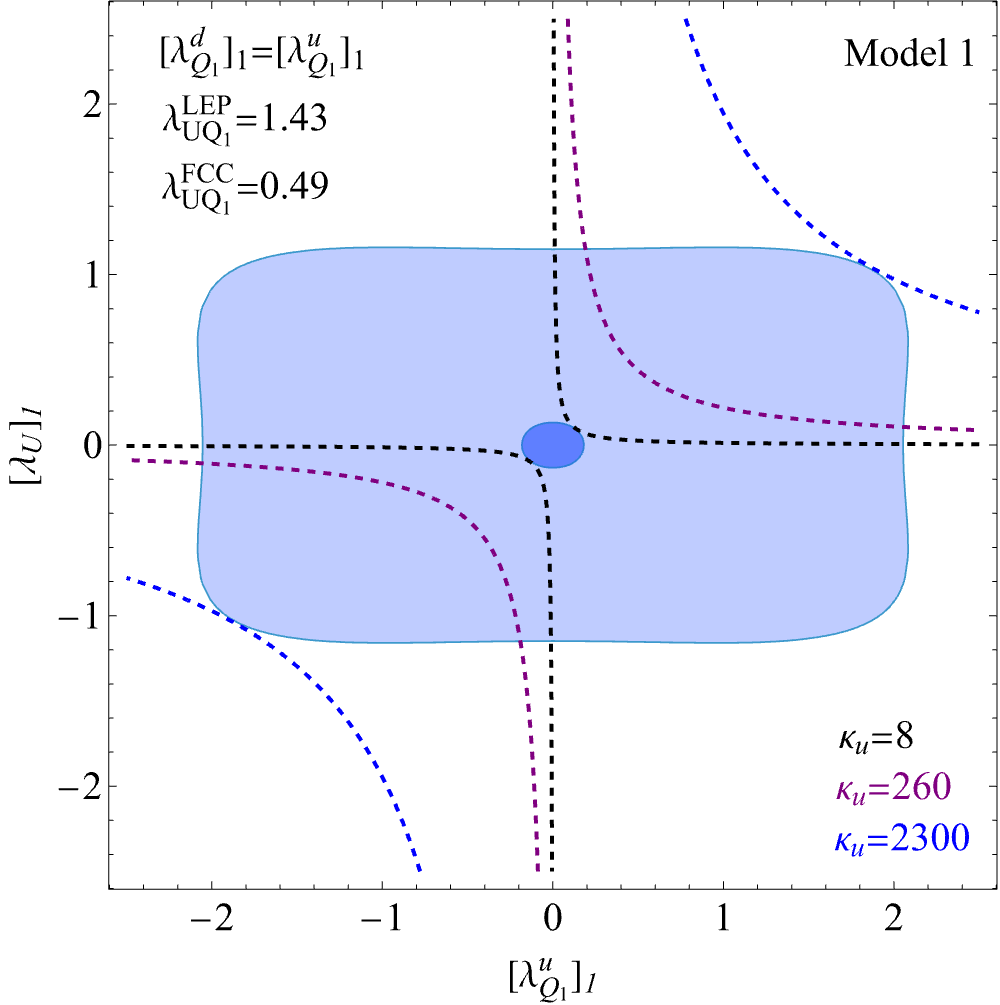}\quad
    \includegraphics[width=0.47\linewidth]{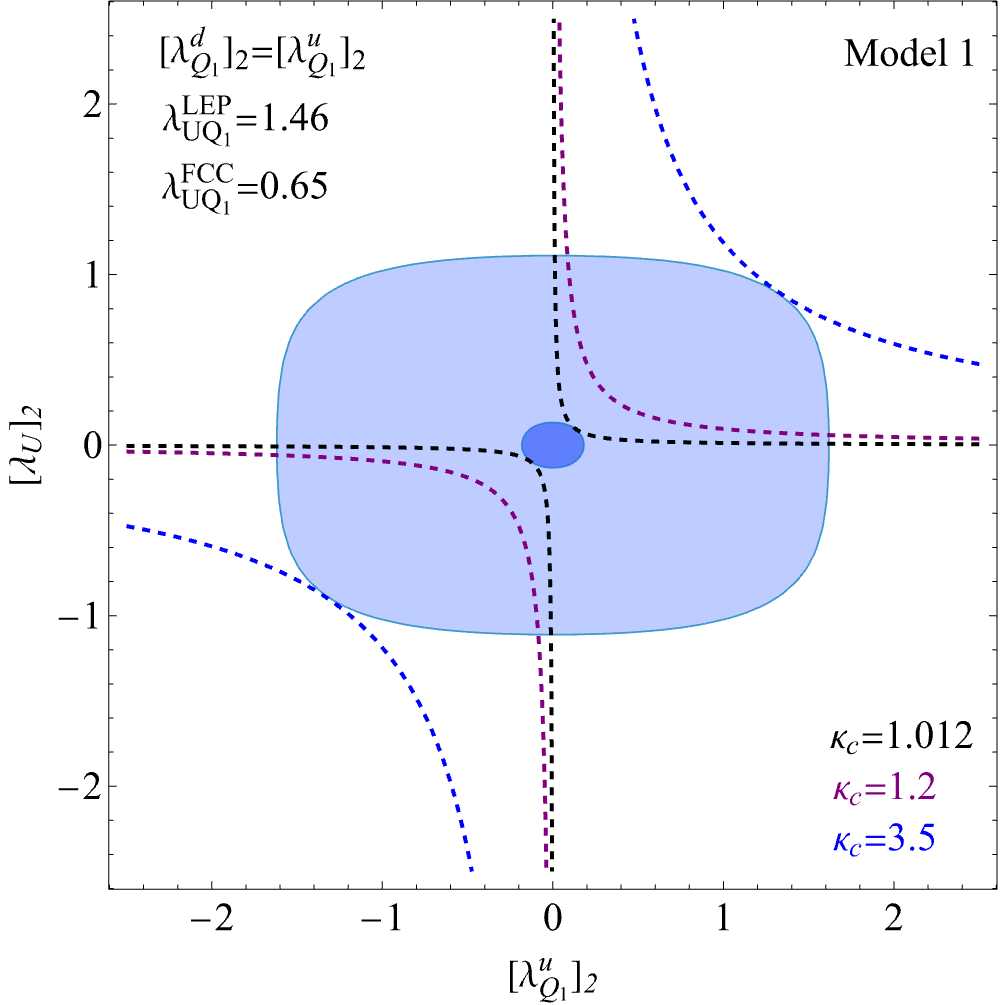}
    \caption{Allowed parameter range from a fit to electroweak data only in Model 1. The lighter (darker) color shows the 95\% CL interval considering the LEP/SLD measurements (FCC-ee projections). The dashed lines show fixed values of $\kappa_q$ in the parameter plane. \textbf{Left:} Vector-like quarks coupled to the first-generation quarks. \textbf{Right:} Vector-like quarks coupled to the second-generation quarks. In both cases the heavy scale was set to $\Lambda=\SI{2.4}{\tera\electronvolt}$ as required by the HL-LHC direct searches.}
    \label{fig:FCCee_M1}
\end{figure}

In order to quantify the FCC-ee sensitivity on the models explored in this work, we reconstruct the electroweak likelihood by setting the experimental value of each observable to its SM prediction~\cite{Breso-Pla:2021qoe} and rescaling the uncertainties as given in Tab.~\ref{tab:FCCee_projections}. In Fig.~\ref{fig:FCCee_M1}, we show the projected FCC-ee constraints on the parameter space of Model 1 and contrast it with the electroweak fit assuming current measurements and $\ldQone=\luQone$. 

In Fig.~\ref{fig:kappa_futurecolliders_comparison}, we show the maximal allowed value of $\kappa_q$ of the FCC-ee Tera-$Z$ projections as green lines. Compared to the HL-LHC run, the improvement is by at least one order of magnitude.

\paragraph{$ZH$ and $t\bar t$ runs.}
\begin{table}[t]
    \centering
    {\renewcommand{\arraystretch}{1.8}
    \begin{tabular}{|c|c||c|c|}
    \hline
    Quantity     &  Precision reach $\left[\%\right]$ & Quantity     &  Precision reach $\left[\%\right]$  \\
    \hline
    $\frac{g_{hZZ}^{\rm SMEFT}}{g_{hZZ}^{\rm SM}} - 1$                      & 0.17 &  $\frac{g_{h\tau\tau}^{\rm SMEFT}}{g_{h\tau\tau}^{\rm SM}} - 1$ & 0.74 \\
    $\frac{g_{hWW}^{\rm SMEFT}}{g_{hWW}^{\rm SM}} - 1$                      & 0.43 &  $\frac{g_{h\mu\mu}^{\rm SMEFT}}{g_{h\mu\mu}^{\rm SM}} - 1$     & 9.0 \\
    $\frac{g_{h\gamma\gamma}^{\rm SMEFT}}{g_{h\gamma\gamma}^{\rm SM}} - 1$  & 3.9  &  $\frac{\Gamma_h^{\rm SMEFT}}{\Gamma_h^{\rm SM}}-1$             & 1.3 \\
    $\frac{g_{hbb}^{\rm SMEFT}}{g_{hbb}^{\rm SM}} - 1$                      & 0.61 &                           &   \\
    \hline
    \end{tabular}
    }
    \caption{Precision reach on the effective Higgs couplings to various states at 240 GeV and 365 GeV center-of-mass runs at the FCC-ee~\cite{FCC:2018evy}.}
    \label{tab:FCC_240-65GeVruns}
\end{table}

To investigate the impact of Higgs physics at the FCC-ee, we construct and add a new Higgs $\chi^2$-function based on the projections for the $\SI{240}{\giga\electronvolt}$ run (optimized for the $ZH$ production) and $\SI{365}{\giga\electronvolt}$ run (optimized for the $t\bar t$ production). In particular, we use the precision on the Higgs couplings provided in Ref.~\cite{FCC:2018evy}, which we report in Tab.~\ref{tab:FCC_240-65GeVruns}. 

The Higgs physics $\chi^2$-function is constructed starting from the signal strengths in Eq.~\eqref{eq:signalstrenght}. Similarly to the electroweak likelihood, we set the experimental value of each observable in Tab.~\ref{tab:FCC_240-65GeVruns} to its SM prediction and use the corresponding uncertainties. 

The purple bars in Fig.~\ref{fig:kappa_futurecolliders_comparison} show the largest allowed values at $95\%$ CL. Interestingly, the results show no significant deviation from those considering the sole Tera-Z run, justifying the statement that electroweak precision observables play a dominant role in constraining UV models with VLQs.

Summarising the results we find for the future projections, we conclude that the Tera-$Z$ run at the FCC-ee provides a unique opportunity to probe the scenarios studied in this work. This comes from the fact that it remains difficult to probe directly light quark Yukawa couplings at future colliders unless it becomes possible to flavour-tag also the light quarks in Higgs boson decays with good precision.\footnote{Even in that case, such decays might not put competitive bounds on the light Yukawa couplings due to Dalitz decays.} 

Moreover, the studied scenarios always imply tree-level modifications of the quark couplings to $Z$ and $W$ bosons, which depend on the mixing between the SM quarks and VLQs. The other couplings entering into the light quark Yukawa modifications, namely the couplings among two different VLQs, can be probed by electroweak precision tests at one-loop level. In contrast, Higgs physics probes the analyzed scenarios only at loop-level, or via the modification of the total width. Finally, precision in EWPOs will be approximately an order of magnitude larger than in the case of Higgs observables. For these reasons, it turns out that the FCC-ee Tera-$Z$ run is more relevant than the 240 GeV and 365 GeV runs for the scenarios considered in our analysis.

Similarly, the significance of electroweak observables and the level of precision attainable in their measurements suggest that the FCC-ee outperforms other lepton colliders, such as CLIC or a high-energy muon collider. This conclusion has been explicitly verified in the case of a muon collider operating at a center-of-mass energy of $\SI{10}{\tera\electronvolt}$ and an integrated luminosity of $\SI{10}{\atto\barn}^{-1}$. Refs.~\cite{Accettura:2023ked, InternationalMuonCollider:2024jyv} indicate that such a collider could probe the positively charged vector-like quark (VLQ) components at an energy scale of approximately $\SI{5}{\tera\electronvolt}$. 
Utilizing the precision reach on effective Higgs couplings as reported in Ref.\cite{deBlas:2022ofj}, we find that the muon collider does not yield competitive constraints on $\kappa_q$ relative to those presented in Fig.\ref{fig:kappa_futurecolliders_comparison}. Instead, its sensitivity to $\kappa_q$ remains comparable to the projected limits of the HL-LHC. In this case the theory bounds considered in Ref.~\cite{Adhikary:2024esf} will become more relevant.

\section{Conclusion}
\label{sec:conclusion}
In this work, we have addressed the question of \textit{how large light quark Yukawa couplings can be}. In order to do so, we systematically identify simplified models that incorporate a pair of vector-like quark representations, which contribute to the effective Yukawa couplings without relying on the insertion of the renormalisable Yukawa coupling, $y_q$. Within the framework of effective field theory, these models generate a dimension-6 Yukawa-like operator, $\phi^\dagger\phi(q_L\phi q_R)/\Lambda^2$, which leads to a significant enhancement compared to the renormalisable term, $y_q(q_L\phi q_R)$, as 
$y_q\ll v^2/\Lambda^2$. Moreover, our analysis is rather comprehensive, as these are \emph{all possible} simplified models that produce such an enhancement without involving any $s$-channel resonances decaying into di-jets.

In addition to modifying the Yukawa interactions, the considered models generate effective operators that affect electroweak precision tests, flavour physics, and Higgs physics. This includes, for instance, tree-level contributions to operators of the class $\psi^2 \phi^2 D$ which modify the couplings of quarks to $Z$ and $W$ bosons which are subject to constraints from electroweak precision tests or flavour physics. Further relevant bounds on the model arise from operators generated at one-loop level, like effective interactions of the Higgs boson with photons and gluons and contributions to operators of type $\phi^4 D^2$ that are constrained by $Z$ pole measurements and Higgs couplings to vector bosons. We carefully analysed the interplay of all these effects by performing a fit to the electroweak precision observables and current Higgs data.

Furthermore, we conducted a detailed analysis of the constraints on vector-like quarks arising from direct searches for their pair production, followed by their decay into $W^\pm q$, where 
$q$ represents a light quark flavour. Based on the limits established by these searches, we set the new physics scale in our analysis to $\Lambda = 1.6$ TeV independent of the new vector-like quark couplings. Concurrently, we demonstrated that flavour physics imposes stringent constraints on the models under consideration. However, these constraints are significantly relaxed if we assume that the vector-like quarks couple exclusively to a single light generation. Under this assumption, flavour physics bounds primarily restrict the couplings of $SU(2)$ singlet and triplet states to Standard Model quarks through the unitarity of the CKM matrix and $\Delta F=2$ transitions (the latter being relevant only for triplet states).

Taking into account all of the constraints we could indeed show that it is still possible to have light quark Yukawa couplings generally enhanced by a factor of $O(600)$ for the up quark, $O(300)$ for the down quark, $O(20)$ for the strange quark, and $O(3)$ for the charm Yukawa coupling. Our findings \emph{motivate} dedicated searches at the HL-LHC for those couplings as proposed for instance in Refs.~\cite{Balzani:2023jas, Brivio:2015fxa, Alasfar:2019pmn,  Vignaroli:2022fqh, Yu:2016rvv, Bishara:2016jga,Aguilar-Saavedra:2020rgo,Soreq:2016rae, Falkowski:2020znk},  which can typically probe enhancement factors of a few hundred for the first generation, values of around 10 for the strange and deviations of order one for the charm Yukawa coupling. 

Given the fact that electroweak precision tests give important constraints on the models, we considered also the future runs at the FCC-ee. We showed an excellent sensitivity reach to the parameter space of all models, allowing maximally $O(20)$ deviations in the up-Yukawa coupling and $O(10)$ deviations in the down-Yukawa coupling. Regarding the second-generation quarks in this setup, the FCC-ee will generally probe contributions to the strange-Yukawa coupling below 1.8 times the SM value, and below 1.04 times the SM value in case of the charm Yukawa coupling. In stating these general statements, the values provided by Model 5 for the up-type coupling modifiers and by Model 7 for the down-type modifiers are considered to be exceptions for the overall behavior. 

Finally, having shown that large enhancements in the light quark Yukawa couplings are possible in light of other experimental probes, we encourage experimental collaborations to perform dedicated searches. We would also like to point out a non-trivial interplay between the light quark Yukawa enhancements and limits on other operators in an EFT~\cite{Alasfar:2022vqw}, such that the possibility of largely enhanced light quark Yukawa couplings should be taken into account when performing fits to data.


\section*{Acknowledgments}
We thank Paride Paradisi for the interesting discussions on flavour bounds on VLQ models. Also, we are grateful to Lukas Allwicher and Javier M. Lizana for providing helpful comments on the electroweak likelihood, and Micha{\l} Ryczkowski for comments about theoretical constraints on the couplings of vector-like quark to the Higgs. Further, we thank John Gargalionis and Tevong You for carefully reading the manuscript. BE would like to thank Stefano Di Noi and Gabriele Levati for helpful discussions.
This work received funding by the INFN Iniziativa Specifica APINE. This work was also partially supported by the Italian MUR Departments of Excellence grant 2023-2027 “Quantum Frontiers”. The work of RG is supported by the University of Padua under the 2023 STARS Grants@Unipd programme (Acronym and title of the project: HiggsPairs – Precise Theoretical Predictions for Higgs pair production at the LHC).

\appendix

\section{Matching of \texorpdfstring{$D^2\phi^4$}{D2phi4} operators}
\label{app:CHD}

Because of their importance in the electroweak and Higgs fits, we show explicitly the matching of the Wilson coefficients of the SMEFT operators in the $D^2\phi^4$ class in the Warsaw basis~\cite{Grzadkowski:2010es}
\begin{align}
    \cO_{\phi\Box} &= (\phi^\dagger \phi)\Box (\phi^\dagger \phi)\,,\\
    \cO_{\phi D} &= |\phi^\dagger D_\mu \phi|^2\,.
\end{align}
The matching is performed by equating the four-point Higgs amplitudes in the SMEFT and the full models. The four-point amplitude in the SMEFT receives contributions from four operators in the Green's basis \cite{Gherardi:2020det}
\begin{align}
     {Q}_{\phi\Box} &= (\phi^\dagger \phi)\Box (\phi^\dagger \phi)\,,\\
     {Q}_{\phi D} &= |\phi^\dagger D_\mu \phi|^2\,,\\
     {Q}_{\phi D}^{\prime} &= (\phi^\dagger \phi)(D_\mu \phi)^\dagger(D^\mu \phi)\,,\\
     {Q}_{\phi D}^{\prime\prime} &= (\phi^\dagger \phi)D_\mu(i \phi^\dagger\overleftrightarrow{D}^\mu \phi)\,.
\end{align}

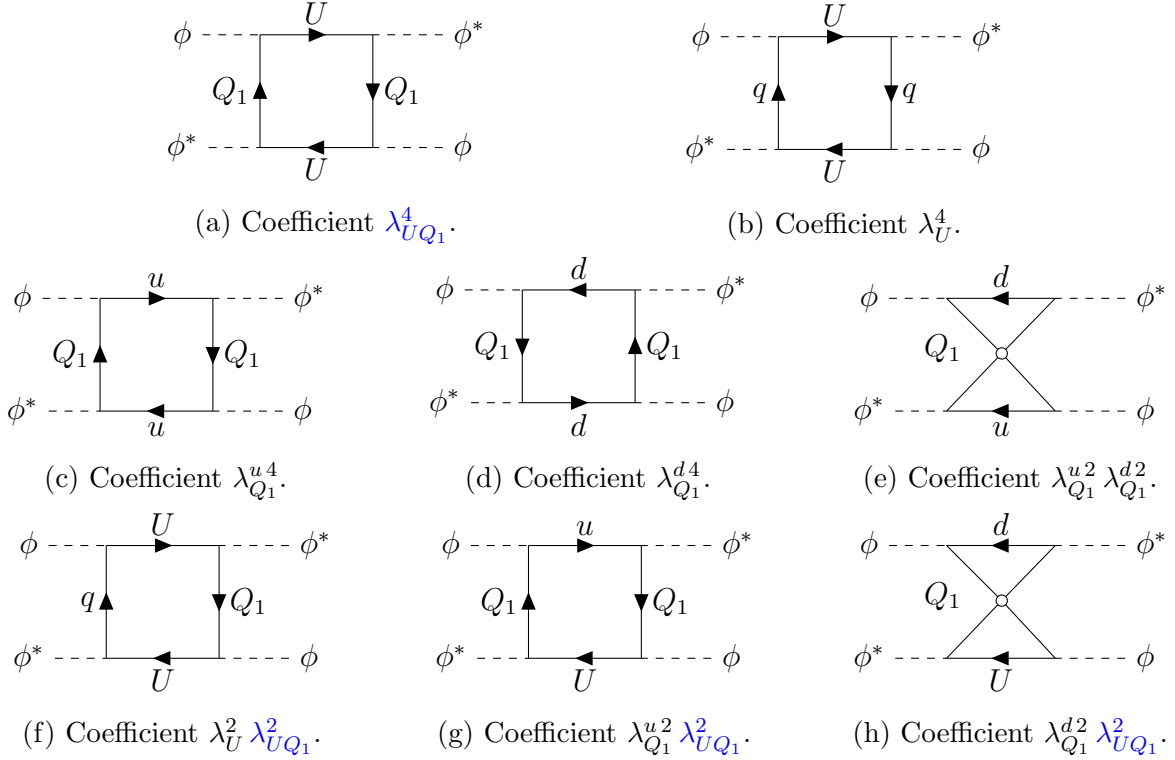
\begin{figure}[t]
\centering
\begin{subfigure}[b]{0.32\linewidth}
    \centering
    \begin{tikzpicture}
        \begin{feynman}
            \vertex (phi1) at (0,0) {\(\phi\)};
            \vertex[right = 2.5em of phi1] (A);
            \vertex[right = of A] (B);
            \vertex[right = 2.25em of B] (phi4) {\(\phi^{*}\)};
            \vertex[below = of phi1] (phi3) {\(\phi^{*}\)};
            \vertex[right= 2.5em of phi3] (D);
            \vertex[right = of D] (C);
            \vertex[right = 2.25em of C] (phi2) {\(\phi\)};
            \diagram*{
                (phi1) -- [scalar] (A);
                (phi2) -- [scalar] (C);
                (phi3) -- [scalar] (D);
                (phi4) -- [scalar] (B);
                (A) -- [fermion, edge label = \(U\)] (B) -- [fermion, edge label = \(Q_1\)] (C) -- [fermion, edge label = \(U\)] (D) -- [fermion, edge label = \(Q_1\)] (A);
            };
        \end{feynman}
    \end{tikzpicture}
    \caption{Coefficient $\textcolor{blue}{\lambda_{UQ_1}^4}$.}
    \label{fig:chduvdiagramsUQ14}
\end{subfigure}
\hspace*{1.25cm}
\begin{subfigure}[b]{0.33\linewidth}
    \centering
    \begin{tikzpicture}
        \begin{feynman}
            \vertex (phi1) at (0,0) {\(\phi\)};
            \vertex[right = 2.5em of phi1] (A);
            \vertex[right = of A] (B);
            \vertex[right = 2.25em of B] (phi4) {\(\phi^{*}\)};
            \vertex[below = of phi1] (phi3) {\(\phi^{*}\)};
            \vertex[right= 2.5em of phi3] (D);
            \vertex[right = of D] (C);
            \vertex[right = 2.25em of C] (phi2) {\(\phi\)};
            \diagram*{
                (phi1) -- [scalar] (A);
                (phi2) -- [scalar] (C);
                (phi3) -- [scalar] (D);
                (phi4) -- [scalar] (B);
                (A) -- [fermion, edge label = \(U\)] (B) -- [fermion, edge label = \(q\)] (C) -- [fermion, edge label = \(U\)] (D) -- [fermion, edge label = \(q\)] (A);
            };
        \end{feynman}
    \end{tikzpicture}
    \caption{Coefficient $\lambda_{U}^4$.}
    \label{fig:chduvdiagramsU4}
\end{subfigure}\\
\begin{subfigure}[b]{0.32\linewidth}
    \centering
    \begin{tikzpicture}
        \begin{feynman}
            \vertex (phi1) at (0,0) {\(\phi\)};
            \vertex[right = 2.5em of phi1] (A);
            \vertex[right = of A] (B);
            \vertex[right = 2.25em of B] (phi4) {\(\phi^{*}\)};
            \vertex[below = of phi1] (phi3) {\(\phi^{*}\)};
            \vertex[right= 2.5em of phi3] (D);
            \vertex[right = of D] (C);
            \vertex[right = 2.25em of C] (phi2) {\(\phi\)};
            \diagram*{
                (phi1) -- [scalar] (A);
                (phi2) -- [scalar] (C);
                (phi3) -- [scalar] (D);
                (phi4) -- [scalar] (B);
                (A) -- [fermion, edge label = \(u\)] (B) -- [fermion, edge label = \(Q_1\)] (C) -- [fermion, edge label = \(u\)] (D) -- [fermion, edge label = \(Q_1\)] (A);
            };
        \end{feynman}
    \end{tikzpicture}
    \caption{Coefficient $\lambda^{u\,4}_{Q_1}$.}
    \label{fig:chduvdiagramsuQ14}
\end{subfigure}
\hfill
\begin{subfigure}[b]{0.32\linewidth}
    \centering
    \begin{tikzpicture}
        \begin{feynman}
            \vertex (phi1) at (0,0) {\(\phi\)};
            \vertex[right = 2.5em of phi1] (A);
            \vertex[right = of A] (B);
            \vertex[right = 2.25em of B] (phi4) {\(\phi^{*}\)};
            \vertex[below = of phi1] (phi3) {\(\phi^{*}\)};
            \vertex[right= 2.5em of phi3] (D);
            \vertex[right = of D] (C);
            \vertex[right = 2.25em of C] (phi2) {\(\phi\)};
            \diagram*{
                (phi1) -- [scalar] (A);
                (phi2) -- [scalar] (C);
                (phi3) -- [scalar] (D);
                (phi4) -- [scalar] (B);
                (A) -- [fermion, edge label' = \(Q_1\)] (D) -- [fermion, edge label' = \(d\)] (C) -- [fermion, edge label' = \(Q_1\)] (B) -- [fermion, edge label' = \(d\)] (A);
            };
        \end{feynman}
    \end{tikzpicture}
    \caption{Coefficient $\lambda^{d\,4}_{Q_1}$.}
    \label{fig:chduvdiagramsdQ14}
\end{subfigure}
\hfill
\begin{subfigure}[b]{0.32\linewidth}
    \centering
    \begin{tikzpicture}
        \begin{feynman}
            \vertex (phi1) at (0,0) {\(\phi\)};
            \vertex[right = 2.5em of phi1] (A);
            \vertex[right = 3.5 em of A] (B);
            \vertex[right = 2.25em of B] (phi4) {\(\phi^{*}\)};
            \vertex[below = of phi1] (phi3) {\(\phi^{*}\)};
            \vertex[right= 2.5em of phi3] (D);
            \vertex[right = 3.5 em of D] (C);
            \vertex[right = 2.25em of C] (phi2) {\(\phi\)};
            \vertex[below right = 2.33 em of A] (empty) [empty dot] {};
            \diagram*{
                (phi1) -- [scalar] (A);
                (phi2) -- [scalar] (C);
                (phi3) -- [scalar] (D);
                (phi4) -- [scalar] (B);
                (B) -- [fermion, edge label' = \(d\)] (A) -- [plain, edge label' = \(Q_1\)] (empty) -- [plain] (C) -- [fermion, edge label = \(u\)] (D) -- [plain] (empty)  -- [plain] (B);
            };
        \end{feynman}
    \end{tikzpicture}
    \caption{Coefficient $\lambda^{u\,2}_{Q_1} \, \lambda^{d\,2}_{Q_1}$.}
    \label{fig:chduvdiagramsuQ12dQ12}
\end{subfigure}
\\
\begin{subfigure}[b]{0.33\linewidth}
    \centering
    \begin{tikzpicture}
        \begin{feynman}
            \vertex (phi1) at (0,0) {\(\phi\)};
            \vertex[right = 2.5em of phi1] (A);
            \vertex[right = of A] (B);
            \vertex[right = 2.25em of B] (phi4) {\(\phi^{*}\)};
            \vertex[below = of phi1] (phi3) {\(\phi^{*}\)};
            \vertex[right= 2.5em of phi3] (D);
            \vertex[right = of D] (C);
            \vertex[right = 2.25em of C] (phi2) {\(\phi\)};
            \diagram*{
                (phi1) -- [scalar] (A);
                (phi2) -- [scalar] (C);
                (phi3) -- [scalar] (D);
                (phi4) -- [scalar] (B);
                (A) -- [fermion, edge label = \(U\)] (B) -- [fermion, edge label = \(Q_1\)] (C) -- [fermion, edge label = \(U\)] (D) -- [fermion, edge label = \(q\)] (A);
            };
        \end{feynman}
    \end{tikzpicture}
    \caption{Coefficient $\lambda_{U}^2\, \textcolor{blue}{\lambda^2_{UQ_1}}$.}
    \label{fig:chduvdiagramsU2UQ12}
\end{subfigure}
\hfill
\begin{subfigure}[b]{0.33\linewidth}
    \centering
    \begin{tikzpicture}
        \begin{feynman}
            \vertex (phi1) at (0,0) {\(\phi\)};
            \vertex[right = 2.5em of phi1] (A);
            \vertex[right = of A] (B);
            \vertex[right = 2.25em of B] (phi4) {\(\phi^{*}\)};
            \vertex[below = of phi1] (phi3) {\(\phi^{*}\)};
            \vertex[right= 2.5em of phi3] (D);
            \vertex[right = of D] (C);
            \vertex[right = 2.25em of C] (phi2) {\(\phi\)};
            \diagram*{
                (phi1) -- [scalar] (A);
                (phi2) -- [scalar] (C);
                (phi3) -- [scalar] (D);
                (phi4) -- [scalar] (B);
                (A) -- [fermion, edge label = \(u\)] (B) -- [fermion, edge label = \(Q_1\)] (C) -- [fermion, edge label = \(U\)] (D) -- [fermion, edge label = \(Q_1\)] (A);
            };
        \end{feynman}
    \end{tikzpicture}
    \caption{Coefficient $\lambda_{Q_1}^{u\,2}\, \textcolor{blue}{\lambda^2_{UQ_1}}$.}
    \label{fig:chduvdiagramsuQ12UQ12}
\end{subfigure}
\hfill
\begin{subfigure}[b]{0.32\linewidth}
    \centering
    \begin{tikzpicture}
        \begin{feynman}
            \vertex (phi1) at (0,0) {\(\phi\)};
            \vertex[right = 2.5em of phi1] (A);
            \vertex[right = 3.5 em of A] (B);
            \vertex[right = 2.25em of B] (phi4) {\(\phi^{*}\)};
            \vertex[below = of phi1] (phi3) {\(\phi^{*}\)};
            \vertex[right= 2.5em of phi3] (D);
            \vertex[right = 3.5 em of D] (C);
            \vertex[right = 2.25em of C] (phi2) {\(\phi\)};
            \vertex[below right = 2.33 em of A] (empty) [empty dot] {};
            \diagram*{
                (phi1) -- [scalar] (A);
                (phi2) -- [scalar] (C);
                (phi3) -- [scalar] (D);
                (phi4) -- [scalar] (B);
                (B) -- [fermion, edge label' = \(d\)] (A) -- [plain, edge label' = \(Q_1\)] (empty) -- [plain] (C) -- [fermion, edge label = \(U\)] (D) -- [plain] (empty)  -- [plain] (B);
            };
        \end{feynman}
    \end{tikzpicture}
    \caption{Coefficient $\lambda^{d\,2}_{Q_1} \, \textcolor{blue}{\lambda^2_{UQ_1}}$.}
    \label{fig:chduvdiagramsdQ12UQ12}
\end{subfigure}
\caption{One Feynman diagram for each of the VLQ coupling combinations appearing in the matched expressions of $\cC_{\phi D}$ and $\cC_{\phi\Box}$ for Model 1. The presence of the VLQ-VLQ-Higgs coupling is highlighted in blue. The empty dot is used to denote the one-loop diagrams in which the propagators cross over each other without intercepting.}
\label{fig:chduvdiagrams}
\end{figure}

The contribution to the four-point Higgs amplitude in the full models originates from the one-loop box diagrams. In Model 1, the one-loop diagrams are shown in Fig.~\ref{fig:chduvdiagrams}. We distinguish the contributions with two, three, or four heavy states propagating in the loop. Furthermore, we choose four different momentum configurations for the external Higgs states and set up the system for the Wilson coefficients in the Green's basis $G_{\phi \Box}\,,G_{\phi D}\,,G_{\phi D}^{\prime}$, and $G_{\phi D}^{\prime\prime}$. Finally, we project onto the Warsaw basis with
\begin{align}
    \cC_{\phi D} &= G_{\phi D}\,,\\
    \cC_{\phi \Box} &= G_{\phi\Box}+\frac{1}{2} G_{\phi D}^\prime\,.
\end{align}

\subsection*{Model 1: \texorpdfstring{$U+Q_1$}{U+Q1}}

In the limit of equal $Q_1$ and $U$ masses, $M_{Q_1}=M_U=\Lambda$, we obtain the following results for Model 1
\begin{align}
    16\pi^2\Lambda^2\cC_{\phi\Box} = &-\frac{6}{5}|\lambda_{UQ_1}|^4 - \frac{5}{4}|\lambda_{UQ_1}|^2\left(\left[\lambda_{Q_1}^d\right]_p^*\left[\lambda_{Q_1}^d\right]_p + 2\left[\lambda_{Q_1}^u\right]_p^*\left[\lambda_{Q_1}^u\right]_p+\left[\lambda_{U}\right]_p^*\left[\lambda_{U}\right]_p\right)\nonumber\\
    & -\frac{3}{4}\left[\lambda_{Q_1}^d\right]_p^*\left[\lambda_{Q_1}^d\right]_p \left[\lambda_{Q_1}^d\right]_q^*\left[\lambda_{Q_1}^d\right]_q -\frac{3}{4}\left[\lambda_{Q_1}^u\right]_p^*\left[\lambda_{Q_1}^u\right]_p \left[\lambda_{Q_1}^u\right]_q^*\left[\lambda_{Q_1}^u\right]_q\nonumber\\
    &- 2\left[\lambda_{Q_1}^d\right]_p^*\left[\lambda_{Q_1}^d\right]_p \left[\lambda_{Q_1}^u\right]_q^*\left[\lambda_{Q_1}^u\right]_q-  \left[\lambda_{U}\right]_p^*\left[\lambda_{U}\right]_p \left[\lambda_{U}\right]_q^*\left[\lambda_{U}\right]_q\,,
\end{align}
\begin{align}
    16\pi^2\Lambda^2\cC_{\phi D} = &-\frac{12}{5}|\lambda_{UQ_1}|^4 + \frac{5}{2}|\lambda_{UQ_1}|^2\left(\left[\lambda_{Q_1}^d\right]_p^*\left[\lambda_{Q_1}^d\right]_p-\left[\lambda_{Q_1}^u\right]_p^*\left[\lambda_{Q_1}^u\right]_p-2\left[\lambda_{U}\right]_p^*\left[\lambda_{U}\right]_p\right)\nonumber\\
    & -2\left[\lambda_{Q_1}^d\right]_p^*\left[\lambda_{Q_1}^d\right]_p \left[\lambda_{Q_1}^d\right]_q^*\left[\lambda_{Q_1}^d\right]_q -2\left[\lambda_{Q_1}^u\right]_p^*\left[\lambda_{Q_1}^u\right]_p \left[\lambda_{Q_1}^u\right]_q^*\left[\lambda_{Q_1}^u\right]_q\nonumber\\
    &+ 4\left[\lambda_{Q_1}^d\right]_p^*\left[\lambda_{Q_1}^d\right]_p \left[\lambda_{Q_1}^u\right]_q^*\left[\lambda_{Q_1}^u\right]_q - \frac{3}{2} \left[\lambda_{U}\right]_p^*\left[\lambda_{U}\right]_p \left[\lambda_{U}\right]_q^*\left[\lambda_{U}\right]_q\,.
\end{align}

\subsection*{Model 2: \texorpdfstring{$D+Q_1$}{D+Q1}}
Setting $M_D = M_{Q_1} = \Lambda$, the two coefficients take the form
\begin{align}
    16\pi^2 \Lambda^2 \cC_{\phi \Box} = &-\frac{6}{5} |\lambda_{DQ_1}|^4 - \frac{5}{4} |\lambda_{DQ_1}|^2 \left(\left[\lambda_{Q_1}^u \right]^*_p\left[\lambda_{Q_1}^u \right]_p + 2 \left[\lambda_{Q_1}^d \right]_p^* \left[\lambda_{Q_1}^d \right]_p + \left[\lambda_D \right]_p^*\left[\lambda_D \right]_p \right)   \nonumber\\
    & -\frac{3}{4}\left[\lambda^d_{Q_1}\right]^*_p\left[\lambda^d_{Q_1}\right]_p \left[\lambda^d_{Q_1}\right]^*_q \left[\lambda^d_{Q_1}\right]_q - \frac{3}{4}\left[\lambda^u_{Q_1}\right]^*_p\left[\lambda^u_{Q_1}\right]_p \left[\lambda^u_{Q_1}\right]^*_q \left[\lambda^u_{Q_1}\right]_q \nonumber\\
    & -2 \left[\lambda^d_{Q_1}\right]_p^* \left[\lambda^d_{Q_1}\right]_p \left[\lambda^u_{Q_1}\right]_q^* \left[\lambda^u_{Q_1}\right]_q - \left[\lambda_D\right]_p^* \left[\lambda_D\right]_p \left[\lambda_D\right]_q^* \left[\lambda_D\right]_q\,,
\end{align}
\begin{align}
    16\pi^2 \Lambda^2 \cC_{\phi D} = & -\frac{12}{5} |\lambda_{DQ_1}|^4 + \frac{5}{2}|\lambda_{DQ_1}|^2 \left( \left[\lambda^u_{Q_1}\right]_p^* \left[\lambda^u_{Q_1}\right]_p - \left[\lambda^d_{Q_1}\right]^*_p \left[\lambda^d_{Q_1}\right]_p - 2 \left[\lambda_D\right]^*_p \left[\lambda_D\right]_p  \right) \nonumber\\
    & -2 \left[\lambda^d_{Q_1}\right]^*_p\left[\lambda^d_{Q_1}\right]_p \left[\lambda^d_{Q_1}\right]^*_q \left[\lambda^d_{Q_1}\right]_q -2 \left[\lambda^u_{Q_1}\right]^*_p\left[\lambda^u_{Q_1}\right]_p \left[\lambda^u_{Q_1}\right]_q^* \left[\lambda^u_{Q_1}\right]_q \nonumber\\
    & + 4 \left[\lambda^d_{Q_1}\right]_p^* \left[\lambda^d_{Q_1}\right]_p \left[\lambda^u_{Q_1}\right]_q^* \left[\lambda^u_{Q_1}\right]_q - \frac{3}{2} \left[\lambda_D\right]_p^* \left[\lambda_D\right]_p \left[\lambda_D\right]_q^* \left[\lambda_D\right]_q\,.
\end{align}

\subsection*{Model 3: \texorpdfstring{$U+Q_7$}{U+Q7}}
For $M_U=M_{Q_7}=\Lambda$, we find
\begin{align}
    16\pi^2 \Lambda^2 \cC_{\phi\Box} =& - \frac{6}{5} |\lambda_{UQ_7}|^4 -\frac{5}{4} |\lambda_{UQ_7}|^2 \left(- \left[\lambda_{U}\right]_p^* \left[\lambda_U\right]_p +  2 \left[\lambda_{Q_7}\right]^*_p \left[\lambda_{Q_7}\right]_p \right) \nonumber\\
    & - \left[\lambda_{U}\right]_p^* \left[\lambda_U\right]_p \left[\lambda_{U}\right]_r^* \left[\lambda_U\right]_r - \frac{3}{4} \left[\lambda_{Q_7}\right]^*_p \left[\lambda_{Q_7}\right]_p \left[\lambda_{Q_7}\right]^*_r \left[\lambda_{Q_7}\right]_r\,,
\end{align}
\begin{align}
    16\pi^2 \Lambda^2 \cC_{\phi D} = & - \frac{12}{5}|\lambda_{UQ_7}|^4 + \frac{5}{2} |\lambda_{UQ_7}|^2 \left( 2 \left[\lambda_U\right]^*_p \left[\lambda_U\right]_p  - \left[\lambda_{Q_7}\right]^*_p \left[\lambda_{Q_7}\right]_p \right) \nonumber\\
    &- \frac{3}{2} \left[\lambda_U\right]^*_p \left[\lambda_U\right]_p \left[\lambda_U\right]^*_r \left[\lambda_U\right]_r - 2 \left[\lambda_{Q_7}\right]^*_p \left[\lambda_{Q_7}\right]_p\left[\lambda_{Q_7}\right]^*_r \left[\lambda_{Q_7}\right]_r\,.
\end{align}

\subsection*{Model 4: \texorpdfstring{$D+Q_5$}{D+Q5}}
For $M_D = M_{Q_5} = \Lambda$, we find
\begin{align}
    16\pi^2 \Lambda^2 \cC_{\phi\Box} = & -\frac{6}{5} |\lambda_{DQ_5}|^4 + \frac{5}{4}|\lambda_{DQ_5}|^2 \left( \left[\lambda_D\right]_p \left[\lambda_D\right]^*_p - 2 \left[\lambda_{Q_5}\right]_p \left[\lambda_{Q_5}\right]_p^* \right)\nonumber\\
    & - \frac{3}{4}\left[\lambda_{Q_5}\right]_p \left[\lambda_{Q_5}\right]_p^*\left[\lambda_{Q_5}\right]_r \left[\lambda_{Q_5}\right]_r^* - \left[\lambda_D\right]_p \left[\lambda_D\right]_p^*\left[\lambda_D\right]_r \left[\lambda_D\right]_r^* \,,
\end{align}
\begin{align}
    16\pi^2 \Lambda^2 \cC_{\phi D} =& -\frac{12}{5} |\lambda_{DQ_5}|^4 + \frac{5}{2}|\lambda_{DQ_5}|^2 \left( 2 \left[\lambda_D\right]_p \left[\lambda_D\right]_p^* - \left[\lambda_{Q_5}\right]_p \left[\lambda_{Q_5}\right]_p^* \right) \nonumber\\
    & - 2\left[\lambda_D\right]_p \left[\lambda_D\right]_p^*\left[\lambda_D\right]_r \left[\lambda_D\right]_r^* - \frac{3}{2}\left[\lambda_{Q_5}\right]_p \left[\lambda_{Q_5}\right]_p^*\left[\lambda_{Q_5}\right]_r \left[\lambda_{Q_5}\right]_r^* \,.
\end{align}

\subsection*{Model 5: \texorpdfstring{$T_1+Q_1$}{T1+Q1}}
In the equal mass limit, $M_{T_1} = M_{Q_1} = \Lambda$, we find
\begin{align}
    16\pi^2 \Lambda^2 \cC_{\phi\Box} = & -\frac{3}{8} |\lambda_{T_1Q_1}|^4 + \frac{15}{64} |\lambda_{T_1Q_1}|^2 \left(4\left[\lambda^u_{Q_1} \right]_p\left[\lambda^u_{Q_1} \right]_p^* - 3 \left[\lambda_{T_1} \right]_p\left[\lambda_{T_1} \right]_p^*  \right) \nonumber\\
    & -2 \left[\lambda^u_{Q_1} \right]_p\left[\lambda^u_{Q_1} \right]_p^* \left[\lambda^d_{Q_1} \right]_r\left[\lambda^d_{Q_1} \right]_r^* - \frac{3}{4}\left[\lambda^u_{Q_1} \right]_p\left[\lambda^u_{Q_1} \right]_p^*\left[\lambda^u_{Q_1} \right]_r\left[\lambda^u_{Q_1} \right]_r^* \nonumber\\
    & - \frac{3}{4}\left[\lambda^d_{Q_1} \right]_p\left[\lambda^d_{Q_1} \right]_p^*\left[\lambda^d_{Q_1} \right]_r\left[\lambda^d_{Q_1} \right]_r^* - \frac{1}{4} \left[\lambda_{T_1} \right]_p\left[\lambda_{T_1} \right]_p^*\left[\lambda_{T_1} \right]_r\left[\lambda_{T_1} \right]_r^*\, ,
\end{align}
\begin{align}
    16\pi^2 \Lambda^2 \cC_{\phi D} = & -\frac{3}{4} |\lambda_{T_1Q_1}|^4 - \frac{15}{16} |\lambda_{T_1Q_1}|^2 \left( 2\left[\lambda^d_{Q_1}\right]_p \left[\lambda^d_{Q_1}\right]_p^* - 2 \left[\lambda^u_{Q_1}\right]_p \left[\lambda^u_{Q_1}\right]_p^*  + \left[\lambda_{T_1}\right]_p \left[\lambda_{T_1}\right]_p^*\right) \nonumber\\
    & + 4 \left[\lambda^u_{Q_1}\right]_p \left[\lambda^u_{Q_1}\right]_p^* \left[\lambda^d_{Q_1}\right]_r \left[\lambda^d_{Q_1}\right]_r^* - 2 \left[\lambda^u_{Q_1}\right]_p \left[\lambda^u_{Q_1}\right]_p^* \left[\lambda^u_{Q_1}\right]_r \left[\lambda^u_{Q_1}\right]_r^* \nonumber\\
    &- 2 \left[\lambda^d_{Q_1}\right]_p \left[\lambda^d_{Q_1}\right]_p^* \left[\lambda^d_{Q_1}\right]_r \left[\lambda^d_{Q_1}\right]_r^* - \frac{19}{32} \left[\lambda_{T_1}\right]_p \left[\lambda_{T_1}\right]_p^* \left[\lambda_{T_1}\right]_r \left[\lambda_{T_1}\right]_r^*\,.
\end{align}

\subsection*{Model 6: \texorpdfstring{$T_1+Q_5$}{T1+Q5}}
In the equal mass limit, $M_{T_1} = M_{Q_5} = \Lambda$, we find
\begin{align}
    16\pi^2 \Lambda^2 \cC_{\phi\Box} = & - \frac{3}{8} |\lambda_{T_1Q_5}|^4 - \frac{15}{64} |\lambda_{T_1 Q_5}|^2 \left[ \lambda_{T_1} \right]_p \left[ \lambda_{T_1} \right]_p^* - \frac{1}{4}\left[ \lambda_{T_1} \right]_p \left[ \lambda_{T_1} \right]_p^*\left[ \lambda_{T_1} \right]_r \left[ \lambda_{T_1} \right]_r^* \nonumber\\
    & -\frac{3}{4} \left[ \lambda_{Q_5} \right]_p \left[ \lambda_{Q_5} \right]_p^* \left[ \lambda_{Q_5} \right]_r \left[ \lambda_{Q_5} \right]_r^* \,,
\end{align}
\begin{align}
    16\pi^2 \Lambda^2 \cC_{\phi D} = & -\frac{3}{4} |\lambda_{T_1Q_5}|^4 - \frac{15}{16} |\lambda_{T_1Q_5}|^2 \left( 2\left[ \lambda_{Q_5} \right]_p \left[ \lambda_{Q_5} \right]_p^* - \left[ \lambda_{T_1} \right]_p \left[ \lambda_{T_1} \right]_p^*\right) \nonumber\\
    & - \frac{19}{32} \left[ \lambda_{T_1} \right]_p \left[ \lambda_{T_1} \right]_p^*\left[ \lambda_{T_1} \right]_r \left[ \lambda_{T_1} \right]_r^* - 2 \left[ \lambda_{Q_5} \right]_p \left[ \lambda_{Q_5} \right]_p^* \left[ \lambda_{Q_5} \right]_R \left[ \lambda_{Q_5} \right]_r^*
\end{align}

\subsection*{Model 7: \texorpdfstring{$T_2+Q_1$}{T2+Q1}}
Setting $M_{Q_1} = M_{T_2} = \Lambda$, we find
\begin{align}
    16\pi^2 \Lambda^2 \cC_{\phi\Box} = & - \frac{3}{8} |\lambda_{T_2Q_1}|^4 + \frac{15}{64} |\lambda_{T_2Q_1}|^2 \left( 4\left[\lambda^d_{Q_1}\right]_p \left[\lambda^d_{Q_1}\right]_p^* - 3 \left[\lambda_{T_2}\right]_p\left[\lambda_{T_2}\right]_p^* \right) \nonumber\\
    &-\frac{3}{4}\left[\lambda^u_{Q_1}\right]_p \left[\lambda^u_{Q_1}\right]^*_p \left[\lambda^u_{Q_1}\right]_r \left[\lambda^u_{Q_1}\right]_r^* -\frac{3}{4}\left[\lambda^d_{Q_1}\right]_p \left[\lambda^d_{Q_1}\right]^*_p \left[\lambda^d_{Q_1}\right]_r \left[\lambda^d_{Q_1}\right]_r^* \nonumber\\
    & -2\left[\lambda^u_{Q_1}\right]_p \left[\lambda^u_{Q_1}\right]^*_p \left[\lambda^d_{Q_1}\right]_r \left[\lambda^d_{Q_1}\right]_r^* - \frac{1}{4} \left[\lambda_{T_2} \right]_p \left[\lambda_{T_2} \right]_p^* \left[\lambda_{T_2} \right]_r \left[\lambda_{T_2} \right]_r^* \,,
\end{align}
\begin{align}
    16 \pi^2 \Lambda^2 \cC_{\phi D} = & - \frac{3}{4}|\lambda_{T_2Q_1}|^4 + \frac{15}{16} |\lambda_{T_2Q_7}|^2 \left( 2\left[\lambda^d_{Q_1}\right]_p \left[\lambda^d_{Q_1}\right]_p^* - 2 \left[\lambda^u_{Q_1}\right]_p \left[\lambda^u_{Q_1}\right]_p^* - \left[\lambda_{T_2}\right]_p \left[\lambda_{T_2}\right]_p^* \right) \nonumber\\
    & - 2\left[\lambda^u_{Q_1}\right]_p \left[\lambda^u_{Q_1}\right]_p^* \left[\lambda^u_{Q_1}\right]_r \left[\lambda^u_{Q_1}\right]_r^* - 2\left[\lambda^d_{Q_1}\right]_p \left[\lambda^d_{Q_1}\right]_p^* \left[\lambda^d_{Q_1}\right]_r \left[\lambda^d_{Q_1}\right]_r^* \nonumber \\
    & + 4 \left[\lambda^u_{Q_1}\right]_p \left[\lambda^u_{Q_1}\right]_p^* \left[\lambda^d_{Q_1}\right]_r \left[\lambda^d_{Q_1}\right]_r^* - \frac{19}{32} \left[\lambda_{T_2}\right]_p \left[\lambda_{T_2}\right]_p^* \left[\lambda_{T_2}\right]_r \left[\lambda_{T_2}\right]_r^* \,.
\end{align}

\subsection*{Model 8: \texorpdfstring{$T_2 + Q_7$}{T2+Q7}}
Setting $M_{T_2} = M_{Q_7} = \Lambda$ we find
\begin{align}
    16 \pi^2 \Lambda^2 \cC_{\phi\Box}=& - \frac{3}{8} |\lambda_{T_2Q_7}|^4 - \frac{15}{64}|\lambda_{T_2Q_7}|^2 \left[\lambda_{Q_7} \right]_p \left[\lambda_{Q_7} \right]_p^* - \frac{1}{4} \left[\lambda_{T_2} \right]_p \left[\lambda_{T_2} \right]_p^* \left[\lambda_{T_2} \right]_r \left[\lambda_{T_2} \right]_r^* \nonumber\\
    & - \frac{3}{4} \left[\lambda_{Q_7} \right]_p \left[\lambda_{Q_7} \right]_p^* \left[\lambda_{Q_7} \right]_r \left[\lambda_{Q_7} \right]_r^* \,,
\end{align}
\begin{align}
    16 \pi^2 \Lambda^2 \cC_{\phi D} =& -\frac{3}{4} |\lambda_{T_2Q_7}|^4 - \frac{15}{16} |\lambda_{T_2Q_7}|^2 \left( 2 \left[\lambda_{Q_7} \right]_p \left[\lambda_{Q_7} \right]_p^* - \left[\lambda_{T_2} \right]_p \left[\lambda_{T_2} \right]_p^* \right)\nonumber\\
    & -\frac{19}{32} \left[\lambda_{T_2} \right]_p \left[\lambda_{T_2} \right]_p^* \left[\lambda_{T_2} \right]_r \left[\lambda_{T_2} \right]_r^* - 2 \left[\lambda_{Q_7} \right]_p \left[\lambda_{Q_7} \right]_p^* \left[\lambda_{Q_7} \right]_r \left[\lambda_{Q_7} \right]_r^*\,.
\end{align}
The matching coefficients when setting the VLQ-VLQ coupling to zero can also be found in Ref.~\cite{Crivellin:2022fdf}.
\section{Global fit}
\label{app:global fit}

\subsection{Electroweak fit}
\label{app:EWPO}

The vertex modifications of the $Z$-boson presented in Eq.~\eqref{eq:EW_eff_Lag} are
\begin{align}
\delta g_{L\,ij}^{Z\nu}=&-\frac{v^2}{2}\left([\cC_{\phi l}^{(1)}]_{ij}-[\cC_{\phi l}^{(3)}]_{ij}\right)+\delta^U(1/2,0)\, \delta_{ij}\,,\\
\delta g_{L\,ij}^{Ze}=&-\frac{v^2}{2}\left([\cC_{\phi l}^{(1)}]_{ij}+[\cC_{\phi l}^{(3)}]_{ij}\right)+\delta^U(-1/2,-1)\delta_{ij}\,,\\
\delta g_{R\,ij}^{Ze}=&\,-\frac{v^2}{2}[\cC_{\phi e}]_{ij}+\delta^U(0,-1)\, \delta_{ij}\,,\\
\delta g_{L\,ij}^{Zu}=&\,-\frac{v^2}{2}V_{ik} \left(
[\cC_{\phi q}^{(1)}]_{kl}-[\cC_{\phi q}^{(3)}]_{kl}
\right) V_{lj}^{\dagger}+\delta^U(1/2,2/3)\, \delta_{ij}\,,\\
\delta g_{R\,ij}^{Zu}=&\,-\frac{v^2}{2}[\cC_{\phi u}]_{ij}+\delta^U(0,2/3)\, \delta_{ij}\,,\\
\delta g_{L\,ij}^{Zd}=&-\frac{v^2}{2}\left([\cC_{\phi q}^{(1)}]_{ij}+[\cC_{\phi q}^{(3)}]_{ij}\right)+\delta^U(-1/2,-1/3)\, \delta_{ij}\,,\\
\delta g_{R\, ij}^{Zd}=&\,-\frac{v^2}{2}[\cC_{\phi d}]_{ij}+\delta^U(0,-1/3)\, \delta_{ij}\,,
\end{align}
with 
\begin{align}
\delta^U(T^3,Q)=&
-v^2\left(T^3+Q\frac{g_1^2}{g_2^2-g_1^2} \right)
\left(\frac{1}{4}\cC_{\phi D}+\frac{1}{2}[\cC^{(3)}_{\phi\ell}]_{22}+\frac{1}{2}[\cC^{(3)}_{\phi\ell}]_{11}-\frac{1}{4}[\cC_{\ell\ell}]_{1221}\right)
\nonumber\\
&-v^2Q\frac{g_2g_1}{g_2^2-g_1^2} \cC_{\phi WB}\,.
\end{align}
Similarly, the modification of the $W$ couplings read
\begin{align}
\delta g^{W\ell}_{ij}=&\delta g_{L\,ij}^{Z\nu}-\delta g_{L\, ij}^{Ze}\,, 
\label{eq:deltaWL}\\
\delta g^{Wq}_{ij}
=&\,\delta g_{L\,ik}^{Zu}V_{kj}-V_{ik}\delta g_{L\,kj}^{Zd}\,.
\end{align}
The variation of the observables, $\Delta O =O_{\rm NP}-O_{\rm SM}$, at leading order in the EW boson vertex modifications and $\delta m_W$ has been derived in Ref.~\cite{Allwicher:2023aql}. We use those expressions in the construction of $\chi^2_{\rm EWPO}$ in Eq.~\eqref{eq:chi2_EWPO}.

\subsection{Higgs fit}
\label{app:Hfit}
The total Higgs decay width in the SMEFT used in the construction of the Higgs fit based on the data from \cite{CMS:2022dwd} reads, introducing a factor of $1/3$ since the results of Ref.~\cite{Brivio:2019myy} were derived in the $U(3)^5$ flavour symmetric limit,
\begin{align}
    \Gamma_h^{SMEFT} =&\,\,\Gamma_h^{SM} \left(1 - 1.50v^2 \cC_{\phi B} - 1.21 v^2 \cC_{\phi W} + 1.21 v^2 \cC_{\phi WB} + 50.6 v^2 \cC_{\phi G} + 1.83 v^2 \cC_{\phi\Box}+ \right. \nonumber\\
    & \quad\quad\quad  - 0.43 v^2 \cC_{\phi D } + \frac{1}{3}\left( 0.002 v^2\cC_{\phi q}^{(1)} +  0.06 v^2 \cC^{(3)}_{\phi q} + 0.001v^2 \cC_{\phi u} - 0.0007 v^2\cC_{\phi d} \right)+ \nonumber\\
    & \quad\quad\quad \left. +2v^2 \left(\text{BR}(h\rightarrow\gamma\gamma) + \text{BR}(h\rightarrow gg) \right) \cC^h_{kin}\right) + \nonumber\\
    & \, + \Gamma^{\rm r}_d \kappa^2_d+ \Gamma_u^{\rm r}\kappa_u^2+ \Gamma_c^{SM} (\kappa_c^2-1) + \Gamma_s^{SM} (\kappa_s^2-1)  \,, \label{eq:higgsdecaywidthsmeft} 
\end{align}
where the last four terms take into account the modification to the Higgs decay into light quark pair. In particular, the first two terms $\Gamma_u^{\rm r}$ and $\Gamma^{\rm r}_d $ arise from the appropriate rescaling of the $h\rightarrow b\bar{b}$. Their numerical values are:
\begin{equation}
    \Gamma_d^{\rm r} = 6.73 \cdot 10^{-9}\,{\rm GeV} \,, \quad \Gamma_u^{\rm r} = 1.51\cdot10^{-9}\,{\rm GeV}\,.
\end{equation}
In the SM limit, where the Wilson coefficients are null and the coupling modifiers are $\kappa_{u,\,d}=1$, these two partial decay widths would be summed to the SM total width, but given their small values they are neglected. The presence of enhanced light quark Yukawa couplings makes their contribution increasingly relevant.

As for the contributions due to the decays of the Higgs boson into charm or strange pairs, taken to be
\begin{equation}
    \Gamma_c^{SM}=1.18\cdot 10^{-4}\text{ GeV}\,,\quad \Gamma_s^{SM}=1.0\cdot 10^{-6}\text{ GeV}\,,
\end{equation}
from Refs.~\cite{LHCHiggsCrossSectionWorkingGroup:2016ypw, HBRs}, they appear multiplied by $\kappa_{c,\,s}^2-1$. This accounts for the fact that they are already present in the SM decay width, hence in the SM limit they should not be added to $\Gamma^{SM}_h$.\\
As for the partial decay widths, those for the fermionic final states are ($\psi= b,\, \tau, \, \mu$)
\begin{equation}
    \Gamma^{SMEFT}(h\rightarrow \psi\bar{\psi})= \Gamma^{SM}(h\rightarrow \psi\bar\psi) \left(1-0.50v^2\cC_{\phi D} + 2v^2 \cC_{\phi\Box}\right)\,,    
\end{equation}
while those with a boson final state are
\begin{align}
    \Gamma^{SMEFT}(h\rightarrow \gamma\gamma)=& \Gamma^{SM}(h\rightarrow \gamma\gamma) \left(1 - 231 \,v^2 \, \cC_{\phi W} - 805\, v^2  \cC_{\phi B} + 431 \, v^2 \cC_{\phi WB} + \right.\nonumber\\
    &\left.\qquad -0.5\,v^2\cC_{\phi D} + 
 2\,v^2\cC_{\phi \Box}\right)\,, \\
    \Gamma^{SMEFT}(h\rightarrow W^+W^-) = &\Gamma^{SM}(h\rightarrow W^+W^-) \left( 1 + \frac{4}{3}\,v^2 \cC_{\phi q}^{(3)} \right) \,,\\
    \Gamma^{SMEFT}(h\rightarrow ZZ) = & \Gamma^{SM}(h\rightarrow ZZ) \left(1 + 0.46\,v^2 \cC_{\phi WB} - 0.07\, v^2 \cC_{\phi D} + \right.\nonumber \\
    &\qquad\left. +0.47\,v^2  \cC_{\phi q}^{(1)} + 1.61\, v^2 \cC_{\phi q}^{(3)} + 0.24 \, v^2 \cC_{\phi u} - 0.18\, v^2 \cC_{\phi d}\right)\,.
\end{align}

\subsection{Additional results}
\label{sec:add_results}

In Tabs.~\ref{tab:pulls_M1-M4} and~\ref{tab:pulls_M5-M8} we show the pulls with respect to the SM for the most influential EWPOs together with the Higgs signal strengths. Having $\Delta P_M$ negative (positive)
signals that Model M reduces (increases) the tension with the experiment compared to the SM for a given observable.

\begin{table}[t]
\centering
    \begin{tabular}{|c|c||c|c||c|c||c|c|}\hline
            M$_{1}^{1^{\rm st}}$ & $\Delta P_1$ & M$_{2}^{1^{\rm st}}$ & $\Delta P_2$ & M$_{3}^{1^{\rm st}}$ & $\Delta P_3$ & M$_{4}^{1^{\rm st}}$ & $\Delta P_4$ \\\hline\hline
            $m_W$  & $-1.01$ & $m_W$ & $-1.00$ & $m_W$ & $-0.99$ & $m_W$ & $-1.04$ \\
            $A_e$  & $-0.14$ & $A_e$ & $-0.14$ & $\Gamma_Z$ & $-0.22$ & $\Gamma_Z$ & $-0.41$ \\
            $R_\mu$  & $0.11$ & $A_{b}^{\rm FB}$ & $0.13$ & $A_e$ & $-0.18$ & $R_\mu$ & $-0.22$ \\
            $A_{b}^{\rm FB}$  & $0.13$ & $R_\mu$ & $0.14$ & $A_{b}^{\rm FB}$ & $0.17$ & $A_{b}^{\rm FB}$ & $0.17$ \\\hline
            $\mu_{\tau\tau}$  & $-0.50$ & $\mu_{WW}$ & $-0.31$ & $\mu_{\tau\tau}$ & $-0.49$ & $\mu_{WW}$ & $-0.18$ \\
            $\mu_{WW}$  & $-0.12$ & $\mu_{\tau\tau}$ & $-0.28$ & $\mu_{WW}$ & $-0.13$ & $\mu_{\tau\tau}$ & $-0.17$ \\ 
            $\mu_{ZZ}$  & $-0.09$ & $\mu_{ZZ}$ & $-0.25$ & $\mu_{ZZ}$ & $-0.10$ & $\mu_{ZZ}$ & $-0.13$ \\
            $\mu_{\mu\mu}$  & $0.11$ & $\mu_{\mu\mu}$ & $0.06$ & $\mu_{\mu\mu}$ & $0.11$ & $\mu_{\mu\mu}$ & $0.04$ \\
            $\mu_{bb}$  & $0.23$ & $\mu_{bb}$ & $0.13$ & $\mu_{bb}$ & $0.23$ & $\mu_{bb}$ & $0.08$ \\
            $\mu_{\gamma\gamma}$  & $0.37$ & $\mu_{\gamma\gamma}$ & $0.29$ & $\mu_{\gamma\gamma}$ & $0.36$ & $\mu_{\gamma\gamma}$ & $0.17$ \\\hline\hline
            M$_{1}^{2^{\rm nd}}$ & $\Delta P_1$ & M$_{2}^{2^{\rm nd}}$ & $\Delta P_2$ & M$_{3}^{2^{\rm nd}}$ & $\Delta P_3$ & M$_{4}^{2^{\rm nd}}$ & $\Delta P_4$ \\\hline\hline
            $m_W$  & $-0.89$ & $m_W$ & $-0.96$ & $m_W$ & $-1.01$ & $m_W$ & $-1.07$ \\
            $A_e$  & $-0.12$ & $R_\tau$ & $-0.15$ & $\Gamma_Z$ & $-0.50$ & $\Gamma_Z$ & $-0.52$ \\
            $\Gamma_Z$  & $-0.10$ & $\sigma_{\rm had}$ & $0.17$ & $R_\mu$ & $-0.30$ & $R_\mu$ & $-0.31$ \\
            $A_{b}^{\rm FB}$  & $0.12$ & $R_\mu$ & $0.20$ & $R_\tau$ & $0.23$ & $R_\tau$ & $0.23$ \\\hline
            $\mu_{\tau\tau}$  & $-0.44$ & $\mu_{WW}$ & $-0.11$ & $\mu_{WW}$ & $-0.15$ & $\mu_{\gamma\gamma}$ & $-0.11$ \\
            $\mu_{WW}$  & $-0.18$ & $\mu_{\tau\tau}$ & $-0.10$ & $\mu_{\tau\tau}$ & $-0.14$ & $\mu_{bb}$ & $-0.05$ \\ 
            $\mu_{ZZ}$  & $-0.14$ & $\mu_{ZZ}$ & $-0.10$ & $\mu_{ZZ}$ & $-0.10$ & $\mu_{\mu\mu}$ & $-0.02$ \\
            $\mu_{\mu\mu}$  & $0.10$ & $\mu_{\mu\mu}$ & $0.02$ & $\mu_{\mu\mu}$ & $0.03$ & $\mu_{\tau\tau}$ & $0.10$ \\
            $\mu_{bb}$  & $0.21$ & $\mu_{bb}$ & $0.05$ & $\mu_{\gamma\gamma}$ & $0.06$ & $\mu_{ZZ}$ & $0.12$ \\
            $\mu_{\gamma\gamma}$  & $0.30$ & $\mu_{\gamma\gamma}$ & $0.10$ & $\mu_{bb}$ & $0.07$ & $\mu_{WW}$ & $0.12$ \\\hline
    \end{tabular}
    \caption{Pulls in Models 1-4 for the selected observables: EWPOs with the most influence on the global fit and all Higgs signal strengths. For example, M$_{1}^{1^{\rm st}}$ (M$_{1}^{2^{\rm nd}}$) means Model 1 with vector-like quarks coupled to the first (second) generation. Additionally, $\Delta P_{\rm M} = |P_{\rm M}|-|P_{\rm SM}|$.}
    \label{tab:pulls_M1-M4}
\end{table}

\begin{table}[t]
\centering
    \begin{tabular}{|c|c||c|c||c|c||c|c|}\hline
            M$_{5}^{1^{\rm st}}$ & $\Delta P_5$ & M$_{6}^{1^{\rm st}}$ & $\Delta P_6$ & M$_{7}^{1^{\rm st}}$ & $\Delta P_7$ & M$_{8}^{1^{\rm st}}$ & $\Delta P_8$ \\\hline
             $m_W$ & $-1.02$ & $m_W$ & $-1.14$ & $m_W$ & $-0.92$ & $m_W$ & $-1.11$ \\
             $A_e$ & $-0.15$ & $\Gamma_Z$ & $-0.36$ & $R_\tau$ & $-0.15$ & $\Gamma_Z$ & $-0.32$ \\
             $A_b^{\rm FB}$ & $0.14$ & $A_e$ & $-0.16$ & $\sigma_{\rm had}$ & $0.18$ & $A_e$ & $-0.12$ \\
             $R_\mu$ & $0.15$ & $A_b^{\rm FB}$ & $0.15$ & $R_\mu$ & $0.20$ & $A_b^{\rm FB}$ & $0.11$ \\\hline
             $\mu_{\tau\tau}$ & $-0.34$ & $\mu_{\tau\tau}$ & $-0.57$ & $\mu_{\tau\tau}$ & $-0.45$ & $\mu_{\tau\tau}$ & $-0.84$ \\
             $\mu_{WW}$ & $-0.29$ & $\mu_{ZZ}$ & $-0.04$ & $\mu_{WW}$ & $-0.16$ & $\mu_{\gamma\gamma}$ & $-0.39$ \\
             $\mu_{ZZ}$ & $-0.21$ & $\mu_{WW}$ & $-0.03$ & $\mu_{ZZ}$ & $-0.12$ & $\mu_{\mu\mu}$ & $0.19$ \\
             $\mu_{\mu\mu}$ & $0.08$ & $\mu_{\gamma\gamma}$ & $0.00$ & $\mu_{\mu\mu}$ & $0.10$ & $\mu_{ZZ}$ & $0.19$ \\
             $\mu_{bb}$ & $0.16$ & $\mu_{\mu\mu}$ & $0.13$ & $\mu_{bb}$ & $0.21$ & $\mu_{WW}$ & $0.26$ \\
             $\mu_{\gamma\gamma}$ & $0.26$ & $\mu_{bb}$ & $0.26$ & $\mu_{\gamma\gamma}$ & $0.44$ & $\mu_{bb}$ & $0.39$ \\\hline\hline
            M$_{5}^{2^{\rm nd}}$ & $\Delta P_5$ & M$_{6}^{2^{\rm nd}}$ & $\Delta P_6$ & M$_{7}^{2^{\rm nd}}$ & $\Delta P_7$ & M$_{8}^{2^{\rm nd}}$ & $\Delta P_8$ \\\hline
            $m_W$  & $-0.96 $& $m_W$ & $-1.15$ & $m_W$ & $-0.82$ & $m_W$ & $-1.12$ \\
            $R_\tau$  & $-0.16$ & $\Gamma_Z$ & $-0.47 $& $R_\tau$ & $-0.31$ & $\Gamma_Z$ & $-0.52 $\\
            $\sigma_{\rm had}$  & 0.18 & $R_\mu$ &$ -0.25$ & $\Gamma_Z$ & 0.37 & $R_\mu$ & $-0.28$ \\
            $R_\mu$  & 0.21 & $A_b^{\rm FB}$ & 0.16 & $R_\mu$ & 0.42 & $R_\tau$ & 0.21 \\\hline
             $\mu_{WW}$ & $-0.14$ & $\mu_{WW}$ & $-0.23$ & $\mu_{\gamma\gamma}$ &$ -0.26$ & $\mu_{\tau\tau}$ & $-0.50$ \\
             $\mu_{ZZ}$ & $-0.13$ & $\mu_{\tau\tau}$ &$ -0.20$  & $\mu_{bb}$ & $-0.11$ & $\mu_{\gamma\gamma}$ & $-0.48 $\\
             $\mu_{\tau\tau}$ & $-0.13$ & $\mu_{\gamma\gamma}$ & $-0.20$ & $\mu_{\mu\mu}$ & $-0.05$ & $\mu_{WW}$ & $-0.11$ \\
             $\mu_{\mu\mu}$ & 0.03 & $\mu_{ZZ}$ & $-0.16$ & $\mu_{ZZ}$ & 0.17 & $\mu_{ZZ}$ & $-0.10$ \\
             $\mu_{bb}$ & 0.06 & $\mu_{\mu\mu}$ & 0.05 & $\mu_{\tau\tau}$ & 0.24 & $\mu_{\mu\mu}$ &  0.11 \\
             $\mu_{\gamma\gamma}$ & 0.06 & $\mu_{bb}$ & 0.09 & $\mu_{WW}$ & 0.27 & $\mu_{bb}$ & 0.23\\\hline
    \end{tabular}
    \caption{Pulls in Models 5-8 for the selected observables: EWPOs with the most influence on the global fit and all Higgs signal strengths. For example, M$_{5}^{1^{\rm st}}$ (M$_{5}^{2^{\rm nd}}$) means Model 5 with vector-like quarks coupled to the first (second) generation. Additionally, $\Delta P_{\rm M} = |P_{\rm M}|-|P_{\rm SM}|$.}
    \label{tab:pulls_M5-M8}
\end{table}

\section{Generalised Branching Ratios}
\label{app:GeneralisedBR}

\begin{table}[t]
    \centering
    \renewcommand{\arraystretch}{1.5}
    \begin{tabular}{|cc|ccccccc|}
    \hline
     &    & $U$ & $D$ & $ Q_1 = \begin{pmatrix} T \\ B\end{pmatrix}$ & $Q_5 = \begin{pmatrix} B \\ Y\end{pmatrix}$ & $Q_7 = \begin{pmatrix} X \\ T\end{pmatrix}$ & $T_1 = \begin{pmatrix} T_1^1 \\ T_1^2 \\ T_1^3 \end{pmatrix}$ & $T_2 = \begin{pmatrix} T_2^1 \\ T_2^2 \\ T_2^3\end{pmatrix}$ \\
    \hline
    \multirow{4}{*}{\rotatebox[origin=c]{90}{Electric Charges}} & $-4/3$ & - & - & - &$Y$& - & $T_1^Y = \frac{T_1^1 +i T_1^2}{\sqrt{2}}$ & - \\
    &$-1/3$ & - &$D$&$B$&$B$& - & $T_1^B = T_1^3$ & $T^B_2 = \frac{T_2^1 + iT_2^2}{\sqrt{2}}$ \\
    &$+2/3$ &$U$& - &$T$& - &$T$& $T_1^T = \frac{T_1^1 - i T_1^2}{\sqrt{2}}$ & $T_2^T = T_2^3$ \\
    &$+5/3$ & - & - & - & - &$X$& - & $T_2^X = \frac{T_2^1 - i T_2^2}{\sqrt{2}}$ \\
    \hline
    \end{tabular}
    \caption{Electric charges for the VLQ components along with their definitions. The absence of a component with the assigned electric charge is denoted with -. }
    \label{tab:VLQcomponentcharges}
\end{table}
The derivation of the branching ratios for the VLQ decays into $W^\pm q$, $hq$ and $Zq$ can be completely generalised for all the eight models we considered. 
The goal of this is to obtain general expressions for the decay widths (and hence the branching ratios) and to provide appropriate substitution rules so that the results of a specific model can be uniquely recovered. The starting point is the NP Lagrangian in the broken phase, separated into sectors as
\begin{equation}
    \mathcal{L} \supset \mathcal{L}^q_\text{mass} + \mathcal{L}^q_h + \mathcal{L}^q_{W} + \mathcal{L}^q_Z\,.
\end{equation}
Starting from the SM Yukawa sector for quarks and the NP interaction terms in the broken phase, the terms involving the vev $v$ and the physical $h$ are gathered in $\mathcal{L}^q_\text{mass}$ and $\mathcal{L}^q_h$ respectively. The VLQ mass terms are contained in $\mathcal{L}^q_\text{mass}$. The other two terms describe the interactions of the quarks (both VLQ and SMQ) with the $W^\pm$ ($\mathcal{L}^q_{W}$) and the $Z$ ($\mathcal{L}^q_{Z}$) bosons.\\
The generalisation is performed in steps:
\begin{enumerate}
    \item Organisation of the quarks in $\mathcal{L}^q_\text{mass}$ with the same electric charge into multiplets. Auxiliary functions are introduced in the other three sectors to recover the same multiplet organisation. The Lagrangian terms $\mathcal{L}^q_h$, $\mathcal{L}^q_W$ and $\mathcal{L}^q_Z$ hence all contain an interaction matrix $C$ between two quark multiplets;
    \item Identification of the generalised mass matrices in $\mathcal{L}^q_\text{mass}$. These matrices are perturbatively diagonalised up to $\mathcal{O}(v^2/\Lambda^2)$ corrections, which, for our choice of $\Lambda=1.6$ TeV, corresponds to $\approx2\%$ corrections;
    \item Rotation in all four sectors into the multiplet basis where the mass matrices are diagonal. The rotated interaction matrices in $\mathcal{L}^q_h$, $\mathcal{L}^q_W$ and $\mathcal{L}^q_Z$ provide the necessary couplings for the calculation of the decay widths. 
\end{enumerate}
Given the presence of four possible electric charges for the VLQ components as shown in Tab.~\ref{tab:VLQcomponentcharges}, four same-charge multiplets can be constructed:
\begin{align}
    \mathcal{U} = \begin{pmatrix} u \\ T \\ S_u \end{pmatrix}\,, \quad \mathcal{D} = \begin{pmatrix} d \\ B \\ S_d\end{pmatrix} \,, \quad \mathcal{X} = \begin{pmatrix} X \\ T_2^{X}\end{pmatrix}\,, \quad \mathcal{Y} = \begin{pmatrix} Y \\ T_1^Y\end{pmatrix}\,.
\end{align}
Noticing that no model introduces both a triplet VLQ and a singlet VLQ, the terms $S_u$ and $S_d$ are defined to be
\begin{equation}
	S_u = \begin{cases}
		U \quad\,\,\, \text{Models } 1,\,3 \\
		T_1^T \quad \text{Models } 5,\,6\\
		T_2^T \quad \text{Models } 7,\,8 
	\end{cases} \, , \qquad S_d = \begin{cases}
	D \quad\,\,\, \text{Models } 2,\,4 \\
	T_1^B \quad \text{Models } 5,\,6\\
	T_2^B \quad \text{Models } 7,\,8 
	\end{cases} \,.
\end{equation}
In the models that are not mentioned the terms do not appear and either $\mathcal{U}$ or $\mathcal{D}$ contains only two quark components. \\
The Lagrangian sector that allows to identify the generalised mass matrices reads
 \begin{align}
 	-\mathcal{L}^q_\text{mass} =\, \bar{\mathcal{U}}_L M_\mathcal{U} \mathcal{U}_R + \bar{\mathcal{D}}_L M_\mathcal{D}\mathcal{D}_R + \bar{\mathcal{X}}_L M_\mathcal{X}\mathcal{X}_R + \bar{\mathcal{Y}}_L M_\mathcal{Y} \mathcal{Y}_R + \text{h.c.}\,.
 \end{align}
The mass matrices contain auxiliary couplings to achieve generalisation. The replacement rules for the auxiliary couplings are defined in Tab.~\ref{tab:auxcouplings_replacements}, and the expressions of the mass matrices are
 \begin{align}
 	M_\mathcal{U} &= \begin{pmatrix}
 		y_u \frac{v}{\sqrt{2}}   &   0   &   \lambda_{S_u} \frac{v}{\sqrt{2}} \\
 		\lambda^u_{Q} \frac{v}{\sqrt{2}} & \Lambda & \lambda_{S_uQ} \frac{v}{\sqrt{2}} \\
 		0       &       0   &   \Lambda\\
 	\end{pmatrix}\,,\qquad  M_\mathcal{X}= \begin{pmatrix}
 		\Lambda   &   \lambda_{XQ} \frac{v}{\sqrt{2}} \\
 		0   &   \Lambda   \\
 	\end{pmatrix}\,,\label{eq:massmatrices_positivecharge}\\
 	M_\mathcal{D} &= \begin{pmatrix}
 		y_d \frac{v}{\sqrt{2}}   &   0   &   \lambda_{S_d} \frac{v}{\sqrt{2}} \\
 		\lambda^d_{Q} \frac{v}{\sqrt{2}} & \Lambda & \lambda_{S_dQ} \frac{v}{\sqrt{2}} \\
 		0       &       0   &   \Lambda\\
 	\end{pmatrix} \,,\qquad M_\mathcal{Y} = \begin{pmatrix}
 		\Lambda   &   \lambda_{YQ} \frac{v}{\sqrt{2}} \\
 		0   &   \Lambda   \\
 	\end{pmatrix}\,. \label{eq:massmatrices_negativecharge}
 \end{align}
The Yukawa couplings $y_u$ and $y_d$ are henceforth set to zero.\\
As for the remaining three sectors, keeping only the terms that are associated with the decay of a VLQ into a SMQ plus a boson, they are described by
\begin{align}
    -\mathcal{L}_h^q =& \, h\,\bar{\mathcal{U}}_L \,C^h_\mathcal{U} \,\mathcal{U}_R + h\,\bar{\mathcal{D}}_L \, C^h_\mathcal{D} \, \mathcal{D}_R + \text{h.c.}\,,\\
    -\frac{\sqrt{2}}{g_L}\mathcal{L}_W^q =& \, \bar{U}_L \slashed{W}^+ \,C^W_{1L}\, \,\mathcal{D}_L + \bar{U}_R \slashed{W}^+ \,C^W_{1R}\, \,\mathcal{D}_R + \bar{\mathcal{X}}\,\slashed{W}^+\, C^W_2\,\mathcal{U} + \bar{\mathcal{D}}\, \slashed{W}^+\,C^W_3 \, \mathcal{Y} + \text{h.c.}\,, \label{eq:generalisedWsector}\\
    -\frac{c_W}{g_L}\mathcal{L}^q_Z =& \, \bar{\mathcal{U}}_L \slashed{Z}\left(\frac{1}{2}\,C^Z_{UL} - \frac{2}{3}s^2_W   \right) \mathcal{U}_L +  \bar{\mathcal{U}}_R \slashed{Z}\left(\frac{1}{2}\,C^Z_{UR} - \frac{2}{3}s^2_W   \right) \mathcal{U}_R + \nonumber  \\
    &+ \bar{\mathcal{D}}_L \slashed{Z}\left(-\frac{1}{2}\,C^Z_{DL} + \frac{1}{3}s^2_W   \right) \mathcal{D}_L + \bar{\mathcal{D}}_R \slashed{Z}\left(-\frac{1}{2}\,C^Z_{DR} + \frac{1}{3}s^2_W   \right) \mathcal{D}_R\,.
\end{align}
The last two terms of Eq.~\eqref{eq:generalisedWsector} have been written in a compact notation without separating the left-handed and right-handed components. This allowed by the fact that the interaction matrices are the same for both cases.
\begin{table}[t]
    \centering
    \renewcommand{\arraystretch}{1.4}
    \begin{tabular}{|c|cccccccc|}
    \hline
    Model & 1 & 2 & 3 & 4 & 5 & 6 & 7 & 8 \\
    \hline
    $\lambda_{S_u}$ & $\lU$ & 0 & $\lU$ & 0 & $\frac{1}{\sqrt{2}}\lambda_{T_1}$ & $\frac{1}{\sqrt{2}}\lambda_{T_1}$ & $\frac{1}{2}\lambda_{T_2}$ & $\frac{1}{2}\lambda_{T_2}$ \\
    $\lambda^u_Q$ & $\luQone$ & $\luQone$ & $\lambda_{Q_7}$ & 0 & $\luQone$ & 0 & $\luQone$ & $\lambda_{Q_7}$ \\
    $\lambda_{S_u Q}$ & $\lUQone$ & 0 & $\lambda_{UQ_7}$ & 0 & $\frac{1}{\sqrt{2}}\lambda_{T_1Q_1}$ & 0 & $\frac{1}{2} \lambda_{T_2Q_1}$ & $-\frac{1}{2}\lambda_{T_2 Q_7}$\\
    $\lambda_{XQ}$ & 0 & 0 & 0 & 0 & 0 & 0 & 0 & $\frac{1}{\sqrt{2}}\lambda_{T_2Q_7}$ \\
    $\lambda_{S_d}$ & 0 & $\lambda_D$ & 0 & $\lambda_D$ & -$\frac{1}{2}\lambda_{T_1}$ & $-\frac{1}{2}\lambda_{T_1}$ & $\frac{1}{\sqrt{2}}\lambda_{T_2}$ & $\frac{1}{\sqrt{2}}\lambda_{T_2}$ \\
    $\lambda^d_Q$ & $\ldQone$ & $\ldQone$ & 0 & $\lambda_{Q_5}$ & $\ldQone$ & $\lambda_{Q_5}$ & $\ldQone$ & 0 \\
    $\lambda_{S_dQ}$ & 0 & $\lambda_{DQ_1}$ & 0 & $\lambda_{DQ_5}$ & $-\frac{1}{2}\lambda_{T_1Q_1}$ & $\frac{1}{2}\lambda_{T_1Q_5}$ & $\frac{1}{\sqrt{2}}\lambda_{T_2Q_1}$ & 0 \\
    $\lambda_{YQ}$ & 0 & 0 & 0 & 0 & 0 & $\frac{1}{\sqrt{2}}\lambda_{T_1Q_5}$ & 0 & 0 \\
    \hline
    \end{tabular}
    \caption{Replacement rules for the VLQ couplings according to the specific model that is being considered.}
    \label{tab:auxcouplings_replacements}
\end{table}
The interaction matrices are defined as
\begin{equation}
    C^h_\alpha = \frac{\partial}{\partial v}M_\alpha \,,
\end{equation}
\begin{align}
    C^W_{1L} &= \begin{pmatrix}
        1 & 0 & 0 \\
        0 & 1 & 0 \\
        0 & 0 & \sqrt{2}f
    \end{pmatrix} \,,  && C^W_{1R} = \begin{pmatrix}
        0 & 0 & 0 \\
        0 & 1 & 0 \\
        0 & 0 & \sqrt{2}f
    \end{pmatrix} \,,\\
    C_2^W &= \begin{pmatrix}
        0 & 1 & 0 \\
        0 & 0 & -\sqrt{2}f\Theta(T_2) 
    \end{pmatrix}\,, && C_3^W = \begin{pmatrix}
        0 & 0 \\
        1 & 0 \\
        0 & \sqrt{2}f \Theta(T_1)
    \end{pmatrix} \,,
\end{align}
\begin{align}
    C^Z_{UL}  &=  \begin{pmatrix}
        1 & 0 & 0 \\
        0 & x & 0 \\
        0 & 0 & 2 \Theta(T_1)  
    \end{pmatrix} \,,  &&C^Z_{UR} = \begin{pmatrix}
        0 & 0 & 0 \\
        0 & x & 0 \\
        0 & 0 & 2 \Theta(T_1)  
    \end{pmatrix} \,, \\
    C_{DL}^Z &= \begin{pmatrix}
        1 & 0 & 0 \\
        0 & x & 0 \\
        0 & 0 & 2 \Theta(T_2)
    \end{pmatrix} \,,  &&C_{DR}^Z = \begin{pmatrix}
        0 & 0 & 0 \\
        0 & x & 0 \\
        0 & 0 & 2\Theta(T_2)
    \end{pmatrix} \,.
\end{align}
Three auxiliary functions have been introduced; their model-dependent expressions are presented in Tab.~\ref{tab:auxfunctions_replacements}, along with other auxiliary functions that have yet to be introduced. The first function, $f$, accounts for the fact that SU$(2)$ singlets do not interact with the $W^\pm$ bosons while the triplets do interact. The $\Theta$ function, instead, is equal to 1 only in the models where the VLQ between the parentheses is included, otherwise it is null. Finally, $x$ accounts for the doublet structure. Indeed, one can observe that $T$ and $B$ pick up different signs after the action of $\sigma_3$ depending on the multiplet they appear in, as shown below
\begin{equation}
    \sigma_3 Q_1 = \begin{pmatrix}
        T\\ -B
    \end{pmatrix}\,,\quad \sigma_3 Q_5 = \begin{pmatrix}
        B \\ -Y
    \end{pmatrix}\,,\quad \sigma_3 Q_7 = \begin{pmatrix}
        X \\ -T
    \end{pmatrix}\,.
\end{equation}
At this point, the second and third steps can be performed: the multiplet rotations that allow to perturbatively put the matrices in Eqs.~\eqref{eq:massmatrices_positivecharge}--\eqref{eq:massmatrices_negativecharge} into diagonal form are performed also in the $\mathcal{L}^q_h$, $\mathcal{L}^q_W$ and $\mathcal{L}^q_Z$ sectors, hence rotating the coupling matrices $C$ into $\tilde{C}$. \\
The rotation of the charge $+2/3$ multiplet is defined as follows, separating the rotation $U$ involving the VLQ-VLQ coupling from the rotation $V$ involving the SMQ-VLQ coupling,
\begin{equation}
    \mathcal{U}^\prime_L = U_{Lu} \, V_{Lu} \, \mathcal{U}_L\, , \quad \mathcal{U}^\prime_R = U_{Ru} \, V_{Ru} \, \mathcal{U}_R\,.
\end{equation}
The analogous transformations for the $-1/3$ charged multiplet are obtained from the previous by replacing $u$ with $d$. As for the multiplets with non-SM charges, the transformations read (replacing $\mathcal{X}$ with $\mathcal{Y}$):
\begin{equation}
    \mathcal{X}^\prime_L = U_{L\mathcal{X}} \, \mathcal{X}_L\,,\quad \mathcal{X}^\prime_R = U_{R\mathcal{X}} \, \mathcal{X}_R \,.
\end{equation}
The rotation matrices for the SM-charged multiplets, introduce further auxiliary functions $aux$ to guarantee that, in the absence of VLQs, the rotation matrices reduce to the identity, read ($\alpha= u,\,d$)
\begin{align}
    &U_{L\,\alpha} = \begin{pmatrix}
        1 & 0 & 0 \\  
        0 & \frac{aux_\alpha}{\sqrt{2}}-\frac{\lambda_{S\alpha\, Q} v}{8 M} & -\frac{(2-aux_\alpha^2)}{\sqrt{2}} - \frac{\lambda_{S\alpha\, Q} v}{8 M}\\
        0 & \frac{(2-aux_\alpha^2)}{\sqrt{2}} + \frac{\lambda_{S\alpha\, Q} v}{8 M} & \frac{aux_\alpha}{\sqrt{2}}-\frac{\lambda_{S\alpha\, Q} v}{8 M} \\ \end{pmatrix} \,, &&V_{L\alpha} = \begin{pmatrix}
        1 & 0 & -\frac{\lambda_{S_\alpha}\,v}{\sqrt{2}\Lambda} \\
        0 & 1 & 0 \\
        \frac{\lambda_\alpha\,v}{\sqrt{2} \Lambda} & 0 & 1\\
    \end{pmatrix}\,, \\
    &U_{R\,\alpha} =\begin{pmatrix}
        1 & 0 & 0 \\
        0 & \frac{aux_\alpha}{\sqrt{2}} + \frac{\lambda_{S\alpha \, Q} v}{8 M} & -\frac{(2-aux_\alpha^2)}{\sqrt{2}} + \frac{\lambda_{S\alpha \, Q} v}{8 M}\\
        0 & \frac{(2-aux_\alpha^2)}{\sqrt{2}} - \frac{\lambda_{S\alpha \, Q} v}{8 M} & \frac{aux_\alpha}{\sqrt{2}} + \frac{\lambda_{S\alpha \, Q} v}{8 M} \\ \end{pmatrix} \,,
        && V_{R\,\alpha} = \begin{pmatrix}
            1 & -\frac{\lambda^\alpha_Q v}{\sqrt{2} M} & 0 \\
            \frac{\lambda^\alpha_Q v}{\sqrt{2} M} & 1 & 0 \\
            0 & 0 & 1 \\ \end{pmatrix}\,.
\end{align}
Instead, for the multiplets involving non-SM charged quarks, the $2\times2$ rotation matrices are defined as ($\beta=\mathcal{X}, \, \mathcal{Y}$)
\begin{align}
    & U_{L\beta} = \begin{pmatrix}
        \frac{aux_\beta}{\sqrt{2}} - \frac{\lambda_{S\beta \, Q} v}{8 M} & -\frac{(2-aux_\beta^2)}{\sqrt{2}} - \frac{\lambda_{S\beta \, Q} v}{8 M}\\
 \frac{(2-aux_\beta^2)}{\sqrt{2}} + \frac{\lambda_{S\beta \, Q} v}{8 M} & \frac{aux_\beta}{\sqrt{2}} - \frac{\lambda_{S\beta \, Q} v}{8 M}
    \end{pmatrix}\,, \\
    & U_{R\beta} = \begin{pmatrix}
         \frac{aux_\beta}{\sqrt{2}} + \frac{\lambda_{S\beta \, Q} v}{8 M} & -\frac{(2-aux_\beta^2)}{\sqrt{2}} + \frac{\lambda_{S\beta \, Q} v}{8 M}\\
 \frac{(2-aux_\beta^2)}{\sqrt{2}} - \frac{\lambda_{S\beta \, Q} v}{8 M} & \frac{aux_\beta}{\sqrt{2}} + \frac{\lambda_{S\beta \, Q} v}{8 M} \end{pmatrix} \,.
\end{align}
\begin{table}[t]
    \centering
    \renewcommand{\arraystretch}{1.4}
    \begin{tabular}{|c|cccccccc|}
    \hline
    Model & 1 & 2 & 3 & 4 & 5 & 6 & 7 & 8 \\
    \hline
    $aux_u$ & 1 & $\sqrt{2}$ & 1 & $\sqrt{2}$ & 1 & 1 & 1 & 1 \\
    $aux_d$ & $\sqrt{2}$ & 1 & $\sqrt{2}$ & 1 & 1 & 1 & 1 & 1 \\
    $aux_\mathcal{X}$ & $\sqrt{2}$ & $\sqrt{2}$ & $\sqrt{2}$ & $\sqrt{2}$ & $\sqrt{2}$ & $\sqrt{2}$ & $\sqrt{2}$ & 1 \\
    $aux_\mathcal{Y}$ & $\sqrt{2}$ & $\sqrt{2}$ & $\sqrt{2}$ & $\sqrt{2}$ & $\sqrt{2}$ & 1 & $\sqrt{2}$ & $\sqrt{2}$ \\
    $f$ & 0 & 0 & 0 & 0 & 1 & 1 & $-1$ & $-1$ \\
    $x$ & 1 & 1 & $-1$ & $-1$ & 1 & $-1$ & 1 & $-1$ \\
    $\Theta(T_1)$ & 0 & 0 & 0 & 0 & 1 & 1 & 0 & 0 \\
    $\Theta(T_2)$ & 0 & 0 & 0 & 0 & 0 & 0  & 1 & 1\\
    \hline
    \end{tabular}
    \caption{Model-dependent replacement rules for the auxiliary functions introduced during the generalisation.}
    \label{tab:auxfunctions_replacements}
\end{table}
The explicit expressions for the rotated interaction matrices $\tilde{C}$ are not reported here, as the presence of auxiliary couplings and parameters makes the expressions cumbersome. Aside from the length, there are no further difficulties: each entry of the rotated matrices $\tilde{C}$ is associated with an interaction vertex. The leading contributions to the decay widths of the three channels read
\begin{align}
    \Gamma_{VLQ \rightarrow qh} = \frac{1}{32\pi} \left( \sum_i \left[\tilde{C}^h\right]^2_i  \right) \Lambda\,, \label{eq:generaliseddecaywidth_h}\\
    \Gamma_{VLQ \rightarrow qZ} = \frac{1}{8\pi} \left(\sum_i \frac{1}{4} \left[\tilde{C}^Z\right]^2_i \right)\frac{\Lambda^3}{v^2} \,, \label{eq:generaliseddecaywidth_Z}\\
    \Gamma_{VLQ \rightarrow qW} = \frac{1}{16\pi} \left(\sum_i \left[\tilde{C}^W\right]^2_i \right) \frac{\Lambda^3}{v^2}\,. \label{eq:generaliseddecaywidth_W}
\end{align}
The sum over $i$ accounts for all possible contributions to the given process, considering both the left-handed and right-handed interaction matrices. Once Eqs.~\eqref{eq:generaliseddecaywidth_h}--\eqref{eq:generaliseddecaywidth_W} are known in general, the replacement rules in Tabs. \ref{tab:auxcouplings_replacements}--\ref{tab:auxfunctions_replacements} allow to obtain the specific model's prediction.

\newpage
\bibliographystyle{utphys.bst}
\bibliography{bibliography}
\end{document}